\documentclass[12pt]{article}
\usepackage{setspace}
\usepackage{graphicx}
\usepackage{lscape}
\usepackage{pdflscape}
\usepackage{afterpage}
\usepackage{pdfpages}
\usepackage{float}
\usepackage{booktabs}
\usepackage[utf8]{inputenc}
\usepackage{indentfirst}
\usepackage{arydshln}
\usepackage{arydshln}
\usepackage{tikz}
\usepackage{lipsum}
\usepackage{pgffor}
\usepackage{dsfont}
\usepackage{amsfonts}
\usepackage{amsmath}
\allowdisplaybreaks
\usepackage{mathtools}

\usepackage{xfrac}
\usepackage{amssymb}
\usepackage{bigints}
\usepackage{graphicx}
\usepackage{url}				
\usepackage{caption}
\usepackage{subcaption} 
\usepackage{enumitem}
\usepackage{relsize,exscale}
\usepackage{multirow}
\usepackage{natbib}
\setcitestyle{authoryear, open={(},close={)}}
\usepackage{datetime}

\newdateformat{monthyeardate}{%
	\monthname[\THEMONTH] \THEYEAR}

\usepackage{titlesec}
\titleformat*{\section}{\large\bfseries}
\titleformat*{\subsection}{\large\bfseries}


\newcounter{parentnumber}
\usepackage{theorem}

\newtheorem{algorithm}{Algorithm}
\newtheorem{assumption}{Assumption}

\newtheorem{definition}{Definition}

\newtheorem{Lemma}{Lemma}

\newtheorem{proposition}{Proposition}

\newtheorem{remark}{Remark}

\newenvironment{proof}[1][Proof]{\noindent \textbf{#1.} }{\  \rule{0.5em}{0.5em}}

\providecommand{\U}[1]{\protect\rule{.1in}{.1in}}

\usepackage{hyperref} 
\hypersetup{
	colorlinks=true, breaklinks=true, bookmarks=true,bookmarksnumbered,
	urlcolor=blue, linkcolor=blue, citecolor=blue, 
	pdftitle={}, 
	pdfauthor={\textcopyright}, 
	pdfsubject={}, 
	pdfkeywords={}, 
	pdfcreator={pdfLaTeX}, 
	pdfproducer={LaTeX with hyperref and ClassicThesis} 
}

\begin{document}
	\setstretch{1}
	\title{{\Large Potato Potahto in the FAO-GAEZ Productivity Measures? \\ Nonclassical Measurement Error with Multiple Proxies\thanks{We would like to thank Bruno Ferman, Lucas Finamor, John Eric Humphries, Toru Kitagawa, Helena Laneuville, Daniel L. Millimet, Marcelo Moreira, Pierluca Pannella and Marcel Ribeiro for helpful suggestions. We also thank seminar participants at Insper, Sao Paulo School of Economics - FGV, University of Sao Paulo, University of Los Andes - Santiago, University of Los Andes - Bogotá, Bristol University and University of Warwick. We are grateful to Joana Getlinger for providing excellent research assistance.}}}

	\author{
		Rafael Araujo\thanks{Sao Paulo School of Economics - FGV. Email: \href{mailto:rafael.araujo@fgv.br}{rafael.araujo@fgv.br}}  \and Vitor Possebom\thanks{Sao Paulo School of Economics - FGV. Email: \href{mailto:vitor.possebom@fgv.br}{vitor.possebom@fgv.br}. This study was financed, in part, by the São Paulo Research Foundation (FAPESP), Brazil. Process Number \#2025/04857-0.}
	}
	\date{}

	\maketitle

	\newsavebox{\tablebox} \newlength{\tableboxwidth}


	\begin{center}

		First Draft: February 2025; This Draft: \monthyeardate\today


		%
		%
		\href{https://sites.google.com/site/vitorapossebom/working-papers}{Please click here for the most recent version}

		\

		\large{\textbf{Abstract}}
	\end{center}

	The FAO-GAEZ productivity data are widely used in Economics. However, the empirical literature rarely discusses measurement error. We use two proxies to derive analytical bounds around the effect of agricultural productivity in a setting with nonclassical measurement error. These bounds rely on assumptions weaker than those imposed in empirical studies and exhaust the information contained in the first two data moments. We reevaluate three influential studies, finding wide intervals around the effects of agricultural productivity. These results call for caution, highlighting the limits of our knowledge about these effects. Our methodology has broad applications in empirical research involving mismeasured variables.

	\



	\newpage

	\doublespacing

	\section{Introduction}\label{Sintro}

	The impact of agricultural productivity on economic outcomes is paramountly important in many fields, including Economic History \citep{nunn2011}, Development Economics \citep{bustos2016agricultural}, and Political Economy \citep{acharya2016political}. The most common measure of agricultural productivity is the Food and Agriculture Organization Global Agro-Ecological Zones (FAO-GAEZ) crop productivity measures. However, this measure is based on a prediction model and, as with any forecasting model, is subject to measurement error \citep{Rhode2024}. This type of complex challenge is acknowledged by FAO, which frequently updates its model and has done so four times \citep{Fischer2021GAEZ}. However, the existence of measurement error in the FAO-GAEZ productivity data is rarely recognized in the empirical literature and is not properly addressed by current statistical methods.



	To address this issue, we propose a novel analytical method that uses multiple proxies to partially identify the effect of agricultural productivity, accounting for measurement error. To illustrate the empirical relevance of our approach, we revisit three influential studies: \citet{nunn2011}, \citet{bustos2016agricultural}, and \citet{acharya2016political}. Applying our method with two available proxies, we show that once researchers relax the (often implicit) no-measurement-error assumption, the implied magnitudes of agricultural productivity effects can change dramatically, calling for substantial caution in interpreting existing results and implying that much of what we think we know about the magnitude of these effects is far more uncertain than previously recognized. Focusing on our lower bound and accounting for sample uncertainty, we find (i) that the data are consistent with an effect of agricultural productivity on population that is smaller than that reported by \citet{nunn2011}; (ii) that the data are consistent with an effect of soy productivity on manufacture employment that is smaller than that reported by \citet{bustos2016agricultural}; and (iii) that the data are consistent with no effect of cotton productivity on political behavior, in contrast to the findings of \citet{acharya2016political}. When we focus on our estimated upper bounds we find that the model uncertainty implied by the data is large in all three studies.


	We start by documenting the possibility of nonclassical measurement error in the FAO-GAEZ productivity variables. The FAO-GAEZ project combines data on climate, soil, and plant growth parameters to predict potential productivity of various crops. However, these measures of agricultural productivity face challenges of model and data uncertainty. For example, FAO-GAEZ's versions 3 and 4 differ with respect to calibrated parameters (e.g., the ability of different plants to capture sunlight) and sources of soil and climate data. These differences cause large changes in measured productivity for many crops and regions. The discrepancies between versions 3 and 4 rule out the possibility that both are measured without error. Consequently, treating either version as error-free amounts to imposing an untestable assumption about the data-generating process, one that our approach explicitly relaxes.\footnote{Additionally, nonclassical measurement error would be present in the following scenario: low-resource countries may install weather stations only in richer areas whose climate is appropriate to agriculture, implying that, in these countries, measured climate variables will be biased towards values that are more beneficial to agriculture productivity. Consequently, measurement error will tend to be positive in countries with low true productivity.}


	Motivated by our assessment of the empirical studies using the FAO-GAEZ productivity measures, we propose to partially identify the effect of agricultural productivity in a linear model with two proxies that are subject to nonclassical measurement error. We present a menu of assumptions and discuss how their inclusion or exclusion affects the estimated effects. In our baseline model, we impose a sign restriction on the effect of productivity and exogeneity of measurement errors --- which relaxes what has been used in the literature so far --- while allowing for measurement errors to be nonclassical, only restricting them to be positively correlated with each other. Measurement error exogeneity and knowledge of the coefficient's direction may be motivated by economic theory, but, if the researcher is not willing to make these assumptions, we derive wider bounds under a weaker set of assumptions. Similarly, positive correlation between the measurement errors in the two proxies follows from the assumption that unobservable factors in the FAO-GAEZ version 4 model are also unobservable in the FAO-GAEZ version 3 model. Again, if the researcher is not willing to make this assumption, we derive wider bounds under a weaker set of assumptions.


	Under four different sets of assumptions, we derive analytical bounds around the true linear effect of agricultural productivity on the outcome of interest. Importantly, our assumptions are no stronger than those commonly applied when ignoring measurement error, and our bounds exhaust all the information contained in the first two moments of the data distribution. That is, our bounds create the shortest possible interval that contains the true target parameter and is constructed through restrictions on only the first two moments of the latent variables. We focus on this type of restriction because most applied researchers use linear regression models and do not restrict higher-order moments.

	Moreover, our bounds have easy-to-understand closed-form solutions and present two interesting features. First, they offer a test for the identifying assumptions. Second, they compare favorably against two alternative strategies in the literature: a reduced-form approach that ignores measurement error and a bias-minimizing tool recommended by \cite{Lubotsky2006}. In particular, there exist data-generating processes where our bounds are closer to the true target parameter than the alternative tools.


	We also propose estimation and inference methods for the identified bounds. The estimation method is parametric and easy to implement through linear regressions. The inference procedure adapts the tool proposed by \cite{Chernozhukov2013} and constructs confidence bounds around the identified set with a known confidence level.



	To illustrate our methodology, we reevaluate the results of three empirical studies: \citet{nunn2011}, \citet{bustos2016agricultural} and \cite{acharya2016political}. To do so, we replicate their findings using the reduced-form approach, which ignores measurement error, and compare these results with those obtained through our partial identification method. When we relax the no-measurement-error assumption, we estimate wide intervals around the true parameter of interest, suggesting that model uncertainty regarding the existence of nonclassical measurement error is a relevant issue in applied practice. In particular, \cite{acharya2016political} finds that cotton productivity significantly increases Republican vote share at the 10\%-confidence level. However, when we account for nonclassical measurement error, we do not reject the null hypothesis that the lower bound of the true effect is zero.



	Lastly, we highlight that our proposed methodology can be used in many empirical contexts. To employ our partial identification strategy, we only require an exogenous mismeasured treatment variable and two exogenous proxies. As examples of possible applications, we mention the impact of media access on economic outcomes \citep{Olken2009,Adena2015} and the effect of social heterogeneity on socioeconomic outcomes \citep{Montalvo2005,Desmet2017,Desmet2025}. In the first case, media access is measured using an irregular terrain model that has been modified over time. In the second case, measures of social heterogeneity may depend on the aggregation level or on the differences between fractionalization and polarization. Appendix \ref{AppOther} explains seven other examples in the fields of Macroeconomics, Behavioral Economics, Economic History, and Political Economy.


	\textbf{Related Literature.} This article contributes to two different branches of literature. Concerning its empirical contribution, many papers in economics have used the FAO-GAEZ data. \cite{nunn2011}, in pioneering work, uses the potential productivity of potatoes to explain how its introduction in the Old World affected population and urbanization. \cite{bustos2016agricultural, bustos2019industrialization, bustos2020capital} use the productivity of soybeans to measure the impact of introducing genetically modified seeds on structural transformation. \cite{bentzen2017irrigation} uses productivity under rain-fed and irrigation conditions to study the impacts of irrigation on the origins of autocracies. \cite{acharya2016political} uses cotton productivity as an instrument for slavery to explain contemporary differences in political attitudes in the U.S.   \cite{carillo2021agricultural} uses wheat productivity to study the impact of agricultural policies on industrialization in Italy.\footnote{Similar papers include \citet{moorthy2022agricultural}, \citet{dias2023} and \citet{carreira2024deforestation} on soybean productivity and gender disparity, water pollution, and deforestation; \cite{galor2016agricultural} on the productivity of crops available before and after the Columbian Exchange and time preferences; \cite{alesina2011fertility, alesina2013origins} on the productivity of various crops, plough use, and gender disparities; \cite{dube2016maize} on maize productivity and the Mexican drug sector; \cite{gollin2021two} on the productivity of various crops and the impacts of the Green Revolution;  \cite{mayshar2022origin} on the relative productivity of cereals over roots and tubers and the origin of states; \cite{fiszbein2022agricultural} on the productivity of multiple crops and its effects on development. These papers use FAO-GAEZ data to estimate linear specifications of the impacts of crop productivity on various outcomes of interest. However, it is worth noting that there is a range of papers employing FAO-GAEZ data in different types of models, such as trade models \citep{costinot2016evolving,sotelo2020domestic, pellegrina2022trade, farrokhi2023trade}, deforestation models \citep{sant2024green, araujo2020efficient, dominguez2021efficiency}, and other types of spatial models \citep{adamopoulos2022geography,adamopoulos2022misallocation,imbert2022migrants,fernandez2023fractured}.}

	We contribute to this literature by reevaluating the results of \cite{nunn2011}, \cite{bustos2016agricultural} and \cite{acharya2016political} while accounting for measurement error in the main explanatory variable. When we relax the assumption of no measurement error, the implied magnitudes of agricultural productivity effects can change dramatically, implying that much of what we think we know about the magnitude of these effects is far more uncertain than previously recognized.

	Concerning its theoretical contribution, our work is inserted in the literature about identifying treatment effect parameters with measurement error. As illustrated by \cite{Schennach2016} and \cite{Hu2017}, this literature is vast and growing. The closest paper to ours is \cite{Lubotsky2006}. While we derive an identified set that contains the true target parameter, they propose a single estimand that minimizes bias. Moreover, they impose that the measurement error terms are uncorrelated with the true variable of interest, while we impose no restriction on this correlation. This extra flexibility is particularly important in our empirical application because measurement error terms must be correlated with the true variable of interest when the latter is bounded, as is the case with crop productivity.

	We also highlight that our work is connected to the studies developed by \cite{Hausman1998}, \citet{Black2000}, \cite{Hu2008}, \citet{Cunha2010}, \cite{Schennach2014}, \cite{Hu2015}, \citet{Alix-Garcia2023}, \cite{Kim2024}, and \citet{Millimet2024}. \cite{Black2000} and \cite{Millimet2024} differ from our work because they impose that the true variable of interest and the measurement error must be negatively correlated with each other. \cite{Kim2024} differ from our work because they impose that the proxies must be correlated with the true variable of interest while we allow them to contain only noise. Furthermore, \cite{Hausman1998} and \cite{Alix-Garcia2023} focus on mismeasured binary outcome variables. Moreover, \cite{Hu2008}, \cite{Cunha2010} and \cite{Hu2015} require the measurement error terms to be mutually independent. Lastly, \cite{Schennach2014} proposes a general method that encompasses our setting. However, her tool requires simulation-based optimization algorithms while our tool is based on an analytic solution that is easily implemented using standard statistical software. The importance of analytical solutions such as ours is emphasized by \citet[page 366]{Schennach2014}.

	\textbf{Paper Organization.} The remainder of this paper is organized as follows. Section \ref{Sdata} describes the FAO-GAEZ data, documents the existence of measurement error in its variables, and intuitively explains the use of partial identification as a solution to measurement error. Section \ref{Sidentification} presents our econometric model and states our main identification results. Section \ref{Sestimation} briefly describes the estimation and inference methods connected to our partial identification strategy. Section \ref{Sresults} illustrates the proposed tools by reevaluating the impact of agricultural productivity on three empirical contexts, and Section \ref{Srecommendation} explains how to use our tools in a practical way in other applications. Lastly, Section \ref{Sconclusion} concludes.

	\section{Measurement Error in the FAO-GAEZ Data and Partial Identification as a Solution}
	\label{Sdata}

	In this section, we explain the empirical context of our methodology and how it is connected with the proposed identification strategy. Section \ref{Scontext} explains the FAO-GAEZ data and the three empirical examples we discuss in this paper. Section \ref{Swhatandwhy} explains partial identification and its usefulness in the applied researcher's toolkit.

	\subsection{Empirical Context and Data}\label{Scontext}

	The Food and Agriculture Organization's Global Agro-Ecological Zones project (FAO-GAEZ) produces a widely used measure of crop productivity. To do so, it combines climate and soil data with plant growth parameters to simulate optimal growth cycles of crops to predict potential crop productivity globally at a resolution of approximately 10 km by 10 km \citep{Fischer2012GAEZ, Fischer2021GAEZ}. The FAO-GAEZ generates productivity measures of crops under different scenarios of water source --- rainfed or irrigation --- and use of inputs --- low use of inputs and high use of inputs, such as machinery, fertilizer, seeds and others.\footnote{Before version 4, FAO-GAEZ contained a scenario with intermediate use of inputs. This option is no longer available in version 4.}

	There are two main advantages that led researchers to widely adopt the FAO-GAEZ data. First, it is independent of observed productivity, allowing it to capture potential productivity even in regions where specific crops are not cultivated. Second, it is not influenced by economic factors such as credit constraints, transportation infrastructure, and subsidies, which likely affect observed agricultural productivity. Thus, FAO-GAEZ is often interpreted as exogenous to institutional factors, greatly simplifying the interpretation of estimated parameters as being causal.

	As the FAO-GAEZ data combine climate, soil, and crop growth parameters to build potential productivity measures, they are susceptible to nonclassical measurement error. For example, low-resource countries may install weather stations only in richer areas, whose climate is appropriate to agriculture, implying that, in these countries, measured climate variables will be biased towards values that are more beneficial to agriculture productivity. Consequently, measurement error will tend to be positive in countries with low true productivity, creating nonclassical measurement error. Although this issue has been ignored in the literature, the FAO-GAEZ data offer a unique opportunity to better understand the structure of such measurement errors since the data have been updated and are currently on their fourth version.

	There is no precise documentation of the differences between versions. Upon inspection of versions 3 and 4, we identified many differences in parameters and raw data.  To illustrate this issue, we provide four examples. First, the ``harvest index" parameter essentially measures the portion of the weight of the crop that is actually of economic interest --- e.g., the beans of soybeans --- and is different for a set of crops between versions. Second, for another set of crops, FAO-GAEZ updated the ``leaf area index", which measures the plant's canopy and its ability to capture sunlight. Third, precipitation data in version 3 is from the Variability Analysis of Surface Climate Observations 1.1 (VASClimO), while version 4's precipitation data is from the Global Precipitation Climatology Centre (GPCC) Full Data Reanalysis Product Version 6. Fourth, data that still uses the same source are using newer versions. For example, temperature, sunshine fraction, and wind speed were from CRU 1.1 in version 3 and from CRU 3.21 in version 4,  which entails a range of internal modifications of each data source.\footnote{\cite{Rhode2024} discusses differences in the model for the productivity of cotton.} Despite these differences, all versions of the FAO-GAEZ intend to measure the same potential productivity, i.e., the agricultural productivity for a given place, climate, technology, and water supply. Appendix \ref{AppSourceError} further discusses sources of measurement error in FAO-GAEZ data.

	Rather than focusing on improving individual versions of the FAO-GAEZ data, our approach centers on combining two available proxies (versions 3 and 4)  to identify the impact of the underlying agricultural productivity on economic outcomes. To do so, we use assumptions that are weaker than those commonly applied when using only one version.\footnote{Version 1 was released in 2000, version 2 in 2002, version 3 in 2012, and version 4 in 2020. Although our framework can be extended to accommodate more than two proxies, we do not pursue this path for the FAO-GAEZ application because we do not have access to the data from versions 1 and 2. Additionally, we want our method to be easily implemented in other settings, which may have only two proxies (Appendix \ref{AppOther}).}

	To illustrate the importance of measurement error in FAO-GAEZ data and motivate our econometric framework in the next section, we highlight three applications in the literature: \cite{nunn2011}, \cite{bustos2016agricultural}, \cite{acharya2016political}. In the remainder of this section, we discuss some of the differences in agricultural productivity measured by different versions of FAO-GAEZ for the crops studied in these three articles. Importantly, the productivity measures used by \cite{nunn2011} and \cite{acharya2016political} are bounded, implying that any measurement error they may contain must be nonclassical.

\begin{figure}[!htbp]
		\begin{center}
			\caption{Log of area suitable for potato in FAO-GAEZ versions 3 and 4 \label{fig:potato}}
			\includegraphics[width=\linewidth]{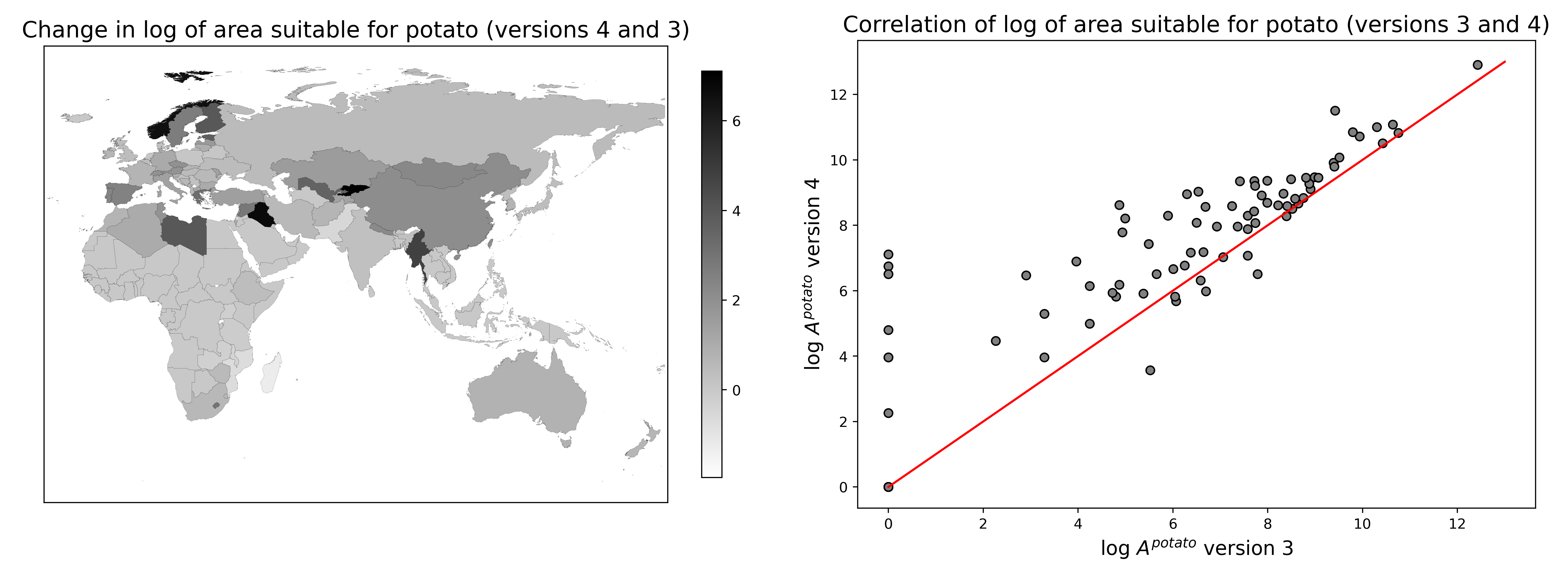}
		\end{center}
		\footnotesize{This figure shows the spatial distribution of the differences in FAO-GAEZ versions 3 and 4 for the log of the area suitable for potato (low input) on the left. The plot on the right shows the correlation between versions 3 and 4 for the log of area suitable for potato $(A^{potato})$. Notice we follow \cite{nunn2011} in using the transformation $\log(1+x)$ since our objective is that our results are comparable with the ones in the original paper. The 45$^\circ$ line is shown in red.}
	\end{figure}

	\cite{nunn2011} study the impact of the introduction of the potato in the Old World on population, using the FAO-GAEZ data on suitability of white potato. Figure \ref{fig:potato} shows the spatial differences between versions 3 and 4 for potato productivity. Some countries --- such as Norway and Iraq --- had little to no area suitable for potato in version 3 but a large suitable area in version 4, creating a large change in logs. Another example is China, which has an increase in area suitable for potatoes of 700\% across versions.

\begin{figure}[!htbp]
		\begin{center}
			\caption{Soybean productivity in FAO-GAEZ versions 3 and 4 \label{fig:soy}}
			\includegraphics[width=\linewidth]{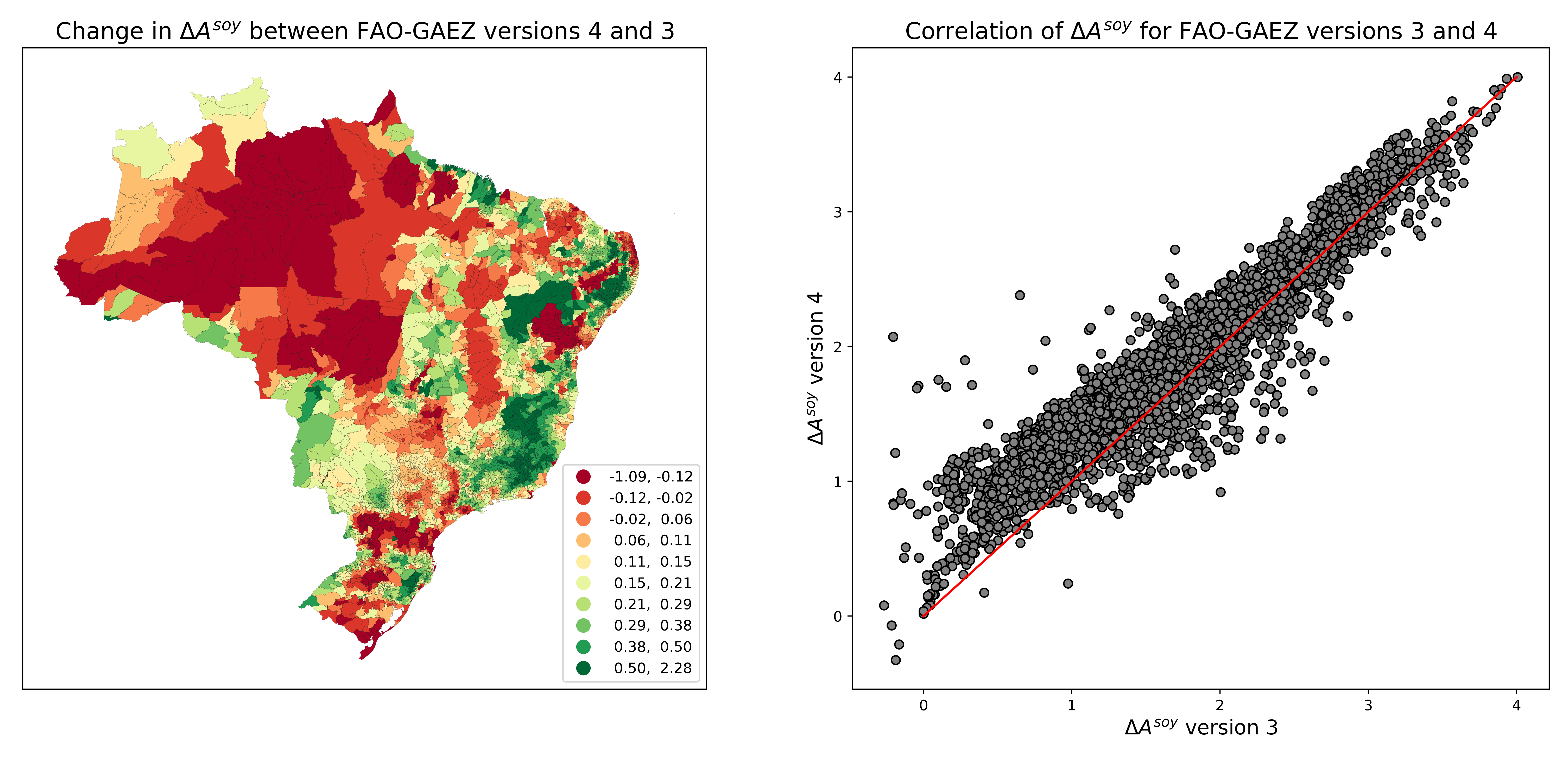}
		\end{center}
		\footnotesize{This figure shows the spatial distribution of the differences in versions 3 and 4 for the change in soybean productivity (high input minus low input) on the left. Each color represents a decile of the data distribution. The plot on the right shows the correlation between versions 3 and 4 for the change in soybean productivity $(\Delta A^{soy})$. The 45$^\circ$ line is shown in red.}
	\end{figure}

	\cite{bustos2016agricultural} studies the impact of the introduction of genetically modified soybean seeds in Brazil on structural transformation. To do so, the authors use the difference between soybean productivity under high and low inputs as a measure of productivity shock. Figure \ref{fig:soy} shows the spatial differences between versions 3 and 4 colored by deciles of this productivity shock. The differences are highly heterogeneous in space: 10\% of regions lose more than 120 kg of production per hectare while another 10\% gain more than 500 kg per hectare from an average productivity before the shock of 300 kg per hectare in version 3.

\begin{figure}[!htbp]
		\begin{center}
			\caption{Cotton productivity in FAO-GAEZ versions 3 and 4 \label{fig:cotton}}
			\includegraphics[width=\linewidth]{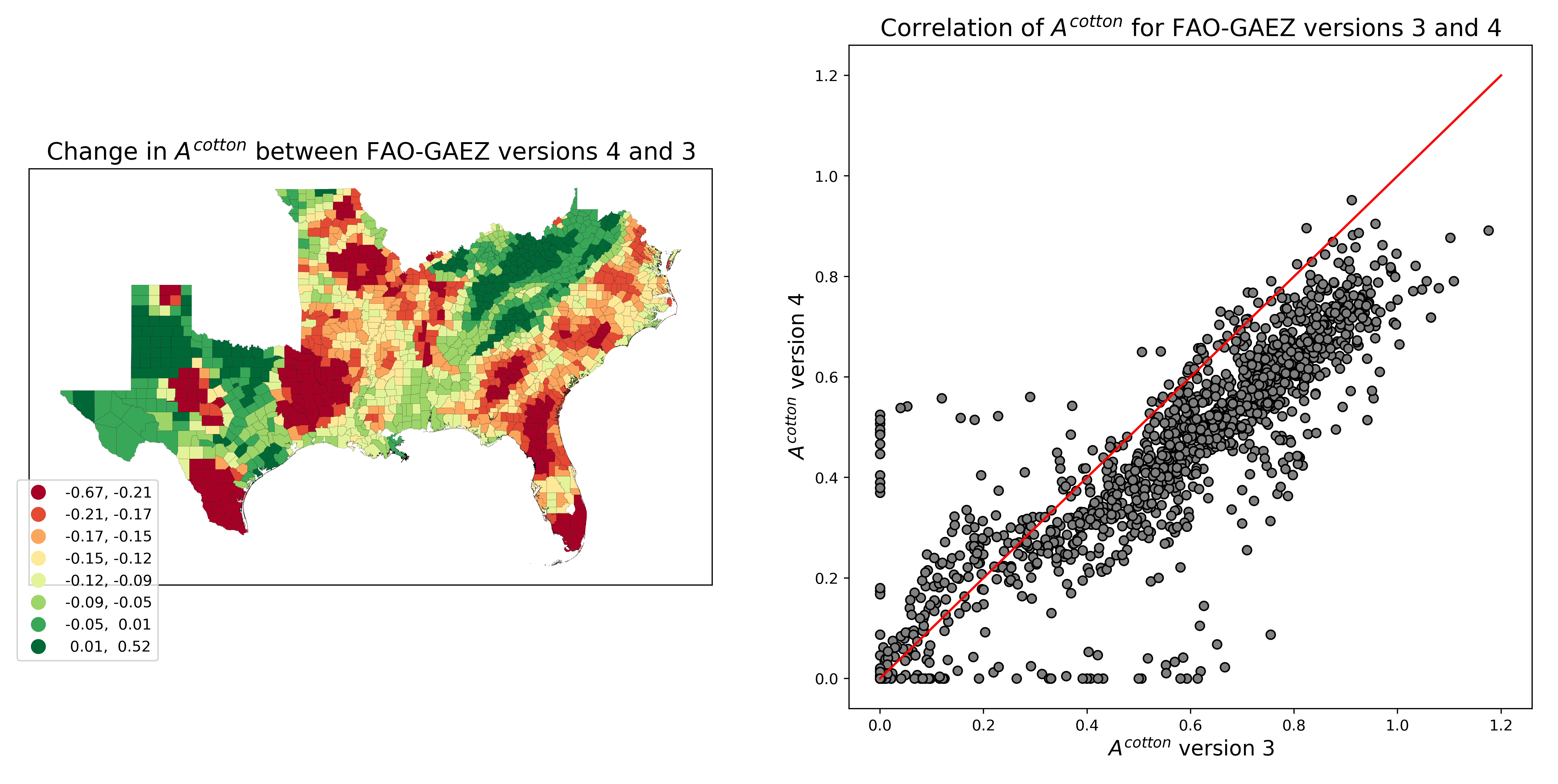}
		\end{center}
		\footnotesize{This figure shows the spatial distribution of the differences in versions 3 and 4 for the cotton productivity high input on the left. Each color represents a quantile of the data distribution. The plot on the right shows the correlation between versions 3 and 4 for cotton productivity $(A^{cotton})$. The 45$^\circ$ line is shown in red.}
	\end{figure}

	Lastly, \cite{acharya2016political} studies the impact of slavery on contemporary political attitudes in the South of the United States, using cotton productivity as an instrument for slavery. Figure \ref{fig:cotton} shows the spatial differences between versions 3 and 4 colored by quintiles of cotton productivity with high input.
	The differences are highly heterogeneous in space: 10\% of regions lose more than
	180 kg of production per hectare, which is close to 10\% of the average productivity in version 3.

	In summary, there is significant variation in the measured productivity of crops across FAO-GAEZ versions 3 and 4. These differences suggest the existence of nonclassical measurement error in the FAO-GAEZ productivity variables and motivate our econometric framework.

	\subsection{What is Partial Identification? Why is it useful?}\label{Swhatandwhy}

	To address the measurement error issue explained in the last section, we propose a novel partial identification strategy. Since this approach is not widely used in empirical studies, this section briefly discusses the concept of partial identification while Appendix \ref{AppPartialIdentification} discusses it in more detail. The reader who is familiar with partial identification may skip this section entirely.

	Empirical research relies on assumptions about a data-generating process, which models the relationship between latent and observable variables. Point identification imposes assumptions on this model to express the target parameter as a function of observable variables. This approach often requires strong restrictions that researchers may diverge about \citep{Manski2011}. Partial identification addresses this type of model uncertainty. A researcher may adopt a weaker set of assumptions, which is not sufficient for point-identification but is sufficient for partial identification. In this case, the researcher builds an interval that contains the target parameter and is expressed in terms of observables.

	Different assumptions yield different interval lengths. In our context, one researcher may follow the literature and assume no measurement error; another (Section \ref{Sidentification}) is not willing to ignore measurement error but is willing to impose restrictions on the sign of the true effect, on the correlation between the measurement error and the outcome shock, and on the sign of the correlation between the measurement error terms; a third (Appendices \ref{AppNoSign}-\ref{AppFreeCov}) is not willing to make one of these assumptions. The identified set of the first researcher is a point, of the second one is an interval, and of the third one is a wider interval. Importantly, the length of the interval is a result of model uncertainty. Sample uncertainty is accounted for by building a confidence set around the identified interval, similar to the point-identification approach that builds confidence intervals around the identified point.

	\section{Econometric Framework and Identification}\label{Sidentification}

	Most papers using the FAO-GAEZ productivity measures \citep{nunn2011,bustos2016agricultural, acharya2016political} are interested in a linear model similar to the following one:\footnote{Even though including covariates at this point requires a more cumbersome notation, doing so is useful to connect the identification approach and the estimation method (Section \ref{Sestimation}) in a straightforward way.}
	\begin{equation}\label{EqModelCovariates}
		Y = \alpha_{0} + \sum_{j = 1}^{J} \alpha_{j} \cdot W_{j} + \beta \cdot X^{*} + \epsilon,
	\end{equation}
	where $J \in \mathbb{N}$, $\alpha_{0},\ldots,\alpha_{J}$ are population parameters, $\beta$ is the population parameter of interest, $Y$ is the outcome variable (e.g., the change over time of the share of workers employed in manufacture in \cite{bustos2016agricultural}), $W_{1},\ldots,W_{J}$ are the covariates (e.g., a time trend and municipality characteristics in \cite{bustos2016agricultural}), $X^{*}$ is the true unobserved variable of interest (e.g., the true change in potential yield of soy after the introduction of new seeds in \cite{bustos2016agricultural}) and $\epsilon$ is an unobservable economic shock that satisfies $Cov\left(A,\epsilon\right) = 0$ for any $A \in \left\lbrace W_{1},\ldots,W_{J},X^{*} \right\rbrace$.

	For brevity, we apply the Frish-Waugh-Lovell Theorem to rewrite Equation \eqref{EqModelCovariates} as
	\begin{equation}\label{EqModel}
		Y_{res} = \beta \cdot X_{res}^{*} + \epsilon_{res},
	\end{equation}
	where $Y_{res} \coloneqq Y - W \cdot \left(\mathbb{E}\left[W^{T} \cdot W\right]\right)^{-1} \cdot \mathbb{E}\left[W^{T} \cdot Y\right]$, $X_{res}^{*} \coloneqq X^{*} - W \cdot \left(\mathbb{E}\left[W^{T} \cdot W\right]\right)^{-1} \cdot \mathbb{E}\left[W^{T} \cdot X^{*}\right]$, $\epsilon_{res} \coloneqq \epsilon - W \cdot \left(\mathbb{E}\left[W^{T} \cdot W\right]\right)^{-1} \cdot \mathbb{E}\left[W^{T} \cdot \epsilon\right] = \epsilon$, and $W = \left[1, W_{1}, \ldots, W_{J} \right]$ is the row vector of covariates including the constant term.

	Since $X^{*}$ is unobservable, we cannot identify $\beta$ based on Equation \eqref{EqModel} alone. To circumvent this issue, empirical researchers have employed a proxy variable: $Z_{1}$, which is usually constructed from the FAO-GAEZ data version 3 and is described in Section \ref{Scontext}. One important point is that this proxy variable has been refined over time, resulting in the proxy constructed from the FAO-GAEZ data version 4. For this reason, we consider the use of two proxies --- $Z_{1}$ and $Z_{2}$, representing FAO-GAEZ data versions 3 and 4 --- that satisfy
	\begin{equation}\label{EqProxiesCovariates}
		\begin{matrix}
			Z_{1,res} = X_{res}^{*} + U_{1} \\
			Z_{2,res} = X_{res}^{*} + U_{2}
		\end{matrix},
	\end{equation}
	where $Z_{1,res} = Z_{1} - W \cdot \left(\mathbb{E}\left[W^{T} \cdot W\right]\right)^{-1} \cdot \mathbb{E}\left[W^{T} \cdot Z_{1}\right]$ and $Z_{2,res} = Z_{2} - W \cdot \left(\mathbb{E}\left[W^{T} \cdot W\right]\right)^{-1} \cdot \mathbb{E}\left[W^{T} \cdot Z_{2}\right]$. Intuitively, $Z_{1,res}$ and $Z_{2,res}$ are two imperfect measures of $X_{res}^{*}$ where $U_{1}$ and $U_{2}$ capture the measurement error.

	Our goal is to use these two proxies to partially identify $\beta$. Section \ref{Sassumptions} states and explains our modeling assumptions, while Section \ref{SidResults} presents our main identification result.

	\subsection{Assumptions}\label{Sassumptions}

	To partially identify $\beta$, in our baseline specification, we impose six assumptions, contextualizing them to our empirical setting. For brevity, we impose assumptions directly on the residualized variables in Equations \eqref{EqModel} and \eqref{EqProxiesCovariates}. In Appendix \ref{AppPrimitive}, we impose assumptions on the primitives of our model (Equation \eqref{EqModelCovariates}) that imply the assumptions described in this section.

	\begin{assumption}[Exogeneity]\label{AsExogeneous}
		After partialling out the covariate vector $W$, the true unobserved measure $X^{*}$ is exogenous, i.e., $Cov\left(X_{res}^{*}, \epsilon_{res}\right) = 0$.
	\end{assumption}

	Assumption \ref{AsExogeneous} is a standard exogeneity condition in any linear model. For example, it is implied by the assumption that $Cov\left(A,\epsilon\right) = 0$ for any $A \in \left\lbrace W_{1},\ldots,W_{J},X^{*} \right\rbrace$ and is imposed in the empirical studies discussed in this article \citep{nunn2011, bustos2016agricultural, acharya2016political}. In our empirical application, it imposes that the natural forces that determine crop productivity (e.g., climate and soil nutrients captured by $X_{res}^{*}$) are uncorrelated with the unobserved economic forces that determine the outcome variable (e.g., financial shocks captured by $\epsilon_{res}$) after controlling for observable economic forces (e.g., educational levels captured by $W$).

	\begin{assumption}[Scale Normalization]\label{AsScale}
		The scale of the true unobserved measure is normalized to have unit variance, i.e., $Var\left(X_{res}^{*}\right) = 1$.
	\end{assumption}

	Assumption \ref{AsScale} allows us to interpret $\beta$ as the effect of increasing one standard deviation of $X_{res}^{*}$ on the outcome variable. Since $X_{res}^{*}$ is unobservable, this restriction is a simple normalization: if the true model were $$Y_{res} = \tilde{\beta} \cdot \tilde{X}_{res}^{*} + \epsilon_{res}$$ with $Var\left(\tilde{X}_{res}^{*}\right) \neq 1$, we could rewrite it according to Equation \eqref{EqModel} by defining $$\beta \coloneqq \tilde{\beta} \cdot \sqrt{Var\left(\tilde{X}_{res}^{*}\right)} \text{ and } X_{res}^{*} \coloneqq \dfrac{\tilde{X}_{res}^{*}}{\sqrt{Var\left(\tilde{X}_{res}^{*}\right)}},$$ enforcing Assumption \ref{AsScale}.\footnote{This normalization does not impact Equation \eqref{EqProxiesCovariates} because we allow the measurement error terms to be correlated with the unobserved measure. In this case, we have $Z_{k} = \tilde{X}_{res}^{\star} + \tilde{U}_{1} = X_{res}^{\star} -  X_{res}^{\star} + \tilde{X}_{res}^{\star} + \tilde{U}_{1} = X_{res}^{\star} + U_{1}$, where $U_{1} \coloneqq  -  X_{res}^{\star} + \tilde{X}_{res}^{\star} + \tilde{U}_{1}$.} One may view this normalization as changing the target parameter from $\tilde{\beta}$ to $\beta = \tilde{\beta} \cdot \sqrt{Var\left(\tilde{X}_{res}^{*}\right)}$ since it is not possible to separately identify $\tilde{\beta}$ and $\sqrt{Var\left(\tilde{X}_{res}^{*}\right)}$.

	\begin{assumption}[Known Direction of the True Effect]\label{AsPosEff}
		The direction of the effect of the true unobservable variable of interest on the outcome variable is known, i.e., $\beta \geq 0$.
	\end{assumption}

	Assumption \ref{AsPosEff} imposes that the effect of the true variable of interest on the outcome variable is non-negative.\footnote{The case that imposes that the true target parameter is non-positive is symmetric.}\textsuperscript{,}\footnote{This type of assumption is common in the literature about partial identification of treatment effects. For example, \cite{Manski1997} and \cite{Kreider2007} refer to this type of assumption as Monotone Treatment Response.}  Importantly, this assumption does not rule out that the effect might be zero. In our empirical application, it imposes that the effect of the true crop productivity on the outcome variable is non-negative. This assumption is motivated by the empirical literature because we assume that the articles using one proxy --- the FAO-GAEZ version 3 --- are at least identifying the correct sign of the true effect. If a researcher is not willing to impose this type of assumption, Appendix \ref{AppNoSign} proposes an alternative way to partially identify this coefficient without making assumptions about its sign.

	\begin{assumption}[Exogeneity of the Measurement Errors]\label{AsMEexogenous}
		The measurement error variables are exogenous, i.e., $Cov\left(U_{1}, \epsilon_{res}\right) = 0$ and $Cov\left(U_{2}, \epsilon_{res}\right) = 0$.
	\end{assumption}

	Assumption \ref{AsMEexogenous} imposes that the measurement error terms in our proxies of the true variable of interest are uncorrelated with the unobservable economic shock after controlling for observable covariates.\footnote{This assumptions is implied by $Cov\left(A,\epsilon\right) = 0$ for any $A \in \left\lbrace W_{1},\ldots,W_{J},X^{*}, Z_{1}, Z_{2} \right\rbrace$ as discussed in Appendix \ref{AppPrimitive}.} This assumption weakens the implicit identifying assumption in the empirical literature about agricultural productivity \citep{nunn2011,  bustos2016agricultural, acharya2016political} since applied researchers impose that $U_{1} = U_{2} = 0$.

	In our empirical application, Assumption \ref{AsMEexogenous} imposes that prediction errors in the FAO-GAEZ measures are uncorrelated with the unobserved economic forces that determine the outcome variable. This assumption would be violated if the institutional factors that impact $\epsilon_{res}$ positively also affect the quality of the environmental data that is used when constructing the FAO-GAEZ measure. In this case, regions with large values of $\epsilon_{res}$ would have a small amount of measurement error. Consequently, $Cov\left(U_{1}, \epsilon_{res}\right) = 0$ would be implausible in this hypothetical scenario.\footnote{Appendix \ref{AppEndoME} offers an alternative partial identification strategy that does not impose exogeneity of the measurement error terms. As a consequence of endogenous measurement error, the lower bound becomes uninformative and entirely determined by prior knowledge of the direction of the true effect (Assumption \ref{AsPosEff}).}

	\begin{assumption}[Nonclassical Measurement Error]\label{AsNonClassical}
		The measurement error terms are positively correlated with each other, i.e., $Cov\left(U_{1}, U_{2}\right) \geq 0$. We name this assumption nonclassical measurement error because the covariances between the true variable of interest and the measurement error terms are unrestricted, i.e., $Cov\left(X_{res}^{*},U_{k}\right) \in \mathbb{R}$ for any $k \in \left\lbrace 1, 2 \right\rbrace$.
	\end{assumption}

	Assumption \ref{AsNonClassical} imposes that the covariance between the measurement error terms is positive. In our empirical application, it imposes that areas with large prediction errors in the FAO-GAEZ version 3 model tend to have large prediction errors in the FAO-GAEZ version 4 model too. This assumption is plausible because unobservable natural productivity factors in the FAO-GAEZ version 4 model are also unobservable in the FAO-GAEZ version 3 model.\footnote{Appendix \ref{AppFreeCov} offers an alternative partial identification strategy that does not restrict the covariance between the two measurement error terms. As a consequence of imposing fewer assumptions, the identified interval around the parameter of interest is weakly wider.} In Appendix \ref{AppSourceError}, we discuss sufficient conditions in the FAO-GAEZ data under which this assumption holds.

	Moreover, Assumption \ref{AsNonClassical} allows nonclassical measurement error. Importantly, it does not restrict the relationship between the true variable of interest and the measurement error terms. Such flexibility is empirically relevant in many possible scenarios. For example, our framework allows the proxies to underpredict areas with high productivity and overpredict areas with low productivity. In this scenario, there is nonclassical measurement error because $Cov\left(X_{res}^{*},U_{k}\right) < 0$ for any $k \in \left\lbrace 1, 2 \right\rbrace$.\footnote{This scenario is possible if low-resources countries install weather stations only in richer areas, whose climate is appropriate to agriculture, implying that, in these countries, measured climate variables will be biased towards values that are more beneficial to agriculture productivity. Consequently, measurement error will tend to be positive in countries with low true productivity.}

	\begin{assumption}[Well-behaved Data Distributions]\label{AsDataRestrictions}
		The variance of the outcome and proxy variables are strictly positive, the covariances between each proxy and the outcome variable are positive, the covariance of the two proxy variables is strictly greater than the negative of the variable of interest's variance, and the proxies are not linearly dependent, i.e., $Var\left(Y_{res}\right) > 0$, $Var\left(Z_{1,res}\right) > 0$, $Var\left(Z_{2,res}\right) > 0$, $Cov\left(Z_{1,res}, Y_{res}\right) \geq 0$, $Cov\left(Z_{2,res}, Y_{res}\right) \geq 0$, $Cov\left(Z_{1,res}, Z_{2,res}\right) > -1$, and $Var\left(Z_{1,res}\right) \cdot Var\left(Z_{2,res}\right) \neq \left[Cov\left(Z_{1,res}, Z_{2,res}\right)\right]^{2}$.
	\end{assumption}

	Assumption \ref{AsDataRestrictions} imposes four restrictions on the distribution of the observable variables. The first restriction imposes that the outcome and proxy variables are not constants. If we do not impose $Var\left(Y_{res}\right) > 0$, we must have that $\beta = 0$ by construction and the identification problem in Equation \eqref{EqModel} is uninteresting. If we do not impose that the proxies have strictly positive variance, then they are uninformative about the true variable of interest. The second condition imposes that the reduced form effects have the same sign as the true target parameter, while the third condition imposes that the correlation between the two proxies is sufficiently large. These restrictions are plausible in our empirical application because both proxies are trying to predict the same underlying true productivity. The fourth restriction imposes that the proxies do not contain identical information. Furthermore, note that all restrictions in Assumption \ref{AsDataRestrictions} are easily testable.

	\subsection{Identification}\label{SidResults}
	In this section, we state our main identification result. Before doing so, we define two objects
	\begin{equation}\label{EqUB}
		UB \coloneqq \sqrt{Var\left(Y_{res}\right)},
	\end{equation}
	and
	\begin{equation}\label{EqLB}
		LB \coloneqq \max \left\lbrace \begin{matrix}
			\dfrac{\sigma_{1,Y} + \sigma_{2,Y}}{1 + \sigma_{1,2}}, \hspace{10pt} \dfrac{- B + \sqrt{B^{2} - 4 \cdot A \cdot C}}{2 \cdot C}, \\
			\\
			\sqrt{\dfrac{
					\sigma_{1,Y}^2 \cdot \sigma_{2}^{2} + \sigma_{2,Y}^2 \cdot \sigma_{1}^{2} - 2\cdot \sigma_{1,Y} \cdot \sigma_{2,Y} \cdot \sigma_{1,2}}{\sigma_{1}^{2} \cdot \sigma_{2}^{2} - \sigma_{1,2}^2}}
		\end{matrix}  \right\rbrace,
	\end{equation}
	where $\sigma_{1,2} \coloneqq Cov\left(Z_{1,res}, Z_{2,res}\right)$, $\sigma_{k,Y} \coloneqq Cov\left(Z_{k,res}, Y_{res}\right)$, $\sigma_{k}^{2} \coloneqq Var(Z_{k,res})$ for any $k \in \left\lbrace 1, 2\right\rbrace$, $A \coloneqq -\left[\sigma_{1,Y} - \sigma_{2,Y}\right]^{2},$ $B \coloneqq -2 \cdot \left\lbrace \sigma_{1,Y} \cdot \left[\sigma_{2}^{2} - \sigma_{1,2}\right] + \sigma_{2,Y} \cdot \left[\sigma_{1}^{2} - \sigma_{1,2}\right]  \right\rbrace,$ and $C \coloneqq \left[\sigma_{1}^{2} + 1\right] \cdot \left[\sigma_{2}^{2} + 1\right] - \left[\sigma_{1,2} + 1\right]^{2}$. Note that $LB \geq 0$ according to Assumption \ref{AsDataRestrictions} and the first term in the lower bound.

	We now state our main result: partial identification of $\beta$ (Equation \eqref{EqModel}) using two proxies (Equation \eqref{EqProxiesCovariates}) with nonclassical measurement error.\footnote{ As discussed in Section \ref{Swhatandwhy}, the bounds in Proposition \ref{PropMainId} account for model uncertainty in the sense that the researcher is not sure about the magnitude of the correlation between the true unobservable variable of interest and the proxies' measurement error terms. Due to the uncertainty behind the underlying model, the researcher may prefer to use the bounds in Proposition \ref{PropMainId} or the ones in Appendices \ref{AppNoSign}--\ref{AppFreeCov}, which use different sets of assumptions. Importantly, these bounds treat the populational distribution of the observable variables as known objects. Consequently, they do not account for sample uncertainty. This type of uncertainty will be addressed in Section \ref{Sestimation}.}
	\begin{proposition}\label{PropMainId}
		Under Assumptions \ref{AsExogeneous}-\ref{AsDataRestrictions}, we have that:
		\begin{equation}\label{EqBounds}
			\beta \in \left\lbrace \begin{array}{cl}
				\left\lbrace \emptyset \right\rbrace & \text{ if } UB < LB \\
				\left[LB, UB\right] & \text{ if } UB \geq LB
			\end{array} \right..
		\end{equation}

		Moreover, the bounds in Equation \eqref{EqBounds} exhaust all the information contained in the first two moments of the observable variables $\left(Y_{res}, Z_{1,res},Z_{2,res}\right)$.\footnote{See Appendix \ref{AppProofAllInfo} for the formal definition of ``exhausting all the information contained in the first two moments of the observable variables.'' Below, we provide an intuitive definition for this concept.}
	\end{proposition}
	\begin{proof}
		See Appendix \ref{AppProofMainId}.
	\end{proof}

	The identified set $\left[LB, UB\right]$ in Equation \eqref{EqBounds} contains all the values of the target parameter that are consistent with the data and the model assumptions. Its width captures the amount of model uncertainty that arises when we go from a setting with perfectly measured variables to one with nonclassical measurement error. Consequently, a wide identified set calls for substantial caution in interpreting results and indicates that, once we account for measurement error, our knowledge of the world might be more uncertain than thought in a point-identification scenario. In contrast, a narrow identified set boosts our confidence in our results because it allows us to draw precise conclusions even with weak assumptions regarding our proxy variables.

	Additionally, Proposition \ref{PropMainId} has two interpretations. One is connected to the lower bound while the other one is connected to the upper bound.

	First, the initial term in the lower bound in Equation \eqref{EqLB} collapses to the standard OLS estimand when there is no measurement error. Consequently, similarity between the first term in the lower bound and the estimated OLS coefficient indicates that the data are compatible with no measurement error and may be used as an indirect test of whether measurement error might be present. Importantly, this test relies on a necessary condition of ``no measurement error'' instead of a sufficient condition. For this reason, it should be used only as a suggestive assessment and combined with an economic argument explaining why measurement error is irrelevant in a specific application.

	Second, the upper bound captures an extreme scenario for measurement error. It is connected to the case when the proxy variables are composed of noise only. The upper bound reflects this uninformative scenario because it does not depend on the proxy variables.

	Furthermore, Proposition \ref{PropMainId} has three important consequences that need further discussion: (i) it offers a test for our identifying assumptions' validity, (ii) it exhausts all the information contained in the first two moments of the data distribution, and (iii) it compares favorably against alternative strategies in the literature.

	First, if $UB < LB$, Proposition \ref{PropMainId} implies that there is no data-generating process that satisfies Assumptions \ref{AsExogeneous}-\ref{AsDataRestrictions} and the restrictions imposed by the first and second moments of the observable data distribution. Consequently, $UB \geq LB$ is a testable necessary condition for the validity of our identifying assumptions.

	Second, the bounds in Equation \eqref{EqBounds} exploit all the restrictions imposed by $\mathbb{E}\left[Y_{res}\right]$, $\mathbb{E}\left[Z_{1,res}\right]$, $\mathbb{E}\left[Z_{1,res}\right]$, $Var\left(Y_{res}\right)$, $Var\left(Z_{1,res}\right)$, $Var\left(Z_{2,res}\right)$, $Cov\left(Y_{res}, Z_{1,res}\right)$, $Cov\left(Y_{res}, Z_{2,res}\right)$, $Cov\left(Z_{1,res}, Z_{2,res}\right)$ on the data generating process. Since they do not use restrictions imposed by all the moments of the data distribution, they are not sharp as defined by \cite{Canay2017}. However, in a linear model context, most researchers are willing to impose restrictions, at most, on the second moments of the underlying data-generating process, and our identification strategy exhausts the information contained in these types of moments.\footnote{A similar approach was adopted by \cite{Mogstad2017} in a different setting. Focusing on an IV problem, their partial identification results \citep[Proposition 3]{Mogstad2017} exploit all the information contained in the conditional expectation of the outcome variable given the instrument, but do not exploit higher-order moments.}

	Third, our method compares favorably against two alternative strategies in the literature. The first alternative method is a reduced-form approach that ignores measurement error. For example, suppose a researcher performs the following regression:
	\begin{equation}\label{EqReducedForm}
		Y_{res} = b_{0} + b_{1} \cdot Z_{1,res} + V_{1},
	\end{equation}
	where $Cov\left(Z_{res},V_{1}\right) = 0$ by construction. The second alternative method, recommended by \cite{Lubotsky2006}, uses an optimally weighted convex combination of the proxy variables to obtain an OLS coefficient as close as possible to the target parameter. To implement their least biased approach, \citet[Equation (4)]{Lubotsky2006} propose the following regression model
	\begin{equation}\label{EqLubotsky}
		Y_{res} = c_{0} + c_{1} \cdot Z_{1,res} + c_{2} \cdot Z_{2,res} + V,
	\end{equation}
	and shows that $c_{1} + c_{2}$ equals the coefficient obtained on the optimally weighted combination of proxies. In Appendix \ref{AppExample}, we use three simple data-generating processes to show that (i) $b_{1}$ may be strictly smaller than our lower bound, (ii) $b_{1}$ may be strictly larger than our upper bound, and (iii) our lower bound may achieve a smaller bias than the procedure proposed by \cite{Lubotsky2006}, which is possible because our lower bound estimand is not within the class of OLS estimands that use convex combinations of both proxies.

	Lastly, Remarks \ref{REupperbound}-\ref{REpositivecorrelation} explain the identifying power of our key assumptions, while Remark \ref{REbetterproxy} proposes an extra assumption that tightens our bounds in some data-generating processes.

	\begin{remark}\label{REupperbound}
		The upper bound is driven solely by Assumptions \ref{AsExogeneous}-\ref{AsPosEff}. Under Assumption \ref{AsExogeneous}, we have that $\beta = \frac{Cov\left(X_{res}^{*}, Y_{res}\right)}{Var\left(X_{res}^{*}\right)}$, implying, by Assumption \ref{AsScale}, that $\beta = Cov\left(X_{res}^{*}, Y_{res}\right)$. According to the Cauchy-Schwarz Inequality, we have that $\beta^{2} = \left[Cov\left(X_{res}^{*}, Y_{res}\right)\right]^{2} \leq Var\left(Y_{res}\right)$,  which implies $\beta \leq \sqrt{Var\left(Y_{res}\right)}$ under Assumption \ref{AsPosEff}. Consequently, Assumptions \ref{AsMEexogenous} and \ref{AsNonClassical} have no identification power regarding the largest possible effect that is compatible with the data distribution.
	\end{remark}

	\begin{remark}\label{REendogenousME}
		Assumption \ref{AsMEexogenous} is paramount to ensure a non-trivial lower bound. In Appendix \ref{AppEndoME}, we show that the lower bound equals zero when we do not impose Assumption \ref{AsMEexogenous}. In this case, the data is uninformative about the smallest possible effect, and the lower bound is entirely determined by prior knowledge of the direction of the true effect (Assumption \ref{AsPosEff}). Consequently, the identification power of Assumption \ref{AsMEexogenous} is captured by a non-trivial lower bound in Proposition \ref{PropMainId}.
	\end{remark}

	\begin{remark}\label{REpositivecorrelation}
		The identification power of Assumption \ref{AsNonClassical} is captured by the term $$\dfrac{Cov\left(Z_{1,res}, Y_{res}\right) + Cov\left(Z_{2,res}, Y_{res}\right)}{1 + Cov\left(Z_{1,res}, Z_{2,res}\right)}$$ in the lower bound in Proposition \ref{PropMainId}. In Appendix \ref{AppFreeCov}, we show that this term disappears from the lower bound when we remove Assumption \ref{AsNonClassical}. Due to its importance, Appendix \ref{AppSourceError} discusses the plausibility of this assumption in our empirical application.
	\end{remark}

	\begin{remark}\label{REbetterproxy}
		If a researcher is willing to impose more assumptions on the data-generating process, she may tighten the bounds derived in Proposition \ref{PropMainId}. For example, imagine that the second proxy, $Z_{2,res}$, is better than the first proxy, $Z_{1,res}$, in the sense that $Var\left(U_{1}\right) \geq Var\left(U_{2}\right)$. Under this additional assumption, we tighten the bounds in Equation \eqref{EqBounds} with the following restriction:
		\begin{equation*}
			\left\lbrace \begin{array}{rclcl}
				\beta & \geq & 2 \cdot \left[\frac{Cov\left(Z_{1,res}, Y_{res}\right) - Cov\left(Z_{2,res}, Y_{res}\right)}{Var\left(Z_{1,res}\right) - Var\left(Z_{2,res}\right)}\right] & \text{if} & Var\left(Z_{1,res}\right) > Var\left(Z_{2,res}\right) \\
				\beta & \leq & 2 \cdot \left[\frac{Cov\left(Z_{1,res}, Y_{res}\right) - Cov\left(Z_{2,res}, Y_{res}\right)}{Var\left(Z_{1,res}\right) - Var\left(Z_{2,res}\right)}\right] & \text{if} & Var\left(Z_{1,res}\right) < Var\left(Z_{2,res}\right)
			\end{array} \right..
		\end{equation*}
		This result illustrates that, even when one proxy is better than the other proxy, we still want to use information from both proxies to bound the true coefficient. That is, when we have nonclassical measurement error, the error term in the worst proxy still contains information about the true variable of interest that may not be contained in the best proxy.
	\end{remark}

	\section{Estimation and Inference}\label{Sestimation}

	In this section, we briefly describe a simple parametric estimator for the bounds described in Proposition \ref{PropMainId} and a method to compute $\alpha$-confidence sets that contain the identified set with a probability at least as large as the chosen confidence level $\alpha \in \left(0.5, 1\right)$. To do so, we assume that we observe an independently distributed sample $\left\lbrace Y_{i}, W_{1,i}, \ldots, W_{J,i}, Z_{1,i}, Z_{2,i} \right\rbrace_{i = 1}^{N}$, where $N \in \mathbb{N}$ is the sample size.

	To estimate the bounds described in Proposition \ref{PropMainId}, we recommend taking the sample analogs of Equations \eqref{EqUB} and \eqref{EqLB}. To do so, residualize the outcome and proxy variables by regressing them on the covariates. Then, use these residuals to compute the desired sample covariances and sample variances. See Appendix \ref{AppEstimator} for a detailed explanation of our estimation procedure.

	To construct an $\alpha$-confidence interval around the identified set, we combine the confidence bounds proposed by \cite{Chernozhukov2013} with a Bonferroni-style correction. Since the upper bound does not rely on intersection bounds, we use a simple bootstrap procedure to compute a $\frac{\left(1 + \alpha\right)}{2}$-confidence bound above it. Since the lower bound relies on intersection bounds, we use the method proposed by \cite{Chernozhukov2013} to compute a $\frac{\left(1 + \alpha\right)}{2}$-confidence bound below it. By connecting the lower confidence bound below our estimated lower bound to the upper confidence bound above our estimated upper bound, we construct an $\alpha$-confidence interval around the identified set. See Appendix \ref{AppInference} for a detailed explanation of our inference procedure.

	\section{Empirical Results}\label{Sresults}

	To illustrate the use of our partial identification strategy, we reevaluate the findings of \cite{nunn2011},  \cite{bustos2016agricultural}, and \cite{acharya2016political}. When we relax the no-measurement-error assumption, we estimate wide intervals around the true parameter of interest of these three studies, suggesting that model uncertainty regarding the existence of nonclassical measurement error is a relevant issue. Focusing on our lower bound and accounting for sample uncertainty, we find (i) that the data are consistent with an effect of agricultural productivity on population that is smaller than that reported by \citet{nunn2011}; (ii) that the data are consistent with an effect of soy productivity on manufacture employment that is smaller than that reported by \citet{bustos2016agricultural}; and (iii) that the data are consistent with no effect of cotton productivity on political behavior, in contrast to the findings of \citet{acharya2016political}.

	To map our assumptions to these three papers, we note the following. Assumption \ref{AsExogeneous} (Exogeneity) is imposed in all of them. Assumption \ref{AsScale} (Scale Normalization) only affects the interpretation of results, which are expressed in terms of standard deviations of $X^*$. Assumption \ref{AsPosEff} (Known Direction of the True Effect) requires that each application correctly identifies at least the sign of the effect.\footnote{Appendix \ref{AppNoSign} shows how this assumption can be relaxed in our setting.} Assumption \ref{AsMEexogenous} (Exogeneity of Measurement Errors) weakens the identification requirements relative to those in the replicated papers. Assumption \ref{AsNonClassical} (Nonclassical Measurement Error) is motivated by the persistence of measurement error in FAO-GAEZ data and is further discussed in Appendix \ref{AppSourceError}. Finally, Assumption \ref{AsDataRestrictions} (Well-behaved Data Distributions) is a standard technical condition.

	For other possible applications using different datasets, see Appendix \ref{AppOther}.

	\subsection{Nunn and Qian (2011)}

	\cite{nunn2011} study the impact of the introduction of the potato in the Old World on population. The empirical specification uses panel data to estimate Equation \eqref{eq:nunn}:

	\begin{equation}
		\label{eq:nunn}
		y_{it} =  \alpha_i + \alpha_t +\beta \cdot I^{Post} \cdot \log  (1+A_{i}^{potato}) + \sum_{j = 1}^{J} \alpha_{j} \cdot w_{i,j} + \epsilon_{it},
	\end{equation}
	where $y_{it}$ is population in country $i$ and year $t$, $A_{i}^{potato}$ is the area of country $i$ that is suitable for the cultivation of white potato,\footnote{A pixel is defined as being suitable if it has more than $40 \%$ of the maximum attainable yield for the crop.} $I^{Post}$ is a dummy variable that equals one for years after 1700, i.e., after the introduction of the potato. In our setting, the interaction $I^{Post} \cdot \log  (1+A_{i}^{potato})$ is the variable of interest $X^{*}$.

	A practical problem in replicating the work of \cite{nunn2011} is that the original results of \cite{nunn2011} use the intermediate input setting of FAO-GAEZ version 2. The FAO-GAEZ version 2 is not available anymore and the intermediate input specification is not available for version 4. For this reason, we replicate the results using the low input setting of versions 3 and 4 to guarantee a better comparison across versions.

	Table \ref{tab:potato} shows descriptive statistics for the interaction between the Post-1700 indicator variable and the natural logarithm of the area suitable for potatoes as measured by the FAO-GAEZ proxies. In version 4, the mean, standard deviation and maximum of this interaction increase. Consequently, there is more spatial heterogeneity in version 4 than in version 3.

	\begin{table}[!htb]
		\begin{center}
			\caption{Descriptive Statistics: Area Suitable for Potato (Interaction) \label{tab:potato}}
			\begin{tabular}{lccc}
				\toprule \toprule
				& \multicolumn{2}{c}{$I^{Post} \cdot \log (1+ A_i^{potato})$} & $\Delta \left\lbrace I^{Post} \cdot \log (1 + A_i^{potato}) \right\rbrace$\\
				& V 3 & V 4 & V 4 - V 3 \\
				& (1) & (2) & (3) \\
				\midrule
				Mean & 1.31 & 1.56 & 0.25 \\
				S.D. & 2.92 & 3.27 & 0.89 \\\\
				Min & 0 & 0 & -1.96 \\
				1\textsuperscript{st} Quartile & 0 & 0 & 0 \\
				Median & 0 & 0 & 0 \\
				3\textsuperscript{rd} Quartile & 0 & 0 & 0 \\
				Max & 12.42 & 12.91 & 7.11 \\
				\bottomrule
			\end{tabular}
		\end{center}
		\footnotesize{Notes: This table shows descriptive statistics for the interaction between the Post-1700 indicator variable and the natural logarithm of potato suitable areas for FAO-GAEZ versions 3 and 4 with the low input specification. The unit of observation is the country-year pair. ``S.D.'' stands for standard deviation.}
	\end{table}

	Table \ref{tab:results_nunn} shows the estimated results of Equation \eqref{eq:nunn}. Column (1) shows the result as reported by \cite{nunn2011}, which the authors interpret as an increase of 1\% in suitable land for potatoes generates an increase in population by 0.04\% after year 1700. Column (2) is our baseline and replicates \cite{nunn2011}'s result using the low input specification of FAO-GAEZ version 3. Reassuringly, the result in Column (2) is close to the one reported in Column (1), implying that we can work with low input productivity data version 3 in our next steps.

	\begin{table}[!htb]
		\begin{center}
			\caption{The Impact of the Potato in Total Population \label{tab:results_nunn}}
			\begin{tabular}{lccccccc}
				\toprule \toprule
				& \multicolumn{7}{c}{log total population}\\
				\cline{2-8}
				& (1) & (2) & (3) & (4) & (5) & (6) & (7) \\
				\hline
				$X^{*}$ & 0.043   & 0.039   & 0.038   & 0.126   & 0.114 & 0.116  & [0.044, 0.230] \\
				& (0.014) & (0.013) & (0.013) & (0.043) & (0.038) & & \\
				90\% CI & & & & & & & [0.017, 0.264] \\
				99\% CI & & & & & & & [0.005, 0.278] \\
				t-stats & & & & & & 10.81 & \\
				p-value & & & & & & 0.005 & \\ \\
				Obs.    & 1,552   & 1,552   & 1,552   & 1,552    & 1,552    & 1,552 & 1,552 \\
				Clusters        & 130     & 130     & 130     & 130      & 130      & 130 & 130 \\
				\bottomrule
			\end{tabular}
		\end{center}
		\footnotesize{Notes: We have that $X^{*} \coloneqq I^{Post} \cdot \log (1+A_i^{potato})$. Observations are at the country-year level. All regressions use a baseline sample of 130 Old World countries. Countries in North and South America are excluded. The
			periods are 1000, 1100, 1200, 1300, 1400, 1500, 1600, 1700, 1750, 1800, 1850, and 1900. Column (1) reports the result of \cite{nunn2011}'s table IV column 5 which uses middle input potato suitability from FAO-GAEZ version 2. To make results more comparable, Columns (2) and (3) report the results for using low input FAO-GAEZ data version 3 and 4, respectively. Columns (4) and (5) show the results when $I^{Post} \cdot \log (1+A_i^{potato})$ is normalized to have a variance equal to one. Column (6) implements the method proposed by \cite{Lubotsky2006}, which accounts for classical measurement error by using an optimally weighted convex combination of the proxy variables. To implement their least biased approach, \citet[Equation (4)]{Lubotsky2006} estimate $Y_{res} = c_{0} + c_{1} \cdot Z_{1,res} + c_{2} \cdot Z_{2,res} + V$ and uses $\hat{c}_{1} + \hat{c}_{2}$ as their estimator. The rows ``t-stats'' and ``p-value'' report the t-statistics and the p-value of a test whose null hypothesis imposes that $c_{1} = 0$ and $c_{2} = 0$, as recommended by \citet[p. 555]{Lubotsky2006}. Column (7) presents the result of our partial identification approach. All specifications include year-fixed effects, country-fixed effects, and a range of controls in logs interacted with year dummies: Old World Crops Area, Elevation, Ruggedness, Tropical Area, Maize Area, Silage Maize Area, Sweet Potatoes Area, Cassava Area. Standard errors, t-statistics, p-values, and confidence intervals are clustered at the country level. The confidence interval in Column (7) is built at the 10\% and 1\% levels.}
	\end{table}

	Column (3) shows the estimate of Equation \eqref{eq:nunn} using FAO-GAEZ version 4. We see this specification as naturally taking place in the coming years, with researchers simply substituting version 3 with version 4 while ignoring that there is information to be extracted by combining both versions. Comparing Columns (2) and (3), we see that changing the version of the FAO-GAEZ data from 3 to 4 results in very mild changes.

	We can now discuss results that account for measurement error. The interpretation of these results does not align directly with the results shown so far for two reasons. First, in our approach, we normalize the variance of $I^{Post} \cdot \log (1+A_i^{potato})$ such that it equals one (Assumption \ref{AsScale}). Second, \cite{nunn2011} uses the transformation $\log(1+x)$ which makes it hard to interpret the estimated coefficient as elasticities \citep{Chen2023}. For these two reasons, we discuss our results in terms of standard deviations of $I^{Post} \cdot \log (1+A_i^{potato})$. To make the comparison easier, Columns (4) and (5) show the result of estimating Equation \eqref{eq:nunn} after normalizing $I^{Post} \cdot \log (1+A_i^{potato})$ for each FAO-GAEZ version.  In Columns (4) and (5), the interpretation of the results is that an increase of one standard deviation in $I^{Post} \cdot \log (1+A_i^{potato})$ causes an increase of 12.6\% and 11.4\% in population, respectively.

	Column (6) reports estimates based on the method proposed by \cite{Lubotsky2006}. It accounts for classical measurement error by using an optimally weighted convex combination of the normalized proxy variables. To implement their least biased approach, \citet[Equation (4)]{Lubotsky2006} estimate $Y_{res} = c_{0} + c_{1} \cdot Z_{1,res} + c_{2} \cdot Z_{2,res} + V$ and uses $\hat{c}_{1} + \hat{c}_{2}$ as their estimator. This estimate is close to the estimate that uses only FAO-GAEZ Version 4 as a proxy and suggests that an increase of one standard deviation in $I^{Post} \cdot \log (1+A_i^{potato})$ causes an increase of 11.6\% in population. When we follow the recommendation by \citet[p. 555]{Lubotsky2006} and test the null hypothesis that $c_{1} = 0$ and $c_{2} = 0$, we find a p-value of 0.005, indicating that the effect of potato productivity on population is statistically significant at the 1\%-level.

	Column (7) shows the results associated with our method, which account for nonclassical measurement error. As we develop a partial identification strategy, our result is an estimated interval. This interval shows a wide range of possible effects, with an increase of one standard deviation in $I^{Post} \cdot \log (1+A_i^{potato})$ leading to an increase in population between 4.4\% and 23.0\%.\footnote{The estimated lower bound is achieved by its third term (Equation \eqref{EqLB}). Consequently, allowing for unrestricted correlation between the measurement error terms (i.e., removing Assumption \ref{AsNonClassical}) would not impact the estimated lower bound according to Appendix \ref{AppFreeCov}.} Importantly, this interval exhausts all the information contained in the first and second moments of the data-generating process, despite the large estimate for the upper bound. Such a wide estimated interval, especially compared with the small standard errors reported in Columns (2) and (3), suggests (i) that weakening the no-measurement-error assumption increases uncertainty substantially and (ii) that model uncertainty regarding the presence of non-classical measurement error is more concerning than sample uncertainty.

	Column (7) also reports the confidence intervals at the 10\% and 1\% levels. Even at the 10\%-confidence level, we do not reject the null that our identified set contains effects that are 14\% of the point-estimate associated with the FAO-GAEZ version 3 (Column (4)). When we consider our 99\%-confidence-level interval, we find that the data are consistent with an effect that is close to zero after accounting for non-classical measurement error and sample uncertainty.

	\subsection{Bustos et al. (2016)}

	\cite{bustos2016agricultural} studies the impact of agricultural productivity on structural transformation. The authors focus on the increase of soybean productivity in Brazil after the introduction of genetically modified seeds in 2003. Their empirical specification uses panel data to estimate Equation \eqref{eq:bustos}:
	\begin{equation}
		\label{eq:bustos}
		\Delta y_{it} =  \alpha_t +\beta \cdot \Delta A_{it}^{soy} + \sum_{j = 1}^{J} \alpha_{j} \cdot w_{i,j} + \epsilon_{jt},
	\end{equation}
	where $y_{it}$ is the employment share of the manufacturing sector in municipality $i$ and year $t$, and $\Delta A_{it}^{soy}$ is the soybean productivity shock, which is the variable of interest $X^{*}$. \cite{bustos2016agricultural} exploits an arguably exogenous shock in soybean productivity: the introduction of genetically modified seeds in 2003, i.e., they use this increase in productivity as an exogenous change in soybean productivity.

	The problem is that there is no measure of the increase in productivity generated by genetically modified seeds. Instead, the authors propose to use the difference between soybean productivity for high and low inputs of FAO-GAEZ version 3. This choice naturally raises questions about the existence of measurement error.

	We begin by discussing the differences in soybean productivity in versions 3 and 4. We are able to reproduce the numbers of productivity using \cite{bustos2016agricultural}'s replication package. The correlation between the change in productivity variable in the replication package and the one generated by us is of 0.99.\footnote{There are many reasons for differences in the computed productivity. In the replication package, there is no mention of how the variables were transformed from raster and vector formats to the tabular format the authors use. We use the geometry of the minimum comparable areas that is present in the replication package in order to compute the average productivity within each municipality. The replication package does not have the raw soybean productivity data, but we know of no different vintages of the FAO-GAEZ version 3 and, therefore, we have no reason to believe that the data is different. The differences then are likely to occur from different ways of assigning pixels of soybean productivity to the municipality geometry. For example: are all touched pixels considered? If yes, how? What is the projection used to make compatible the raster file of FAO-GAEZ with the vector file of the municipalities? There are also small holes in some municipalities in the Northeast region that we do not know if it was treated somehow after the data were generated.} To ensure that all the differences we point out are coming from different FAO-GAEZ versions and not different data handling choices, we compare the results across our constructed productivity variables.

	We focus on the difference between soybean productivity with high and low use of inputs ($\Delta A_{i}^{soy} \coloneqq A^{soy}_{i,high} - A^{soy}_{i,low}$), which is interpreted as the productivity shock brought by the genetically modified seeds.  In our setting, the variable $\Delta A_{i}^{soy}$ is the variable of interest $X^{*}$. Table \ref{tab:soy} shows descriptive statistics of the changes in productivity. On average, version 4 measures a higher soybean productivity gain of 170 kg per hectare. Overall, version 4 has lower spatial heterogeneity than version 3.

	\begin{table}[!htbp]
		\begin{center}
			\caption{Descriptive Statistics: Change in Soybean Productivity (ton/ha)\label{tab:soy}}
			\begin{tabular}{lccc}
				\toprule \toprule
				& \multicolumn{3}{c}{$\Delta A_i^{soy}$ (ton/ha)}\\
				& V 3 & V 4 & V 4 - V 3 \\
				& (1) & (2) & (3) \\
				\midrule
				Mean & 1.79  & 1.96  & 0.17 \\
				S.D.  & 0.83  & 0.72  & 0.27 \\\\
				Min  & -0.26 & -0.32 & -1.09 \\
				1\textsuperscript{st} Quartile & 1.17  & 1.40  & 0.02 \\
				Median & 1.82  & 1.94  & 0.15 \\
				3\textsuperscript{rd} Quartile & 2.43  & 2.49  & 0.33 \\
				Max  & 4.00  & 4.00  & 2.27 \\
				\bottomrule
			\end{tabular}
		\end{center}
		\footnotesize{Notes: This table shows descriptive statistics for the difference between soybean productivity with high and low input. The first two columns show the statistics for FAO-GAEZ versions 3 and 4. The third column shows statistics for the difference between versions. The productivity is measured in tons per hectare. ``S.D.'' stands for standard deviation.}
	\end{table}

	Table \ref{tab:results_bustos} shows the estimated results of Equation \eqref{eq:bustos}. We focus on Column (2) of Table 9 of the original article, which reports the impact of agricultural productivity gains on employment share in manufacturing. Column (1) in Table \ref{tab:results_bustos} shows the result as reported by \cite{bustos2016agricultural}, which the authors interpret as an increase of one unit in potential soy productivity leads to an increase of 2.1 p.p. in the share of manufacturing employment.

	\begin{table}[!htb]
		\begin{center}
			\caption{The Effect of Agricultural Technological Change on Manufacturing \label{tab:results_bustos}}
			\begin{tabular}{lccccccc}
				\toprule \toprule
				& \multicolumn{7}{c}{$\Delta$ Employment share}\\
				\cline{2-8}
				& (1) & (2) & (3) & (4) & (5) & (6) & (7) \\
				\hline
				$X^{*}$ & 0.021   & 0.022   & 0.029 & 0.021 & 0.018 & 0.019 & [0.014, 0.056] \\
				& (0.002) & (0.002)  & (0.002) & (0.002) & (0.002) & &  \\
				90\% CI &         &         &          &         &         &&   [0.012, 0.058]  \\
				99\% CI &         &         &          &         &         &&   [0.011, 0.059]  \\
				t-stats &         &         &          &         &         & 246.25 &     \\
				p-value &         &         &          &         &         & 0.000 &     \\\\
				Observations & 4,149   & 4,149   & 4,149 & 4,149 & 4,149 & 4,149    & 4,149 \\
				\bottomrule
			\end{tabular}
		\end{center}
		\footnotesize{Notes: We have that $X^{*} \coloneqq \Delta \text{ soy productivity}$. Observations are at the municipality level. All results use a baseline sample of 4,149 municipalities. Changes in dependent variables are calculated over the years 2000 and 2010. All municipality controls are
			from the population census of 1991: share rural population, log of income per capita, log of population density, literacy rate. Column (1) reports the result of \cite{bustos2016agricultural}'s Column 2 in Table 9, which uses soybean productivity from FAO-GAEZ version 3. To make results more comparable, Columns (2) and (3) report our results using FAO-GAEZ data versions 3 and 4, respectively. Columns (4) and (5) show the results when $\Delta A_{it}^{soy}$ is normalized to have variance equal to one. Column (6) implements the method proposed by \cite{Lubotsky2006}, which accounts for classical measurement error by using an optimally weighted convex combination of the proxy variables. To implement their least biased approach, \citet[Equation (4)]{Lubotsky2006} estimate $Y_{res} = c_{0} + c_{1} \cdot Z_{1,res} + c_{2} \cdot Z_{2,res} + V$ and uses $\hat{c}_{1} + \hat{c}_{2}$ as their estimator. The rows ``t-stats'' and ``p-value'' report the t-statistics and the p-value of a test whose null hypothesis imposes that $c_{1} = 0$ and $c_{2} = 0$, as recommended by \citet[p. 555]{Lubotsky2006}. Column (7) presents our partial identification approach. Standard errors, t-statistics, p-values and confidence intervals are robust to heteroskedasticity. Confidence intervals in Column (7) are built at the 10\% and 1\% levels.}
	\end{table}

	In Columns (2) and (3), we replicate \cite{bustos2016agricultural}'s result using the variables we constructed based on FAO-GAEZ versions 3 and 4. Reassuringly, the result in Column (2) is very close to the one reported in Column (1), since they both use version 3 of the explanatory variable. In Column (3), we estimate Equation \eqref{eq:bustos} using FAO-GAEZ version 4. Comparing Columns (2) and (3), we see a large increase in the estimated effect when we substitute version 3 of the FAO-GAEZ data for version 4. These discrepancies rule out the possibility that both proxies are measured without error. Consequently, treating either version as error-free amounts to imposing a strong and untestable assumption about the data-generating process. This no-measurement-error assumption is relaxed by the method proposed by \cite{Lubotsky2006} and further relaxed by our partial identification approach.

	We can now discuss the results of tools that account for measurement error. The interpretation of these results does not align directly with those shown so far because, in our approach, we normalize the variance of $\Delta A_{it}^{soy}$ to 1 (Assumption \ref{AsScale}). For this reason, when using tools that account for measurement error, we discuss their results in terms of standard deviations of $\Delta A_{it}^{soy}$. To make the comparison easier, Columns (4) and (5) show the result of estimating Expression \eqref{eq:bustos} with FAO-GAEZ versions 3 and 4 after normalizing the main explanatory variable. In Columns (4) and (5), the interpretation of the estimated results is that an increase of one standard deviation in $\Delta A_{it}^{soy}$ causes an increase of 2.1 p.p and 1.8 p.p in the share of manufacturing jobs.

	Column (6) reports estimates based on the method proposed by \cite{Lubotsky2006}. It accounts for classical measurement error by using an optimally weighted convex combination of the normalized proxy variables. To implement their least biased approach, \citet[Equation (4)]{Lubotsky2006} estimate $Y_{res} = c_{0} + c_{1} \cdot Z_{1,res} + c_{2} \cdot Z_{2,res} + V$ and uses $\hat{c}_{1} + \hat{c}_{2}$ as their estimator. This estimate is close to the estimate that uses only FAO-GAEZ Version 4 as a proxy and suggests that an increase of one standard deviation in $\Delta A_{it}^{soy}$ causes an increase of 1.9 p.p. in the share of manufacturing jobs. When we follow the recommendation by \citet[p. 555]{Lubotsky2006} and test the null hypothesis that $c_{1} = 0$ and $c_{2} = 0$, we find a t-statistic of 246.25, indicating that the effect of genetically modified soy is statistically significant at the 1\%-level.

	Column (7) shows the results associated with our method, which accounts for nonclassical measurement error. The estimated interval shows a wide range of possible effects, with an increase of one standard deviation in $\Delta A_{it}^{soy}$ leading to an increase in the share of manufacturing jobs between 1.4 p.p. and 5.6 p.p.\footnote{The estimated lower bound is achieved by its third term (Equation \eqref{EqLB}). Consequently, allowing for unrestricted correlation between the measurement error terms (i.e., removing Assumption \ref{AsNonClassical}) would not impact the estimated lower bound according to Appendix \ref{AppFreeCov}.} Importantly, this interval exhausts all the information contained in the first and second moments of the data-generating process, despite the large estimate for the upper bound. Such a wide estimated interval, especially compared with the small standard errors reported in Columns (2) and (3), suggests (i) that weakening the no-measurement-error assumption increases uncertainty substantially and (ii) that model uncertainty regarding the presence of non-classical measurement error is more concerning than sample uncertainty.

	Column (7) also reports the confidence intervals at the 10\% and 1\% levels. Even at the 10\%-confidence level, we do not reject the null hypothesis that our identified set contains effects that are 59\% of the point-estimate associated with the FAO-GAEZ version 3 (Column (4)). When we consider our 99\%-confidence-level interval, we find that the data are consistent with an effect that is 55\% of the point-estimate associated with the FAO-GAEZ version 3 (Column (4)).

	\subsection{Acharya et al. (2016)}

	\cite{acharya2016political} study the impact of slavery on contemporary political attitudes in the South of the United States. In particular, the authors estimate the following equation:
	\begin{equation*}
		y_{i} = \gamma \cdot S_i + \sum_{j=1}^{J} \delta_j \cdot w_{i,j} + \epsilon_i,
	\end{equation*}
	where $y_i$ is the proportion of republicans in the white population in county $i$ pooled from Cooperative Congressional Election Study for the years 2006, 2008, 2009, 2010 and 2011, and $S_i$ is the proportion of slaves in county $i$ in 1860. The authors point out at least two potential problems with estimating the above expression via OLS: measurement error in the explanatory variable and omitted variable bias. To overcome these concerns the authors propose to use the productivity of cotton from FAO-GAEZ data version 3 as an instrument for slavery $(S_i)$. To fit this approach within our framework, we focus on the reduced-form specification given by Equation \eqref{eq:acharya}:
	\begin{equation}
		\label{eq:acharya}
		y_i = \beta \cdot A_i^{cotton} + \sum_{j=1}^{J} \alpha_j \cdot w_{i,j} + \epsilon_i
	\end{equation}

	The authors use an average of the FAO-GAEZ version 3 with high and intermediate inputs. As FAO-GAEZ version 4 has no intermediate input specification, we reproduce their paper using the high input specification. In our setting, the variable $A_i^{cotton}$ is the variable of interest $X^{*}$.

	We start by discussing the differences in FAO-GAEZ cotton productivity in versions 3 and 4. Table \ref{tab:cotton} shows that, for the counties in the South of the United States, cotton productivity is lower and less heterogeneous in version 4 than in version 3.

	\begin{table}[!htbp]
		\begin{center}
			\caption{Descriptive Statistics: Cotton Productivity (ton/ha)\label{tab:cotton}}
			\begin{tabular}{lccc}
				\toprule \toprule
				& \multicolumn{3}{c}{$A_i^{cotton}$ (ton/ha)}\\
				& V 3 & V 4 & V 4 - V 3 \\
				& (1) & (2) & (3) \\
				\midrule
				Mean & 0.56 & 0.45 & -0.11 \\
				S.D.  & 0.27 & 0.21 & 0.12 \\\\
				Min  & 0.00 & 0.00 & -0.66 \\
				1\textsuperscript{st} Quartile & 0.42 & 0.31 & -0.17 \\
				Median & 0.62 & 0.50 & -0.12 \\
				3\textsuperscript{rd} Quartile & 0.77 & 0.62 & -0.04 \\
				Max  & 1.17 & 0.95 & 0.52 \\
				\bottomrule
			\end{tabular}
		\end{center}
		\footnotesize{Notes: This table shows descriptive statistics for the cotton productivity with high input. The first two columns show the statistics for FAO-GAEZ versions 3 and 4. The third column shows statistics for the difference between versions. The productivity is measured in tons per hectare. ``S.D.'' stands for standard deviation.}
	\end{table}

	In Table \ref{tab:results_acharya}, we estimate Equation \eqref{eq:acharya} with the FAO-GAEZ data versions 3 and 4. Once more, to make the results comparable to our partial identification approach, we normalize the $A^{cotton}_i$ variable to have a variance of 1. The interpretation is, then, that an increase of one standard deviation in cotton productivity generates an increase of the share of Republicans in the White population of 2 p.p using FAO-GAEZ version 3 (Column (1)) and a non-significant effect of 0.9 p.p. using FAO-GAEZ version 4 (Column (2)). These discrepancies rule out the possibility that both proxies are measured without error. Consequently, treating either version as error-free amounts to imposing a strong and untestable assumption about the data-generating process. This argument highlights the importance of using tools that account for measurement error, as done by the method proposed by \cite{Lubotsky2006} and our partial identification approach.

	\begin{table}[!htb]
		\begin{center}
			\caption{The Effect of Cotton Productivity on the Proportion of Republicans in the White Population \label{tab:results_acharya}}
			\begin{tabular}{lcccc}
				\toprule \toprule
				& \multicolumn{4}{c}{Proportion of Republicans}\\
				\cline{2-5}
				& (1) & (2) & (3) & (4) \\
				\hline
				Cotton Productivity & 0.020    & 0.009 & 0.020   & [0.010, 0.245] \\
				& (0.010)  & (0.008)  & &  \\
				90\% CI             &          &          & &   [0.000, 0.257]  \\
				99\% CI             &          &          & &   [0.000, 0.261]  \\
				t-stats             &          &          & 4.701 &     \\
				p-value             &          &          & 0.095 &     \\\\
				Observations        & 1,180    & 1,180    & 1,180 \\
				\bottomrule
			\end{tabular}
		\end{center}
		\footnotesize{Notes: Observations are at the county level. All results include state fixed effects and the following controls as measured in 1860: log of the total county population, proportion of farms in the county smaller than 50 acres, inequality of farmland holdings as measured by the Gini coefficient for landownership, log of total farm value per improved acre of farmland in the county, log of the acres of improved farmland; the proportion of total population that is free black, access to rails and steamboat-navigable rivers or canals, log of the county acreage, ruggedness of the county, latitude and longitude as well as their squared terms. We keep the weighting scheme of the original paper with the within-county sample size (appropriately weighted by the sampling weights) as weights. Columns (1) and (2) report the results of the reduced form counterpart of \cite{acharya2016political}'s Column 2 in Table 2 using cotton productivity from FAO-GAEZ versions 3 and 4, respectively. Column (3) implements the method proposed by \cite{Lubotsky2006}, which accounts for classical measurement error by using an optimally weighted convex combination of the proxy variables. To implement their least biased approach, \citet[Equation (4)]{Lubotsky2006} estimate $Y_{res} = c_{0} + c_{1} \cdot Z_{1,res} + c_{2} \cdot Z_{2,res} + V$ and uses $\hat{c}_{1} + \hat{c}_{2}$ as their estimator. The rows ``t-stats'' and ``p-value'' report the t-statistics and the p-value of a test whose null hypothesis imposes that $c_{1} = 0$ and $c_{2} = 0$, as recommended by \citet[p. 555]{Lubotsky2006}. Column (4) presents results associated with our partial identification approach. Standard errors, t-statistics, p-values and confidence intervals are robust to heteroskedasticity. Confidence intervals in Column (4) are built at the 10\% and 1\% levels.}
	\end{table}

	Column (3) reports estimates based on the method proposed by \cite{Lubotsky2006}. It accounts for classical measurement error by using an optimally weighted convex combination of the normalized proxy variables. To implement their least biased approach, \citet[Equation (4)]{Lubotsky2006} estimate $Y_{res} = c_{0} + c_{1} \cdot Z_{1,res} + c_{2} \cdot Z_{2,res} + V$ and uses $\hat{c}_{1} + \hat{c}_{2}$ as their estimator. This estimate is close to the estimate that uses only FAO-GAEZ Version 3 as a proxy, and suggests that an increase of one standard deviation in cotton productivity increases the share of Republicans in the White population by 2.0 p.p. When we follow the recommendation by \citet[p. 555]{Lubotsky2006} and test the null hypothesis that $c_{1} = 0$ and $c_{2} = 0$, we find a t-statistic of 4.701, indicating that the effect of cotton productivity is statistically significant at the 10\%-level.

	Column (4) shows the results associated with our method, which accounts for nonclassical measurement error. The estimated interval shows a wide range of possible effects, with an increase of one standard deviation in cotton productivity leading to an increase in the share of Republicans in the White population between 1.0 p.p. and 24.5 p.p.\footnote{The estimated lower bound is achieved by its third term (Equation \eqref{EqLB}). Consequently, allowing for unrestricted correlation between the measurement error terms (i.e., removing Assumption \ref{AsNonClassical}) would not impact the estimated lower bound according to Appendix \ref{AppFreeCov}.} Importantly, this interval exhausts all the information contained in the first and second moments of the data-generating process, despite the large estimate for the upper bound. Such a wide estimated interval, especially compared with the small standard errors reported in Columns (2) and (3), suggests (i) that weakening the no-measurement-error assumption increases uncertainty substantially and (ii) that model uncertainty regarding the presence of non-classical measurement error is more concerning than sample uncertainty.

	Column (4) also reports the confidence intervals at the 10\% and 1\% levels. Regardless of the confidence level, we do not reject the null hypothesis that the lower bound of the true effect is zero.\footnote{The estimated 90\%-confidence interval is $\left[-0.003, 0.257\right]$ while the 99\%-confidence interval is $\left[-0.009, 0.262\right]$. In Column (3) of Table \ref{tab:results_acharya}, we report the intersection of these estimated intervals with $\mathbb{R}_{+}$ due to Assumption \ref{AsPosEff}.} Consequently, the data are consistent with a zero effect after we account for non-classical measurement error.

	\section{Recommendations to the Practitioner}\label{Srecommendation}

	In this section, we provide a detailed 4-step guide on how to use our method in any empirical application involving a mismeasured treatment variable and two proxies.

	First, the practitioner must reflect on the exogeneity of the correctly measured treatment variable (Assumption \ref{AsExogeneous}). If the true treatment were observable, could the practitioner identify the treatment effect of interest using a linear regression model? This type of argument is present in any empirical application and practitioners still need to think about it when using our proposed method. If the answer is positive, proceed to the next step. If the answer is negative, the practitioner should use a different approach (e.g., instrumental variables).

	Second, the practitioner must reflect on the exogeneity of the measurement errors (Assumption \ref{AsMEexogenous}), on their priors about the sign of the coefficient of interest (Assumption \ref{AsPosEff}), and on the sign of the correlation between the measurement error terms (Assumption \ref{AsNonClassical}).

	To think about the exogeneity of the measurement errors, the practitioner must think about the causes of the measurement error. Is it due to arbitrary modeling decisions about imputation or interpolation methods? Or is it due to better institutions being able to collect better data? If the first scenario is more likely, then Assumption \ref{AsMEexogenous} is plausible. If the practitioner believes that Assumption \ref{AsMEexogenous} is valid, proceed to the next step.

	To think about the sign of the coefficient of interest, the practitioner may compare different economic theories or analyze previous empirical results using different datasets. Do they agree about the direction of the true effect? If so, Assumption \ref{AsPosEff} is plausible.

	To think about the sign of the correlation between the measurement error terms, the practitioner must, once more, think about the causes of the measurement error. If it is hard to collect high-quality data on some regions, then Assumption \ref{AsNonClassical} is plausible. If measurement error is due to different modeling choices, then Assumption \ref{AsNonClassical} may not be plausible.

	\begin{enumerate}
		\item If the practitioner believes that Assumptions \ref{AsPosEff}, \ref{AsMEexogenous} and \ref{AsNonClassical} are plausible, then proceed with the method described in Section \ref{Sidentification}.

		\item If the practitioner believes that only Assumptions \ref{AsMEexogenous} and \ref{AsNonClassical} are plausible, then proceed with the method described in Appendix \ref{AppNoSign}.

		\item If the practitioner believes that only Assumptions \ref{AsPosEff} and \ref{AsNonClassical} are plausible, then proceed with the method described in Appendix \ref{AppEndoME}.

		\item If the practitioner believes that only Assumptions \ref{AsPosEff} and \ref{AsMEexogenous} are plausible, then proceed with the method described in Appendix \ref{AppFreeCov}.
	\end{enumerate}

	Third, the practitioner must estimate the bounds around the true coefficient of interest using the estimator proposed in Appendix \ref{AppEstimator}. The estimated bounds are interesting for at least three aspects:
	\begin{enumerate}
		\item If the first term in the estimated lower bound is similar to the OLS estimate, then the results are compatible with no measurement error, and the practitioner may focus their analysis on their OLS results.

		\item If the estimated lower bound is numerically close to zero, then the results are compatible with the true coefficient of interest being small. This type of result suggests that the economic relevance of the target parameter may not be large. This scenario is illustrated by our analysis of the dataset used by \cite{acharya2016political}.

		\item If the estimated interval is wide, then the model uncertainty generated by the possible existence of measurement error is relevant. This scenario is illustrated by our analysis of the datasets used by \cite{nunn2011} and \cite{bustos2016agricultural}.
	\end{enumerate}

	Lastly, the practitioner must estimate the confidence intervals around the identified set using the algorithm proposed in Appendix \ref{AppInference}. If the confidence interval contains zero, then the ``null hypothesis that the identified set contains zero'' is not rejected. In this case, the data are consistent with a zero effect of $X^{*}$ on $Y$ when we account for non-classical measurement error.

	\section{Conclusion}\label{Sconclusion}

	In this paper, we develop a novel partial identification method to identify the effect of a mismeasured variable when two proxies are available. Importantly, our estimator relies on assumptions no stronger than those commonly applied in linear models that ignore measurement error. Moreover, our partial identification approach provides bounds that (i) incorporate all available information from the first two moments of the data distribution and (ii) compare favorably against alternative approaches, such as a simple reduced-form estimand and the method proposed by \cite{Lubotsky2006}.

	We further explore the implications of relaxing our assumptions. In particular, we relax restrictions on the sign of the target parameter (Appendix \ref{AppNoSign}), on measurement error exogeneity (Appendix \ref{AppEndoME}) and the correlation of measurement errors (Appendix \ref{AppFreeCov}), deriving wider identified intervals in all cases. Thus, modeling assumptions should be chosen in a way that best fits each application.

	To illustrate our methodology, we reevaluate the results of three empirical studies: \citet{nunn2011}, \citet{bustos2016agricultural} and \cite{acharya2016political}. To do so, we replicate their findings using the reduced-form approach, which ignores measurement error, and compare these results with those obtained through our partial identification method. Once we allow for nonclassical measurement error, we find wide intervals around the target parameters, indicating that the magnitudes of agricultural productivity effects can change dramatically. These findings call for substantial caution in interpreting existing results. In particular, our wide estimated intervals highlight that our knowledge of these effects is limited.

\singlespace

\bibliography{refs1}


\pagebreak

\newpage

\pagebreak

\setcounter{table}{0}
\renewcommand\thetable{A.\arabic{table}}

\setcounter{figure}{0}
\renewcommand\thefigure{A.\arabic{figure}}

\setcounter{equation}{0}
\renewcommand\theequation{A.\arabic{equation}}

\appendix

\begin{center}
	{\Large Potato Potahto in the FAO-GAEZ Productivity Measures? \\ Nonclassical Measurement Error with Multiple Proxies} \medskip \\ \Large (Online Appendix) \bigskip \\
	\large Rafael Araujo  \hspace{0.3cm} Vitor Possebom \medskip\\
	\monthyeardate\today
	\bigskip
\end{center}

This online appendix contains proofs, additional results, and detailed discussions related to the paper ``Potato Potahto in the FAO-GAEZ Productivity Measures? Nonclassical Measurement Error with Multiple Proxies'' by Rafael Araujo and Vitor Possebom. Section \ref{AppProofMainId} contains proofs and auxiliary lemmas for results stated in the main text.  Sections \ref{AppNoSign}, \ref{AppEndoME} and \ref{AppFreeCov} present partial identification results using weaker sets of assumptions than the ones used in the main text. Section \ref{AppEstimation} explains how to estimate our bounds and how to conduct statistical inference about the identified set. Section \ref{AppSourceError} discusses potential sources of measurement error in the FAO-GAEZ data. Section \ref{AppOther} discusses other empirical contexts where our partial identification strategy may be useful. Section \ref{AppExample} presents simple examples comparing our proposed tool against alternative methods. Section \ref{AppPartialIdentification} intuitively explains the concept of partial identification. Lastly, Section \ref{AppPrimitive} presents primitive assumptions that imply the assumptions discussed in the main text.

\doublespacing
\normalsize

\pagebreak

\section{Proof of Proposition \ref{PropMainId}}\label{AppProofMainId}

\setcounter{table}{0}
\renewcommand\thetable{A.\arabic{table}}

\setcounter{figure}{0}
\renewcommand\thefigure{A.\arabic{figure}}

\setcounter{equation}{0}
\renewcommand\theequation{A.\arabic{equation}}

\setcounter{theorem}{0}
\renewcommand\thetheorem{A.\arabic{theorem}}

\setcounter{proposition}{0}
\renewcommand\theproposition{A.\arabic{proposition}}

\setcounter{corollary}{0}
\renewcommand\thecorollary{A.\arabic{corollary}}

\setcounter{assumption}{0}
\renewcommand\theassumption{A.\arabic{assumption}}

\setcounter{definition}{0}
\renewcommand\thedefinition{A.\arabic{definition}}

\setcounter{Lemma}{0}
\renewcommand\theLemma{A.\arabic{Lemma}}

We break the proof of Proposition \ref{PropMainId} in 4 parts:
\begin{enumerate}
	\item Section \ref{AppProofDataRestriction} derives the restrictions imposed by the first two moments of the observable data on the underlying data-generating process.

	\item Section \ref{AppProofLbBetaUB} shows that $LB \leq \beta \leq UB$ if $UB \geq LB$.

	\item Section \ref{AppProofNoDGP} shows that, if $LB > UB$, then there is no data-generating process that satisfies Assumptions \ref{AsExogeneous}-\ref{AsDataRestrictions} and the restrictions imposed by the first two moments of the data distribution.

	\item Section \ref{AppProofAllInfo} defines the meaning of exhausting all the information contained in the first two moments of the data distribution and shows that, if $UB \geq LB$, then the interval $\left[LB,UB \right]$ exhausts all the information contained in the first two moments of the data distribution.
\end{enumerate}

\subsection{Data Restrictions}\label{AppProofDataRestriction}

In this section, we derive the restrictions imposed by the first two moments of the observable data distribution $\left(Y_{res}, Z_{1,res}, Z_{2,res}\right)$ on the distribution of the latent variables $\left(X_{res}^{*}, \epsilon_{res},U_{1},U_{2}\right)$. First, we state all nine data restrictions in a lemma. Afterwards, we prove them.

\begin{Lemma}[Data Restrictions]\label{LemmaDataRestrictions}
	Under Assumptions \ref{AsExogeneous}, \ref{AsScale} and \ref{AsMEexogenous}, we have that
	\begin{align}
		\label{EqExpY} \mathbb{E}\left[Y_{res}\right] & = 0 \\
		\label{EqExpZ1} \mathbb{E}\left[Z_{1,res}\right] & = \mathbb{E}\left[U_{1}\right] \\
		\label{EqExpZ2} \mathbb{E}\left[Z_{2,res}\right] & = \mathbb{E}\left[U_{2}\right] \\
		\label{EqVarY} Var\left(Y_{res}\right) & = \beta^{2} + Var\left(\epsilon_{res}\right) \\
		\label{EqVarZ1} Var\left(Z_{1,res}\right) & = 1 + Var\left(U_{1}\right) + 2 \cdot Cov\left(X_{res}^{*}, U_{1}\right) \\
		\label{EqVarZ2} Var\left(Z_{2,res}\right) & = 1 + Var\left(U_{2}\right) + 2 \cdot Cov\left(X_{res}^{*}, U_{2}\right) \\
		\label{EqCovZ1Y} Cov\left(Z_{1,res}, Y_{res}\right) & = \beta + \beta \cdot Cov\left(X_{res}^{*}, U_{1}\right) \\
		\label{EqCovZ2Y} Cov\left(Z_{2,res}, Y_{res}\right) & = \beta + \beta \cdot Cov\left(X_{res}^{*}, U_{2}\right) \\
		\label{EqCovZ1Z2} Cov\left(Z_{1,res}, Z_{2,res}\right) & = 1 + Cov\left(X_{res}^{*}, U_{1}\right) + Cov\left(X_{res}^{*}, U_{2}\right) + Cov\left(U_{1}, U_{2}\right).
	\end{align}
\end{Lemma}

\begin{proof}
	To prove Equation \eqref{EqExpY}, note that:
	\begin{align*}
		\mathbb{E}\left[Y_{res}\right] & = \mathbb{E}\left[\beta \cdot X_{res}^{*} + \epsilon_{res}\right] & \text{by Equation \eqref{EqModel}} \\
		& = \beta \cdot \mathbb{E}\left[X_{res}^{*}\right] + \mathbb{E}\left[\epsilon_{res}\right] & \text{by linearity of the expectation operator} \\
		& = 0 & \text{because } X_{res}^{*} \text{ and } \epsilon_{res} \text{ are residuals.}
	\end{align*}

	To prove Equations \eqref{EqExpZ1} and \eqref{EqExpZ2}, fix $k \in \left\lbrace 1, 2 \right\rbrace$ arbitrarily and observe that
	\begin{align*}
		\mathbb{E}\left[Z_{k,res}\right] & = \mathbb{E}\left[ X_{res}^{*} + U_{k}\right] & \text{by Equation \eqref{EqProxiesCovariates}} \\
		& = \mathbb{E}\left[X_{res}^{*}\right] + \mathbb{E}\left[U_{k}\right] & \text{by linearity of the expectation operator} \\
		& = \mathbb{E}\left[U_{k}\right] & \text{because } X_{res}^{*} \text{ is a residual.}
	\end{align*}

	To prove Equation \eqref{EqVarY}, notice that:
	\begin{align*}
		Var\left(Y_{res}\right) & = Var\left(\beta \cdot X_{res}^{*} + \epsilon_{res}\right) \\
		& \hspace{20pt} \text{by Equation \eqref{EqModel}} \\
		& = \beta^{2} \cdot Var\left(X_{res}^{*}\right) + Var\left(\epsilon_{res}\right) + 2 \cdot \beta \cdot Cov\left(X_{res}^{*}, \epsilon_{res}\right) \\
		& \hspace{20pt} \text{by the properties of the variance} \\
		& = \beta^{2} + Var\left(\epsilon_{res}\right) \\
		& \hspace{20pt} \text{by Assumptions \ref{AsExogeneous} and \ref{AsScale}.}
	\end{align*}

	To prove Equations \eqref{EqVarZ1} and \eqref{EqVarZ2}, fix $k \in \left\lbrace 1, 2 \right\rbrace$ arbitrarily and observe that
	\begin{align*}
		Var\left(Z_{k,res}\right) & = Var\left(X_{res}^{*} + U_{k}\right) \\
		& \hspace{20pt} \text{by Equation \eqref{EqProxiesCovariates}} \\
		& = Var\left(X_{res}^{*}\right) + Var\left(U_{k}\right) + 2 \cdot Cov\left(X_{res}^{*}, U_{k}\right) \\
		& \hspace{20pt} \text{by the properties of the variance} \\
		& = 1 + Var\left(U_{k}\right) + 2 \cdot Cov\left(X_{res}^{*}, U_{k}\right) \\
		& \hspace{20pt} \text{by Assumption \ref{AsScale}.}
	\end{align*}

	To prove Equations \eqref{EqCovZ1Y} and \eqref{EqCovZ2Y}, fix $k \in \left\lbrace 1, 2 \right\rbrace$ arbitrarily and note that
	\begin{align*}
		& Cov\left(Z_{k,res},Y_{res}\right) \\
		& \hspace{20pt} = Cov\left(X_{res}^{*} + U_{k}, \beta \cdot X_{res}^{*} + \epsilon_{res}\right) \\
		& \hspace{40pt} \text{by Equations \eqref{EqModel} and \eqref{EqProxiesCovariates}} \\
		& \hspace{20pt} = \beta \cdot Var\left(X_{res}^{*}\right) + Cov\left(X_{res}^{*}, \epsilon_{res}\right) + \beta \cdot Cov\left(X_{res}^{*}, U_{k}\right) + Cov\left(U_{k}, \epsilon_{res}\right) \\
		& \hspace{40pt} \text{by the properties of the covariance} \\
		& \hspace{20pt} = \beta + \beta \cdot Cov\left(X_{res}^{*}, U_{k}\right) \\
		& \hspace{40pt} \text{by Assumptions \ref{AsExogeneous}, \ref{AsScale} and \ref{AsMEexogenous}.}
	\end{align*}

	To prove Equation \eqref{EqCovZ1Z2}, observe that:
	\begin{align*}
		& Cov\left(Z_{1,res},Z_{2,res}\right) \\
		& \hspace{20pt} = Cov\left(X_{res}^{*} + U_{1}, X_{res}^{*} + U_{2}\right) \\
		& \hspace{40pt} \text{by Equation \eqref{EqProxiesCovariates}} \\
		& \hspace{20pt} = Var\left(X_{res}^{*}\right) + Cov\left(X_{res}^{*}, U_{1}\right) + Cov\left(X_{res}^{*}, U_{2}\right) + Cov\left(U_{1}, U_{2}\right) \\
		& \hspace{40pt} \text{by the properties of the covariance} \\
		& \hspace{20pt} = 1 + Cov\left(X_{res}^{*}, U_{1}\right) + Cov\left(X_{res}^{*}, U_{2}\right) + Cov\left(U_{1}, U_{2}\right) \\
		& \hspace{40pt} \text{by Assumption \ref{AsScale}.}
	\end{align*}
\end{proof}

\subsection{$LB \leq \beta \leq UB$ if $UB \geq LB$}\label{AppProofLbBetaUB}

In this section, we consider the case when $UB \geq LB$ and prove that (i) $\beta \leq UB$, (ii) $\beta \geq LB$ when $\beta = 0$, and (iii) $\beta \geq LB$ when $\beta > 0$. In this proof, we do not impose that $LB \geq 0$.

\textbf{Part 1: $\beta \leq UB$}

Rearranging Equation \eqref{EqVarY}, we have that
\begin{align*}
	\beta^{2} & = Var\left(Y_{res}\right) - Var\left(\epsilon_{res}\right) \\
	& \leq Var\left(Y_{res}\right)
\end{align*}
because $Var\left(\epsilon_{res}\right) \geq 0$, implying that $$\beta \leq \sqrt{Var\left(Y_{res}\right)} = UB$$ by Assumption \ref{AsPosEff}.

~

\textbf{Part 2: $\beta \geq LB$ when $\beta = 0$}

By combining Equations \eqref{EqCovZ1Y} and \eqref{EqCovZ2Y} with $\beta = 0$, we find that $$Cov\left(Z_{1,res}, Y_{res} \right) = 0$$ and $$Cov\left(Z_{2,res}, Y_{res} \right) = 0,$$ implying that $$LB = 0,$$ according to Equation \eqref{EqLB}, and that $$\beta = 0 = LB.$$

\textbf{Part 3: $\beta \geq LB$ when $\beta > 0$}

Having in mind that $\beta > 0$ and rearranging Equations \eqref{EqCovZ1Y} and \eqref{EqCovZ2Y}, we have that
\begin{equation}\label{EqCovXU1}
	Cov\left(X_{res}^{*}, U_{1}\right) = \dfrac{Cov\left(Z_{1,res}, Y_{res}\right)}{\beta} - 1
\end{equation}
and
\begin{equation}\label{EqCovXU2}
	Cov\left(X_{res}^{*}, U_{2}\right) = \dfrac{Cov\left(Z_{2,res}, Y_{res}\right)}{\beta} - 1.
\end{equation}

Rearranging Equation \eqref{EqVarZ1}, we find that
\begin{align}
	Var\left(U_{1}\right) & = Var\left(Z_{1,res}\right) - 1 - 2 \cdot Cov\left(X_{res}^{*}, U_{1}\right) \nonumber \\
	& = Var\left(Z_{1,res}\right) - 1 - 2 \cdot \left[ \dfrac{Cov\left(Z_{1,res}, Y_{res}\right)}{\beta} - 1 \right] \nonumber \\
	& \hspace{20pt} \text{according to Equation \eqref{EqCovXU1}} \nonumber \\
	& \label{EqVarU1} = 1 + Var\left(Z_{1,res}\right) - 2 \cdot \dfrac{Cov\left(Z_{1,res}, Y_{res}\right)}{\beta}
\end{align}

{
	Before showing that $\beta \geq LB$ when $\beta > 0$, we show that $\beta \geq LB^{\star}$ when $\beta > 0$, where
	\begin{equation}\label{EqLBstar}
		LB^{*} \coloneqq \max \left\lbrace \begin{matrix}
			\dfrac{Cov\left(Z_{1,res}, Y_{res}\right)}{\sqrt{Var\left(Z_{1,res}\right)}}, \hspace{10pt} \dfrac{Cov\left(Z_{2,res}, Y_{res}\right)}{\sqrt{Var\left(Z_{2,res}\right)}}, \\
			\\
			\dfrac{Cov\left(Z_{1,res}, Y_{res}\right) + Cov\left(Z_{2,res}, Y_{res}\right)}{1 + Cov\left(Z_{1,res}, Z_{2,res}\right)}, \hspace{10pt} \dfrac{-B + \sqrt{B^2 - 4 \cdot A \cdot C}}{2 \cdot C}, \\
			\\
			\sqrt{\dfrac{\left\lbrace \begin{matrix}
						Cov(Z_{1,res},Y_{res})^2 \cdot Var(Z_{2,res}) + Cov(Z_{2,res},Y_{res})^2 \cdot Var(Z_{1,res}) \\
						- 2\cdot Cov(Z_{1,res},Y_{res}) \cdot Cov(Z_{2,res},Y_{res}) \cdot Cov(Z_{1,res},Z_{2,res})
					\end{matrix} \right\rbrace}{Var(Z_{1,res}) \cdot Var(Z_{2,res}) - Cov(Z_{1,res},Z_{2,res})^2}}, \\
			\\
			2 \cdot \dfrac{Cov\left(Z_{1,res}, Y_{res}\right)}{1 + Var\left(Z_{1,res}\right)}, \hspace{10pt} 2 \cdot \dfrac{Cov\left(Z_{2,res}, Y_{res}\right)}{1 + Var\left(Z_{2,res}\right)}
		\end{matrix}  \right\rbrace,
	\end{equation}
	Afterward, we show that the first, second, sixth, and seventh terms are non-binding, implying that $\beta \geq LB$ when $\beta > 0$.
}

\textbf{Part 3.1: $\beta \geq LB^{\star}$ when $\beta > 0$}

\textbf{First Term in $LB^{\star}$:} Since $Var\left(U_{1}\right) \geq \left[Cov\left(X_{res}^{*}, U_{1}\right)\right]^{2}$ by the Cauchy–Schwarz inequality and Assumption \ref{AsScale}, Equations \eqref{EqCovXU1} and \eqref{EqVarU1} imply that
\begin{equation*}
	\begin{array}{crcl}
		& 1 + Var\left(Z_{1,res}\right) - 2 \cdot \dfrac{Cov\left(Z_{1,res}, Y_{res}\right)}{\beta} & \geq & \left[\dfrac{Cov\left(Z_{1,res}, Y_{res}\right)}{\beta} - 1\right]^2 \\
		\Leftrightarrow & 1 + Var\left(Z_{1,res}\right) - 2 \cdot \dfrac{Cov\left(Z_{1,res}, Y_{res}\right)}{\beta} & \geq & \dfrac{\left[Cov\left(Z_{1,res}, Y_{res}\right)\right]^{2}}{\beta^{2}} - 2 \cdot \dfrac{Cov\left(Z_{1,res}, Y_{res}\right)}{\beta} + 1 \\
		\Leftrightarrow & \beta^{2} & \geq & \dfrac{\left[Cov\left(Z_{1,res}, Y_{res}\right)\right]^{2}}{Var\left(Z_{1,res}\right)}
	\end{array}
\end{equation*}
implying that
\begin{equation}\label{EqBetaZ1}
	\beta \geq \dfrac{Cov\left(Z_{1,res}, Y_{res}\right)}{\sqrt{Var\left(Z_{1,res}\right)}}
\end{equation}
because $\beta > 0$.

\textbf{Second Term in $LB^{\star}$:} Combining Equations \eqref{EqVarZ2} and \eqref{EqCovXU2} and using the same steps used to derive Equation \eqref{EqBetaZ1}, we have that
\begin{equation}\label{EqBetaZ2}
	\beta \geq \dfrac{Cov\left(Z_{2,res}, Y_{res}\right)}{\sqrt{Var\left(Z_{2,res}\right)}}.
\end{equation}

\textbf{Third Term in $LB^{\star}$:} Substituting Equations \eqref{EqCovXU1} and \eqref{EqCovXU2} in Equation \eqref{EqCovZ1Z2}, we have that
\begin{align}
	Cov\left(Z_{1,res}, Z_{2,res}\right) & = 1 + Cov\left(X_{res}^{*}, U_{1}\right) + Cov\left(X_{res}^{*}, U_{2}\right) + Cov\left(U_{1}, U_{2}\right) \nonumber \\
	& = 1 + \left[\dfrac{Cov\left(Z_{1,res}, Y_{res}\right)}{\beta} - 1\right] + \left[\dfrac{Cov\left(Z_{2,res}, Y_{res}\right)}{\beta} - 1\right] + Cov\left(U_{1}, U_{2}\right) \nonumber \\
	& = -1 + \dfrac{Cov\left(Z_{1,res}, Y_{res}\right) + Cov\left(Z_{2,res}, Y_{res}\right)}{\beta} + Cov\left(U_{1}, U_{2}\right) \label{EqCovU1U2third}\\
	& \geq -1 + \dfrac{Cov\left(Z_{1,res}, Y_{res}\right) + Cov\left(Z_{2,res}, Y_{res}\right)}{\beta} \nonumber \\
	& \hspace{20pt} \text{by Assumption \ref{AsNonClassical}} \nonumber
\end{align}
implying that
\begin{equation}\label{EqBetaZ1Z2}
	\beta \geq \dfrac{Cov\left(Z_{1,res}, Y_{res}\right) + Cov\left(Z_{2,res}, Y_{res}\right)}{1 + Cov\left(Z_{1,res}, Z_{2,res}\right)}
\end{equation}
because $\beta > 0$ and $Cov\left(Z_{1,res}, Z_{2,res}\right) > -1$ according to Assumption \ref{AsDataRestrictions}.

\textbf{Fourth Term in $LB^{\star}$:} Since $Var\left(U_{1}\right) \cdot Var\left(U_{2}\right) \geq \left[Cov\left(U_{1}, U_{2}\right)\right]^{2}$ by the Cauchy–Schwarz inequality, Equations \eqref{EqCovXU1}, \eqref{EqCovXU2} and \eqref{EqCovU1U2third} imply that
\begin{align}
	& \left(Var\left(Z_{1,res}\right) + 1 - 2 \cdot \frac{ Cov\left(Z_{1,res}, Y_{res}\right)}{\beta}\right)\left(Var\left(Z_{2,res}\right) + 1 - 2 \cdot \frac{  Cov\left(Z_{2,res}, Y_{res}\right)}{\beta}\right) \nonumber \\
	& \hspace{20pt} \geq \left(Cov\left(Z_{1,res}, Z_{2,res}\right) + 1 - \frac{Cov\left(Z_{1,res}, Y_{res}\right) + Cov\left(Z_{2,res}, Y_{res}\right)}{\beta}\right)^2.\label{EqLargeInequality1}
\end{align}
To simplify our notation, let $x \coloneqq \dfrac{1}{\beta}$. Expanding the terms in the left-hand side of Equation \eqref{EqLargeInequality1}, we have that it is given by
\begin{align}
	& (Var\left(Z_{1,res}\right) + 1) \cdot (Var\left(Z_{2,res}\right) + 1) \nonumber \\
	& \hspace{20pt} - 2 \cdot x \cdot [Cov\left(Z_{1,res}, Y_{res}\right) \cdot (Var\left(Z_{2,res}\right) + 1) + Cov\left(Z_{2,res}, Y_{res}\right) \cdot (Var\left(Z_{1,res}\right) + 1)] \nonumber \\
	& \hspace{20pt} + 4 \cdot Cov\left(Z_{1,res}, Y_{res}\right) \cdot Cov\left(Z_{2,res}, Y_{res}\right) \cdot x^2 \label{EqLargeInequality2}
\end{align}
Expanding the terms in the right-hand side of Equation \eqref{EqLargeInequality1}, we have that it is given by
\begin{align}
	& (Cov\left(Z_{1,res}, Z_{2,res}\right) + 1)^2 \nonumber \\
	& \hspace{20pt} - 2 \cdot x \cdot (Cov\left(Z_{1,res}, Z_{2,res}\right) + 1) \cdot (Cov\left(Z_{1,res}, Y_{res}\right) + Cov\left(Z_{2,res}, Y_{res}\right)) \nonumber \\
	& \hspace{20pt} + (Cov\left(Z_{1,res}, Y_{res}\right) + Cov\left(Z_{2,res}, Y_{res}\right))^2 \cdot x^2. \label{EqLargeInequality3}
\end{align}

Combining Equations \eqref{EqLargeInequality1}-\eqref{EqLargeInequality3}, we have that
\begin{equation*}
	Ax^2 + Bx + C \geq 0,
\end{equation*}
where $A$, $B$ and $C$ are defined at the beginning of Section \ref{SidResults}. Now, we use the fact that $x = \dfrac{1}{\beta}$ to rewrite the last equation as
\begin{equation}
	\label{EqSquaredX}
	A + B \cdot \beta + C \cdot \beta^{2} \geq 0,
\end{equation}

Note that $C > 0$ by Assumption \ref{AsDataRestrictions}, implying that the parabola in Equation \eqref{EqSquaredX} opens upward.\footnote{\label{FootCS} To see that $C > 0$, note that $\left[Var\left(Z_{1,res}\right) + 1\right] \cdot \left[Var\left(Z_{2,res}\right) + 1\right] - \left[Cov\left(Z_{1,res}, Z_{2,res}\right) + 1\right]^{2} = 1 + Var\left(Z_{1,res}\right) + Var\left(Z_{2,res}\right) + Var\left(Z_{1,res}\right) \cdot Var\left(Z_{2,res}\right) - \left[Cov\left(Z_{1,res}, Z_{2,res}\right)\right]^{2} - 2 \cdot Cov\left(Z_{1,res}, Z_{2,res}\right) - 1 = Var\left(Z_{1,res}\right) + Var\left(Z_{2,res}\right) - 2 \cdot Cov\left(Z_{1,res}, Z_{2,res}\right) + Var\left(Z_{1,res}\right) \cdot Var\left(Z_{2,res}\right) - \left[Cov\left(Z_{1,res}, Z_{2,res}\right)\right]^{2} = Var\left(Z_{1,res} - Z_{2,res}\right) + Var\left(Z_{1,res}\right) \cdot Var\left(Z_{2,res}\right) - \left[Cov\left(Z_{1,res}, Z_{2,res}\right)\right]^{2} \geq Var\left(Z_{1,res}\right) \cdot Var\left(Z_{2,res}\right) - \left[Cov\left(Z_{1,res}, Z_{2,res}\right)\right]^{2} > 0$, where first inequality relies on $Var\left(Z_{1,res} - Z_{2,res}\right) \geq 0$ and the second (strict) inequality relies on the Cauchy–Schwarz inequality and Assumption \ref{AsDataRestrictions}.}

Observe that Equation \eqref{EqSquaredX} is satisfied if $\beta$ is less than or equal to the smallest root of  ``$A + B \cdot \beta + C \cdot \beta^{2} = 0$'' or greater than or equal to the largest root of ``$A + B \cdot \beta + C \cdot \beta^{2} = 0$''.

The smallest root of ``$A + B \cdot \beta + C \cdot \beta^{2} = 0$'' is given by $$\beta_{1} \coloneqq \dfrac{- B - \sqrt{B^2 - 4 \cdot A \cdot C}}{2 \cdot C}.$$ Since $A \leq 0$ and $C > 0$, we have that $\beta_{1} \leq 0$, implying that values less than or equal to $\beta_{1}$ are not possible values for $\beta$ according to Assumption \ref{AsPosEff}.

The largest root of ``$A + B \cdot \beta + C \cdot \beta^{2} = 0$'' is given by  $$\beta_{2} \coloneqq \dfrac{-B + \sqrt{B^2 - 4 \cdot A \cdot C}}{2 \cdot C}.$$ Since $A \leq 0$ and $C > 0$, we have that $\beta_{2} \geq 0$, implying that values greater than or equal to $\beta_{2}$ are possible values for $\beta$. Consequently, $\beta_{2}$ is a lower bound for $\beta$, i.e.,
\begin{equation}
	\label{EqBetaCS}
	\beta \geq \dfrac{-B + \sqrt{B^2 - 4 \cdot A \cdot C}}{2 \cdot C}.
\end{equation}

\textbf{Fifth Term in $LB^{\star}$:} Define the following variance-covariance matrix:
\begin{equation}
	A \coloneqq \left[\begin{matrix}
		a & b & c \\
		b & d & e \\
		c & e & f
	\end{matrix}\right]
\end{equation}
where
\begin{align}
	a & \coloneqq Var\left(X_{res}^{*}\right) = 1 \\
	b & \coloneqq Cov\left(X_{res}^{*}, U_{1}\right) = \dfrac{Cov\left(Z_{1,res}, Y_{res}\right)}{\beta} - 1 \\
	c & \coloneqq Cov\left(X_{res}^{*}, U_{2}\right) = \dfrac{Cov\left(Z_{2,res}, Y_{res}\right)}{\beta} - 1\\
	d & \coloneqq Var\left(U_{1}\right) = Var\left(Z_{1,res}\right) + 1 - 2 \cdot \dfrac{Cov\left(Z_{1,res}, Y_{res}\right)}{\beta} \\
	e & \coloneqq Cov\left(U_{1}, U_{2}\right) = Cov\left(Z_{1,res}, Z_{2,res}\right) + 1 - \dfrac{Cov\left(Z_{1,res}, Y_{res}\right) + Cov\left(Z_{2,res}, Y_{res}\right)}{\beta}\\
	f & \coloneqq Var\left(U_{2}\right) = Var\left(Z_{2,res}\right) + 1 - 2 \cdot \dfrac{Cov\left(Z_{2,res}, Y_{res}\right)}{\beta}
\end{align}
according to Lemma \ref{LemmaDataRestrictions}.

The determinant of this matrix is given by
\begin{equation}\label{EqDet1}
	det(A) = a \cdot (d \cdot f - e^2) + b \cdot (c \cdot e - b \cdot f) + c \cdot (b \cdot e - d \cdot c).
\end{equation}
Since matrix $A$ is a variance-covariance matrix, we must have that $det(A) \geq 0$, which we will use to derive the fifth term in the lower bound.\footnote{This condition is derived from the fact that variance-covariance matrices must be positive-semidefinite. As a consequence, all its principal minors must be positive. One of the principal minors of matrix A is the determinant of matrix A and imposing that it is positive generates the fifth term in $LB^{\star}$. The other principal minors are the elements in the main diagonal, which generate the sixth and seventh terms in $LB^{\star}$, and the expressions associated with the Cauchy-Schwartz inequality, which generate the first, second and fourth terms in $LB^{\star}$.}
Note that Equation \eqref{EqDet1} can be rewritten as
\begin{align}
	det(A) & = f \cdot \left(d - b^{2} \right) - c^{2} \cdot d - e^2 + 2 \cdot b \cdot c \cdot e \nonumber \\
	& = f \cdot \left(d - b^{2} \right) - c^{2} \cdot \left( d - b^{2} \right) - b^{2} \cdot c^{2}  - e^2 + 2 \cdot b \cdot c \cdot e \nonumber  \\
	& = \left(d - b^{2} \right) \cdot \left(f - c^{2} \right) - \left(e - b \cdot c \right)^{2}. \label{EqDet2}
\end{align}

Next, we calculate the individual simplified parenthetical terms:
\begin{align*}
	(d - b^2) & = Var(Z_{1,res}) + 1 - 2 \cdot \dfrac{Cov\left(Z_{1,res}, Y_{res}\right)}{\beta} - \left(\dfrac{Cov\left(Z_{1,res}, Y_{res}\right)}{\beta} - 1\right)^2 \\
	& = Var(Z_{1,res}) - \dfrac{Cov\left(Z_{1,res}, Y_{res}\right)^2}{\beta^2} \\
	(f - c^2) & = Var(Z_{2,res}) + 1 - 2 \cdot \dfrac{Cov\left(Z_{2,res}, Y_{res}\right)}{\beta} - \left(\dfrac{Cov\left(Z_{2,res}, Y_{res}\right)}{\beta} - 1\right)^2 \\
	& = Var(Z_{2,res}) - \dfrac{Cov\left(Z_{2,res}, Y_{res}\right)^2}{\beta^2}, \\
	(e - b \cdot c) & = Cov(Z_{1,res},Z_{2,res}) + 1 - \dfrac{Cov\left(Z_{1,res}, Y_{res}\right)}{\beta} - \dfrac{Cov\left(Z_{2,res}, Y_{res}\right)}{\beta} \\
	& \hspace{20pt}- \left(\dfrac{Cov\left(Z_{1,res}, Y_{res}\right)}{\beta} - 1\right) \cdot \left(\dfrac{Cov\left(Z_{2,res}, Y_{res}\right)}{\beta} - 1\right) \\
	& = Cov(Z_{1,res},Z_{2,res}) - \dfrac{Cov\left(Z_{1,res}, Y_{res}\right)}{\beta} \cdot \dfrac{Cov\left(Z_{2,res}, Y_{res}\right)}{\beta},
\end{align*}

Now, we substitute these back into Equation \eqref{EqDet2}:
\begin{align*}
	det(A) & = \left(Var(Z_{1,res}) - \dfrac{Cov\left(Z_{1,res}, Y_{res}\right)^2}{\beta^2}\right) \cdot \left(Var(Z_{2,res}) - \dfrac{Cov\left(Z_{2,res}, Y_{res}\right)^2}{\beta^2}\right) \\
	& \hspace{20pt} - \left(Cov(Z_{1,res},Z_{2,res}) - \dfrac{Cov\left(Z_{1,res}, Y_{res}\right)}{\beta} \cdot \dfrac{Cov\left(Z_{2,res}, Y_{res}\right)}{\beta}\right)^2 \\
	& = Var(Z_{1,res}) \cdot Var(Z_{2,res}) - Cov(Z_{1,res},Z_{2,res})^2 - \left(\frac{Cov(Z_{1,res},Y_{res})}{\beta}\right)^2 \cdot Var(Z_{2,res}) \\
	& \hspace{20pt} - \left(\frac{Cov(Z_{2,res},Y_{res})}{\beta}\right)^2 \cdot Var(Z_{1,res}) \\
	& \hspace{20pt}+ 2 \cdot \frac{Cov(Z_{1,res},Y_{res}) \cdot Cov(Z_{2,res},Y_{res})}{\beta^2} \cdot Cov(Z_{1,res},Z_{2,res})
\end{align*}
Since $det(A) \geq 0$, the last equation implies that
\begin{align*}
	& \beta^{2} \cdot \left\lbrace Var(Z_{1,res}) \cdot Var(Z_{2,res}) - Cov(Z_{1,res},Z_{2,res})^2 \right\rbrace \\
	& \hspace{20pt} \geq Cov(Z_{1,res},Y_{res})^2 \cdot Var(Z_{2,res}) + Cov(Z_{2,res},Y_{res})^2 \cdot Var(Z_{1,res}) \\
	& \hspace{20pt} - 2\cdot Cov(Z_{1,res},Y_{res}) \cdot Cov(Z_{2,res},Y_{res}) \cdot Cov(Z_{1,res},Z_{2,res}).
\end{align*}
Consequently, we have that
\begin{align*}
	\beta^{2} & \geq \dfrac{\left\lbrace \begin{matrix}
			Cov(Z_{1,res},Y_{res})^2 \cdot Var(Z_{2,res}) + Cov(Z_{2,res},Y_{res})^2 \cdot Var(Z_{1,res}) \\
			- 2\cdot Cov(Z_{1,res},Y_{res}) \cdot Cov(Z_{2,res},Y_{res}) \cdot Cov(Z_{1,res},Z_{2,res})
		\end{matrix} \right\rbrace}{Var(Z_{1,res}) \cdot Var(Z_{2,res}) - Cov(Z_{1,res},Z_{2,res})^2} \\
	& \hspace{20pt} \text{because the Cauchy–Schwarz inequality and Assumption \ref{AsDataRestrictions} ensure } \\ & \hspace{20pt} Var(Z_{1,res}) \cdot Var(Z_{2,res}) > Cov(Z_{1,res},Z_{2,res})^2,
\end{align*}
implying that
\begin{equation}\label{EqLB5}
	\beta \geq \sqrt{\dfrac{\left\lbrace \begin{matrix}
				Cov(Z_{1,res},Y_{res})^2 \cdot Var(Z_{2,res}) + Cov(Z_{2,res},Y_{res})^2 \cdot Var(Z_{1,res}) \\
				- 2\cdot Cov(Z_{1,res},Y_{res}) \cdot Cov(Z_{2,res},Y_{res}) \cdot Cov(Z_{1,res},Z_{2,res})
			\end{matrix} \right\rbrace}{Var(Z_{1,res}) \cdot Var(Z_{2,res}) - Cov(Z_{1,res},Z_{2,res})^2}}
\end{equation}
because $\beta > 0$.

{

	\textbf{Sixth Term in $LB^{\star}$:} Since $Var\left(U_{1}\right) \geq 0$ and $\beta > 0$, Equation \eqref{EqVarU1} implies that
	\begin{equation*}
		\begin{array}{crcl}
			& 1 + Var\left(Z_{1,res}\right) - 2 \cdot \dfrac{Cov\left(Z_{1,res}, Y_{res}\right)}{\beta} & \geq & 0 \\
			\Leftrightarrow & 1 + Var\left(Z_{1,res}\right) & \geq & 2 \cdot \dfrac{Cov\left(Z_{1,res}, Y_{res}\right)}{\beta} \\
			\Leftrightarrow & \beta \cdot \left( 1 + Var\left(Z_{1,res}\right) \right) & \geq &  2 \cdot Cov\left(Z_{1,res}, Y_{res}\right),
		\end{array}
	\end{equation*}
	implying that
	\begin{equation}\label{EqBetaZ1Term6}
		\beta \geq 2 \cdot \dfrac{Cov\left(Z_{1,res}, Y_{res}\right)}{1 + Var\left(Z_{1,res}\right)}.
	\end{equation}

	\textbf{Seventh Term in $LB^{\star}$:} Combining Equations \eqref{EqVarZ2} and \eqref{EqCovXU2} and using the same steps used to derive Equation \eqref{EqBetaZ1Term6}, we have that
	\begin{equation}\label{EqBetaZ2Term7}
		\beta \geq 2 \cdot \dfrac{Cov\left(Z_{2,res}, Y_{res}\right)}{1 + Var\left(Z_{2,res}\right)}.
	\end{equation}

	\phantom{a}

	\textbf{Part 3.2: } Now, we show that the first, second, sixth, and seventh terms in $LB^{\star}$ are non-binding.

	\textbf{Sixth Term in $LB^{\star}$ is non-binding:} We must show that
	\begin{equation}\label{EqNB2}
		\dfrac{Cov\left(Z_{1,res}, Y_{res}\right)}{\sqrt{Var\left(Z_{1,res}\right)}} \geq 2 \cdot \dfrac{Cov\left(Z_{1,res}, Y_{res}\right)}{1 + Var\left(Z_{1,res}\right)}.
	\end{equation} Equivalently, we must show that $$\dfrac{1}{\sqrt{V}} \geq \dfrac{2}{1 + V}$$ for any $V > 0$. Note that
	\begin{align*}
		\begin{array}{crcl}
			& \left(1 - \sqrt{V}\right)^{2} & \geq & 0 \\
			\Leftrightarrow & 1 - 2 \cdot \sqrt{V} + V & \geq & 0 \\
			\Leftrightarrow & 1 + V & \geq & 2 \cdot \sqrt{V} \\
			\Leftrightarrow & \dfrac{1}{\sqrt{V}} & \geq & \dfrac{2}{1 + V},
		\end{array}
	\end{align*}
	implying that the sixth term in $LB^{\star}$ is non-binding.

	\textbf{Seventh Term in $LB^{\star}$ is non-binding:} This step is analogous to the step that proves that the sixth term in $LB^{\star}$ is non-binding.

	\textbf{First Term in $LB^{\star}$ is non-binding:} We must show that
	\begin{equation}\label{EqNB1}
		\sqrt{\dfrac{\left\lbrace \begin{matrix}
					Cov(Z_{1,res},Y_{res})^2 \cdot Var(Z_{2,res}) + Cov(Z_{2,res},Y_{res})^2 \cdot Var(Z_{1,res}) \\
					- 2\cdot Cov(Z_{1,res},Y_{res}) \cdot Cov(Z_{2,res},Y_{res}) \cdot Cov(Z_{1,res},Z_{2,res})
				\end{matrix} \right\rbrace}{Var(Z_{1,res}) \cdot Var(Z_{2,res}) - Cov(Z_{1,res},Z_{2,res})^2}} \geq \dfrac{Cov\left(Z_{1,res}, Y_{res}\right)}{\sqrt{Var\left(Z_{1,res}\right)}}.
	\end{equation}

	Since both sides of this inequality are positive, this inequality is equivalent to
	\begin{equation}\label{EqRedudant1}
		\dfrac{\left\lbrace \begin{matrix}
				Cov(Z_{1,res},Y_{res})^2 \cdot Var(Z_{2,res}) + Cov(Z_{2,res},Y_{res})^2 \cdot Var(Z_{1,res}) \\
				- 2\cdot Cov(Z_{1,res},Y_{res}) \cdot Cov(Z_{2,res},Y_{res}) \cdot Cov(Z_{1,res},Z_{2,res})
			\end{matrix} \right\rbrace}{Var(Z_{1,res}) \cdot Var(Z_{2,res}) - Cov(Z_{1,res},Z_{2,res})^2} - \dfrac{\left[Cov\left(Z_{1,res}, Y_{res}\right)\right]^{2}}{Var\left(Z_{1,res}\right)} \geq 0.
	\end{equation}

	After finding the common denominator of the two terms in Inequality \eqref{EqRedudant1}, we have that its denominator is given by
	$$Var\left(Z_{1,res}\right) \cdot \left[Var(Z_{1,res}) \cdot Var(Z_{2,res}) - Cov(Z_{1,res},Z_{2,res})^2\right],$$ which is strictly positive due to Assumption \ref{AsDataRestrictions}. Consequently, we only have to show that the numerator in Inequality \eqref{EqRedudant1} is positive.

	After finding the common denominator of the two terms in Inequality \eqref{EqRedudant1}, we have that its numerator is given by
	\begin{align*}
		& Var\left(Z_{1,res}\right) \cdot \left\lbrace \begin{matrix}
			Cov(Z_{1,res},Y_{res})^2 \cdot Var(Z_{2,res}) + Cov(Z_{2,res},Y_{res})^2 \cdot Var(Z_{1,res}) \\
			- 2\cdot Cov(Z_{1,res},Y_{res}) \cdot Cov(Z_{2,res},Y_{res}) \cdot Cov(Z_{1,res},Z_{2,res})
		\end{matrix} \right\rbrace \\
		& \hspace{40pt} - \left[Cov\left(Z_{1,res}, Y_{res}\right)\right]^{2} \cdot \left[Var(Z_{1,res}) \cdot Var(Z_{2,res}) - Cov(Z_{1,res},Z_{2,res})^2\right] \\
		& \hspace{20pt} = Cov(Z_{1,res},Y_{res})^2 \cdot Var\left(Z_{1,res}\right) \cdot Var(Z_{2,res}) + Cov(Z_{2,res},Y_{res})^2 \cdot Var(Z_{1,res})^{2} \\
		& \hspace{40pt} - 2\cdot Cov(Z_{1,res},Y_{res}) \cdot Cov(Z_{2,res},Y_{res}) \cdot Cov(Z_{1,res},Z_{2,res}) \cdot Var\left(Z_{1,res}\right) \\
		& \hspace{40pt} - \left[Cov\left(Z_{1,res}, Y_{res}\right)\right]^{2} \cdot Var(Z_{1,res}) \cdot Var(Z_{2,res}) \\
		& \hspace{40pt} + \left[Cov\left(Z_{1,res}, Y_{res}\right)\right]^{2} \cdot Cov(Z_{1,res},Z_{2,res})^2 \\
		& \hspace{20pt} = Cov(Z_{2,res},Y_{res})^2 \cdot Var(Z_{1,res})^{2} \\
		& \hspace{40pt} - 2\cdot Cov(Z_{1,res},Y_{res}) \cdot Cov(Z_{2,res},Y_{res}) \cdot Cov(Z_{1,res},Z_{2,res}) \cdot Var\left(Z_{1,res}\right) \\
		& \hspace{40pt} + \left[Cov\left(Z_{1,res}, Y_{res}\right)\right]^{2} \cdot Cov(Z_{1,res},Z_{2,res})^2 \\
		& \hspace{20pt} = \left[ Cov(Z_{2,res},Y_{res}) \cdot Var(Z_{1,res}) - Cov\left(Z_{1,res}, Y_{res}\right) \cdot Cov(Z_{1,res},Z_{2,res}) \right]^{2} \\
		& \hspace{20pt} \geq 0,
	\end{align*}
	implying that the first term in $LB^{\star}$ is non-binding.

	\textbf{Second Term in $LB^{\star}$ is non-binding:} This step is analogous to the step that proves that the first term in $LB^{\star}$ is non-binding.

}

\phantom{a}

\textbf{Part 3.3: } Combining Equations \eqref{EqLB}, \eqref{EqBetaZ1Z2}, \eqref{EqBetaCS} and \eqref{EqLB5} and the fact that the first, second, sixth, and seventh terms in $LB^{\star}$ are non-binding, we have that $$\beta \geq LB.$$

\subsection{No Valid Data-Generating Process if $LB > UB$}\label{AppProofNoDGP}

In this section, we show that, when $LB > UB$, there is no combination of $\beta \geq 0$ and distribution of latent variables $\left(X_{res}^{*}, \epsilon_{res},U_{1},U_{2}\right)$ that satisfies Assumptions \ref{AsExogeneous}-\ref{AsDataRestrictions} and the data restrictions described in Lemma \ref{LemmaDataRestrictions}. We prove this result by contradiction.

Assume that $LB > UB$ and that there exist parameter $\beta \geq 0$ and variables $\left(X_{res}^{*}, \epsilon_{res},U_{1},U_{2}\right)$ such that Assumptions \ref{AsExogeneous}-\ref{AsDataRestrictions} and Equations \eqref{EqExpY}-\eqref{EqCovZ1Z2} are valid.

Following the same steps described in Parts 2 and 3 in Appendix \ref{AppProofLbBetaUB}, we must have that $\beta \geq LB$. Rearranging Equation \eqref{EqVarY}, we have that
\begin{align*}
	Var\left(\epsilon_{res}\right) & =  Var\left(Y_{res}\right) - \beta^{2} \\
	& \leq Var\left(Y_{res}\right) - \left(LB\right)^{2} \\
	& \hspace{20pt} \text{because } \beta \geq LB > UB > 0 \text{ by Assumption \ref{AsDataRestrictions}} \\
	& = \left(UB\right)^{2} - \left(LB\right)^{2} \\
	& \hspace{20pt} \text{according to Equation \eqref{EqUB}} \\
	& < 0 \hspace{10pt} \text{because } LB > UB,
\end{align*}
which is a contradiction.

\subsection{$\left[LB,UB \right]$ exhausts all the information contained in the first two moments of the data distribution if $UB \geq LB$}\label{AppProofAllInfo}

First, we define the meaning of exhausting all the information contained in the first two moments of the data distribution. Second, we show that, if $UB \geq LB$, then the interval $\left[LB,UB \right]$ exhausts all the information contained in the first two moments of the data distribution.

\begin{definition}
	A set $\mathcal{B} \subseteq \mathbb{R}$ exhausts all the information contained in the first two moments of the data distribution if, for any candidate target parameter $\tilde{\beta} \in \mathcal{B}$, there exist candidate latent variables $\left(\tilde{X}_{res}^{*}, \tilde{\epsilon}_{res}, \tilde{U}_{1}, \tilde{U}_{2}\right)$ and candidate observable variables $\left(\tilde{Y}_{res}, \tilde{Z}_{1,res}, \tilde{Z}_{1,res}\right)$ such that
	\begin{enumerate}
		\item the candidate observable variables are induced by the candidate latent variables according to the model restrictions imposed by Equations \eqref{EqModel} and \eqref{EqProxiesCovariates}, i.e.,
		\begin{align}
			\label{EqCandY} \tilde{Y}_{res} & \coloneqq \tilde{\beta} \cdot \tilde{X}_{res}^{*} + \tilde{\epsilon}_{res} \\
			\label{EqCandZ1} \tilde{Z}_{1,res} & \coloneqq \tilde{X}_{res}^{*} + \tilde{U}_{1} \\
			\label{EqCandZ2} \tilde{Z}_{2,res} & \coloneqq \tilde{X}_{res}^{*} + \tilde{U}_{2};
		\end{align}

		\item the candidate latent variables and the candidate target parameter satisfy the model restrictions imposed by Assumptions \ref{AsExogeneous}-\ref{AsNonClassical}, i.e.,
		\begin{align}
			\label{EqCandCovXeps} Cov\left(\tilde{X}_{res}^{*}, \tilde{\epsilon}_{res}\right) & = 0 \\
			\label{EqCandVarX} Var\left(\tilde{X}_{res}^{*}\right) & = 1 \\
			\label{EqCandBeta} \tilde{\beta} & \geq 0 \\
			\label{EqCandCovU1eps} Cov\left(\tilde{U}_{1}, \tilde{\epsilon}_{res}\right) & = 0 \\
			\label{EqCandCovU2eps} Cov\left(\tilde{U}_{2}, \tilde{\epsilon}_{res}\right) & = 0 \\
			\label{EqCandCovU1U2} Cov\left(\tilde{U}_{1}, \tilde{U}_{2}\right) & \geq 0
		\end{align}

		\item the first two moments of the candidate observable variables are equal to the first two moments of the true observable variables, i.e.,
		\begin{align}
			\label{EqCandExpY} \mathbb{E}\left[\tilde{Y}_{res}\right] & = \mathbb{E}\left[Y_{res}\right] \\
			\label{EqCandExpZ1} \mathbb{E}\left[\tilde{Z}_{1,res}\right] & = \mathbb{E}\left[Z_{1,res}\right] \\
			\label{EqCandExpZ2} \mathbb{E}\left[\tilde{Z}_{2,res}\right] & = \mathbb{E}\left[Z_{2,res}\right] \\
			\label{EqCandVarY} Var\left(\tilde{Y}_{res}\right) & = Var\left(Y_{res}\right) \\
			\label{EqCandVarZ1} Var\left(\tilde{Z}_{1,res}\right) & = Var\left(Z_{1,res}\right) \\
			\label{EqCandVarZ2} Var\left(\tilde{Z}_{2,res}\right) & = Var\left(Z_{2,res}\right) \\
			\label{EqCandCovZ1Y} Cov\left(\tilde{Z}_{1,res}, \tilde{Y}_{res}\right) & = Cov\left(Z_{1,res}, Y_{res}\right) \\
			\label{EqCandCovZ2Y} Cov\left(\tilde{Z}_{2,res}, \tilde{Y}_{res}\right) & = Cov\left(Z_{2,res}, Y_{res}\right) \\
			\label{EqCandCovZ1Z2} Cov\left(\tilde{Z}_{1,res}, \tilde{Z}_{2,res}\right) & = Cov\left(Z_{1,res}, Z_{2,res}\right).
		\end{align}
	\end{enumerate}
\end{definition}

Second, we show that, if $UB \geq LB \geq 0$, then the interval $\left[LB,UB \right]$ exhausts all the information contained in the first two moments of the data distribution.

Assume that $UB \geq LB$ and fix $\tilde{\beta} \in \left[LB,UB\right]$ arbitrarily. We consider two cases: (i) $\tilde{\beta} > 0$, and (ii) $\tilde{\beta} = 0$. In both cases, the proof is by construction.

\subsubsection{Case 1: $\tilde{\beta} > 0$}

We need to create a distribution for the candidate latent variables $\left(\tilde{X}_{res}^{*}, \tilde{\epsilon}_{res}, \tilde{U}_{1}, \tilde{U}_{2}\right)$. The candidate observable variables $\left(\tilde{Y}_{res}, \tilde{Z}_{1,res}, \tilde{Z}_{1,res}\right)$ are created based on Equations \eqref{EqCandY}-\eqref{EqCandZ2}.

We impose that $$\left(\tilde{X}_{res}^{*}, \tilde{\epsilon}_{res}, \tilde{U}_{1}, \tilde{U}_{2}\right) \text{ are jointly distributed with mean } \boldsymbol{\mu} \text{ and variance } \boldsymbol{\Omega}.$$

The vector $\boldsymbol{\mu}$ is given by
\begin{align}
	\label{EqCandXexp} \mathbb{E}\left[\tilde{X}_{res}^{*}\right] & \coloneqq 0 \\
	\label{EqCandEpsexp} \mathbb{E}\left[\tilde{\epsilon}_{res} \right] & \coloneqq 0 \\
	\label{EqCandU1exp} \mathbb{E}\left[\tilde{U}_{1} \right] & \coloneqq \mathbb{E}\left[Z_{1,res} \right] \\
	\label{EqCandU2exp} \mathbb{E}\left[\tilde{U}_{2} \right] & \coloneqq \mathbb{E}\left[Z_{2,res} \right]
\end{align}
and the covariance matrix $\boldsymbol{\Omega}$ is given by
\begin{align}
	Var\left(\tilde{X}_{res}^{*} \right) & \coloneqq 1 \\
	\label{EqCandEpsVar} Var\left(\tilde{\epsilon}_{res} \right) & \coloneqq Var\left(Y_{res}\right) - \tilde{\beta}^{2} \\
	\label{EqCandU1Var} Var\left(\tilde{U}_{1} \right) & \coloneqq 1 + Var\left(Z_{1,res}\right) - 2 \cdot \dfrac{Cov\left(Z_{1,res}, Y_{res}\right)}{\tilde{\beta}} \\
	\label{EqCandU2Var} Var\left(\tilde{U}_{2} \right) & \coloneqq 1 + Var\left(Z_{2,res}\right) - 2 \cdot \dfrac{Cov\left(Z_{2,res}, Y_{res}\right)}{\tilde{\beta}} \\
	\label{EqCandXEps} Cov\left(\tilde{X}_{res}^{*},\tilde{\epsilon}_{res}\right) & \coloneqq 0 \\
	\label{EqCandU1Eps} Cov\left(\tilde{U}_{1},\tilde{\epsilon}_{res}\right) & \coloneqq 0 \\
	\label{EqCandU2Eps} Cov\left(\tilde{U}_{2},\tilde{\epsilon}_{res}\right) & \coloneqq 0 \\
	\label{EqCandXU1Cov} Cov\left(\tilde{X}_{res}^{*},\tilde{U}_{1}\right) & \coloneqq \dfrac{Cov\left(Z_{1,res}, Y_{res}\right)}{\tilde{\beta}} - 1 \\
	\label{EqCandXU2Cov} Cov\left(\tilde{X}_{res}^{*},\tilde{U}_{2}\right) & \coloneqq \dfrac{Cov\left(Z_{2,res}, Y_{res}\right)}{\tilde{\beta}} - 1 \\
	\label{EqCandU1U2Cov} Cov\left(\tilde{U}_{1},\tilde{U}_{2}\right) & \coloneqq 1 + Cov\left(Z_{1,res}, Z_{2,res}\right) - \left[\dfrac{Cov\left(Z_{1,res}, Y_{res}\right) + Cov\left(Z_{2,res}, Y_{res}\right)}{\tilde{\beta}} \right]
\end{align}

First, we must show that the above distribution is a valid distribution, i.e., we must show that
\begin{align}
	\label{EqRes1} Var\left(\tilde{\epsilon}_{res} \right) & \geq 0 \\
	\label{EqRes2} Var\left(\tilde{U}_{1} \right) & \geq 0 \\
	\label{EqRes3} Var\left(\tilde{U}_{2} \right) & \geq 0 \\
	\label{EqRes4} Var\left(\tilde{\epsilon}_{res} \right) & \geq \left[ Cov\left(\tilde{X}_{res}^{*},\tilde{\epsilon}_{res}\right) \right]^{2} \\
	\label{EqRes5}  Var\left(\tilde{U}_{1} \right) & \geq \left[ Cov\left(\tilde{X}_{res}^{*},\tilde{U}_{1}\right) \right]^{2} \\
	\label{EqRes6} Var\left(\tilde{U}_{2} \right) & \geq \left[ Cov\left(\tilde{X}_{res}^{*},\tilde{U}_{2}\right) \right]^{2} \\
	\label{EqRes7} Var\left(\tilde{U}_{1} \right) \cdot Var\left(\tilde{\epsilon}_{res} \right) & \geq \left[ Cov\left(\tilde{U}_{1},\tilde{\epsilon}_{res}\right) \right]^{2} \\
	\label{EqRes8} Var\left(\tilde{U}_{2} \right) \cdot Var\left(\tilde{\epsilon}_{res} \right) & \geq \left[ Cov\left(\tilde{U}_{2},\tilde{\epsilon}_{res}\right) \right]^{2} \\
	\label{EqRes9} Var\left(\tilde{U}_{1} \right) \cdot Var\left(\tilde{U}_{2} \right) & \geq \left[ Cov\left(\tilde{U}_{1},\tilde{U}_{2}\right) \right]^{2},
\end{align}
and
\begin{equation}\label{EqDet1tilde}
	det(\tilde{A}) = \tilde{a} \cdot (\tilde{d} \cdot \tilde{f} - \tilde{e}^2) + \tilde{b} \cdot (\tilde{c} \cdot \tilde{e} - \tilde{b} \cdot \tilde{f}) + \tilde{c} \cdot (\tilde{b} \cdot \tilde{e} - \tilde{d} \cdot \tilde{c}) \geq 0,
\end{equation}
where
\begin{equation}
	\tilde{A} \coloneqq \left[\begin{matrix}
		\tilde{a} & \tilde{b} & \tilde{c} \\
		\tilde{b} & \tilde{d} & \tilde{e} \\
		\tilde{c} & \tilde{e} & \tilde{f}
	\end{matrix}\right]
\end{equation}
and
\begin{align}
	\tilde{a} & \coloneqq Var\left(\tilde{X}_{res}^{*}\right) \\
	\tilde{b} & \coloneqq Cov\left(\tilde{X}_{res}^{*}, \tilde{U}_{1}\right)\\
	\tilde{c} & \coloneqq Cov\left(\tilde{X}_{res}^{*}, \tilde{U}_{2}\right)\\
	\tilde{d} & \coloneqq Var\left(\tilde{U}_{1}\right) \\
	\tilde{e} & \coloneqq Cov\left(\tilde{U}_{1}, \tilde{U}_{2}\right)\\
	\tilde{f} & \coloneqq Var\left(\tilde{U}_{2}\right).
\end{align}

Observe that Inequalities \eqref{EqRes4}, \eqref{EqRes7} and \eqref{EqRes8} are trivially satisfied due to Equations \eqref{EqCandXEps}-\eqref{EqCandU2Eps}.

To prove that Inequality \eqref{EqRes1} is satisfied, note that
\begin{align*}
	Var\left(\tilde{\epsilon}_{res} \right) & \coloneqq Var\left(Y_{res}\right) - \tilde{\beta}^{2} \\
	& \hspace{20pt} \text{according to Equation \eqref{EqCandEpsVar}} \\
	& \geq Var\left(Y_{res}\right) - \left(UB\right)^{2} \\
	& \hspace{20pt} \text{because } \tilde{\beta} \leq UB \\
	& = Var\left(Y_{res}\right) - \left(\sqrt{Var\left(Y_{res}\right)}\right)^{2} \\
	& \hspace{20pt} \text{according to Equation \eqref{EqUB}} \\
	& = 0.
\end{align*}

Now, we show that Inequality \eqref{EqRes2} is satisfied. To do so, we analyze two cases. If $Cov\left(Z_{1,res}, Y_{res}\right) = 0$, then $Var\left(\tilde{U}_{1} \right) = 1 + Var\left(Z_{1,res}\right) > 0$ according to Equation \eqref{EqCandU1Var}. If $Cov\left(Z_{1,res}, Y_{res}\right) > 0$, then
\begin{align*}
	Var\left(\tilde{U}_{1} \right) & = 1 + Var\left(Z_{1,res}\right) - 2 \cdot \dfrac{Cov\left(Z_{1,res}, Y_{res}\right)}{\tilde{\beta}} \\
	& \hspace{20pt} \text{according to Equation \eqref{EqCandU1Var}} \\
	& \geq 1 + Var\left(Z_{1,res}\right) - 2 \cdot \dfrac{Cov\left(Z_{1,res}, Y_{res}\right)}{\sfrac{2 \cdot Cov\left(Z_{1,res}, Y_{res}\right)}{\left(1 + Var\left(Z_{1,res}\right)\right)}} \\
	& \hspace{20pt} \text{because } \tilde{\beta} \geq LB \text{ and by Assumption \ref{AsDataRestrictions} and Equations \eqref{EqLB}, \eqref{EqNB2} and \eqref{EqNB1}} \\
	& = 0.
\end{align*}

We can show that Inequality \eqref{EqRes3} is satisfied using Equation \eqref{EqCandU2Var} and the steps above.

Inequality \eqref{EqRes5} is equivalent to $\tilde{\beta} \geq \dfrac{Cov\left(Z_{1,res}, Y_{res}\right)}{\sqrt{Var\left(Z_{1,res}\right)}}$ according to Equations \eqref{EqCandU1Var} and \eqref{EqCandXU1Cov}. Since $\tilde{\beta} \geq LB \geq \dfrac{Cov\left(Z_{1,res}, Y_{res}\right)}{\sqrt{Var\left(Z_{1,res}\right)}}$ by Equation \eqref{EqNB1}, Inequality \eqref{EqRes5} is satisfied.

Inequality \eqref{EqRes6} is valid according to an analogous argument using Equations \eqref{EqCandU2Var} and \eqref{EqCandXU2Cov}.

According to Equations \eqref{EqCandU1Var}, \eqref{EqCandU2Var} and \eqref{EqCandU1U2Cov}, Inequality \eqref{EqRes9} is equivalent to Equation \eqref{EqSquaredX} where we use $\tilde{\beta}$ instead of $\beta$. In turn, this equation is satisfied because $\tilde{\beta} \geq LB \geq \dfrac{-B + \sqrt{B^2 - 4 \cdot A \cdot C}}{2 \cdot C}$ by definition. Consequently, Inequality \eqref{EqRes9}  is satisfied.

Lastly, we need to check whether Inequality \eqref{EqDet1tilde} is satisfied. This inequality is equivalent to Inequality \eqref{EqLB5} where $\beta$ is replaced by $\tilde{\beta}$. In turn, this condition is satisfied because $$\tilde{\beta} \geq LB \geq \sqrt{\dfrac{\left\lbrace \begin{matrix}
		Cov(Z_{1,res},Y_{res})^2 \cdot Var(Z_{2,res}) + Cov(Z_{2,res},Y_{res})^2 \cdot Var(Z_{1,res}) \\
		- 2\cdot Cov(Z_{1,res},Y_{res}) \cdot Cov(Z_{2,res},Y_{res}) \cdot Cov(Z_{1,res},Z_{2,res})
		\end{matrix} \right\rbrace}{Var(Z_{1,res}) \cdot Var(Z_{2,res}) - Cov(Z_{1,res},Z_{2,res})^2}}$$ by construction. Consequently, Inequality \eqref{EqDet1tilde} is satisfied.

Hence, we can conclude that the proposed distribution for $\left(\tilde{X}_{res}^{*}, \tilde{\epsilon}_{res}, \tilde{U}_{1}, \tilde{U}_{2}\right)$ is a valid distribution.

Second, we must show that the candidate latent variables satisfy Equations \eqref{EqCandCovXeps}-\eqref{EqCandCovU1U2}. Since Restrictions \eqref{EqCandCovXeps}-\eqref{EqCandCovU2eps} are satisfied by construction, we only have to show that Restriction \eqref{EqCandCovU1U2} is satisfied. To do so, we analyze two cases. If $Cov\left(Z_{1,res}, Y_{res}\right) + Cov\left(Z_{2,res}, Y_{res}\right) = 0$, then $Cov\left(\tilde{U}_{1},\tilde{U}_{2}\right) = 1 + Cov\left(Z_{1,res}, Z_{2,res}\right) > 0$ according to Equation \eqref{EqCandU1U2Cov} and Assumption \ref{AsDataRestrictions}. If $Cov\left(Z_{1,res}, Y_{res}\right) + Cov\left(Z_{2,res}, Y_{res}\right) > 0$, then
\begin{align}
	& Cov\left(\tilde{U}_{1},\tilde{U}_{2}\right) \nonumber \\
	& \hspace{20pt} = 1 + Cov\left(Z_{1,res}, Z_{2,res}\right) - \left[\dfrac{Cov\left(Z_{1,res}, Y_{res}\right) + Cov\left(Z_{2,res}, Y_{res}\right)}{\tilde{\beta}} \right] \nonumber  \\
	& \hspace{40pt} \text{according to Equation \eqref{EqCandU1U2Cov}} \nonumber \\
	& \hspace{20pt} \geq 1 + Cov\left(Z_{1,res}, Z_{2,res}\right) - \left[\dfrac{Cov\left(Z_{1,res}, Y_{res}\right) + Cov\left(Z_{2,res}, Y_{res}\right)}{\sfrac{\left(Cov\left(Z_{1,res}, Y_{res}\right) + Cov\left(Z_{2,res}, Y_{res}\right)\right)}{\left(1 + Cov\left(Z_{1,res}, Z_{2,res}\right)\right)}} \right] \nonumber \\
	& \hspace{40pt} \text{because } \tilde{\beta} \geq LB \text{ and by Assumption \ref{AsDataRestrictions} and Equation \eqref{EqLB}} \nonumber \\
	& \hspace{20pt} = 0, \label{EqCovU1U2}
\end{align}
implying that all the model restrictions are satisfied.

Lastly, we must show that the data restrictions (Equations \eqref{EqCandExpY}-\eqref{EqCandCovZ1Z2}) are valid. Since we have shown that candidate latent variables satisfy Equations \eqref{EqCandCovXeps}-\eqref{EqCandCovU1U2}, we can follow the same steps used to prove Lemma \ref{LemmaDataRestrictions} to show that
\begin{align}
	\label{EqCandidateExpY} \mathbb{E}\left[\tilde{Y}_{res}\right] & = 0 \\
	\label{EqCandidateExpZ1} \mathbb{E}\left[\tilde{Z}_{1,res}\right] & = \mathbb{E}\left[\tilde{U}_{1}\right] \\
	\label{EqCandidateExpZ2} \mathbb{E}\left[\tilde{Z}_{2,res}\right] & = \mathbb{E}\left[\tilde{U}_{2}\right] \\
	\label{EqCandidateVarY} Var\left(\tilde{Y}_{res}\right) & = \tilde{\beta}^{2} + Var\left(\tilde{\epsilon}_{res}\right) \\
	\label{EqCandidateVarZ1} Var\left(\tilde{Z}_{1,res}\right) & = 1 + Var\left(\tilde{U}_{1}\right) + 2 \cdot Cov\left(\tilde{X}_{res}^{*}, \tilde{U}_{1}\right) \\
	\label{EqCandidateVarZ2} Var\left(\tilde{Z}_{2,res}\right) & = 1 + Var\left(\tilde{U}_{2}\right) + 2 \cdot Cov\left(\tilde{X}_{res}^{*}, \tilde{U}_{2}\right) \\
	\label{EqCandidateCovZ1Y} Cov\left(\tilde{Z}_{1,res}, \tilde{Y}_{res}\right) & = \tilde{\beta} + \tilde{\beta} \cdot Cov\left(\tilde{X}_{res}^{*}, \tilde{U}_{1}\right) \\
	\label{EqCandidateCovZ2Y} Cov\left(\tilde{Z}_{2,res}, \tilde{Y}_{res}\right) & = \tilde{\beta} + \tilde{\beta} \cdot Cov\left(\tilde{X}_{res}^{*}, \tilde{U}_{2}\right) \\
	\label{EqCandidateCovZ1Z2} Cov\left(\tilde{Z}_{1,res}, \tilde{Z}_{2,res}\right) & = 1 + Cov\left(\tilde{X}_{res}^{*}, \tilde{U}_{1}\right) + Cov\left(\tilde{X}_{res}^{*}, \tilde{U}_{2}\right) + Cov\left(\tilde{U}_{1}, \tilde{U}_{2}\right).
\end{align}

Based on Equation \eqref{EqCandidateExpY}, we have that
\begin{align*}
	\mathbb{E}\left[\tilde{Y}_{res}\right] & = 0 = \mathbb{E}\left[Y_{res}\right]
\end{align*}
according to Equation \eqref{EqExpY}, implying that Equation \eqref{EqCandExpY} is valid.

Based on Equation \eqref{EqCandidateExpZ1}, we have that
\begin{align*}
	\mathbb{E}\left[\tilde{Z}_{1,res}\right] & = \mathbb{E}\left[\tilde{U}_{1}\right] = \mathbb{E}\left[Z_{1,res}\right]
\end{align*}
according to Equation \eqref{EqCandU1exp}, implying that Equation \eqref{EqCandExpZ1} is valid.

Based on Equation \eqref{EqCandidateExpZ2}, we have that
\begin{align*}
	\mathbb{E}\left[\tilde{Z}_{2,res}\right] & = \mathbb{E}\left[\tilde{U}_{2}\right] = \mathbb{E}\left[Z_{2,res}\right]
\end{align*}
according to Equation \eqref{EqCandU2exp}, implying that Equation \eqref{EqCandExpZ2} is valid.

Based on Equation \eqref{EqCandidateVarY}, we have that
\begin{align*}
	Var\left(\tilde{Y}_{res}\right) & = \tilde{\beta}^{2} + Var\left(\tilde{\epsilon}_{res}\right) \\
	& = \tilde{\beta}^{2} + \left[ Var\left(Y_{res}\right) - \tilde{\beta}^{2} \right] \\
	& \hspace{20pt} \text{by Equation \eqref{EqCandEpsVar}} \\
	& = Var\left(Y_{res}\right),
\end{align*}
implying that Equation \eqref{EqCandVarY} is valid.

Based on Equation \eqref{EqCandidateVarZ1}, we have that
\begin{align*}
	& Var\left(\tilde{Z}_{1,res}\right) \\
	& \hspace{20pt} = 1 + Var\left(\tilde{U}_{1}\right) + 2 \cdot Cov\left(\tilde{X}_{res}^{*}, \tilde{U}_{1}\right) \\
	& \hspace{20pt} = 1 + \left[1 + Var\left(Z_{1,res}\right) - 2 \cdot \dfrac{Cov\left(Z_{1,res}, Y_{res}\right)}{\tilde{\beta}}\right] + 2 \cdot \left[ \dfrac{Cov\left(Z_{1,res}, Y_{res}\right)}{\tilde{\beta}} - 1 \right] \\
	& \hspace{40pt} \text{by Equations \eqref{EqCandU1Var} and \eqref{EqCandXU1Cov}} \\
	& \hspace{20pt} = Var\left(Z_{1,res}\right),
\end{align*}
implying that Equation \eqref{EqCandVarZ1} is valid.

Based on Equation \eqref{EqCandidateVarZ2}, we have that
\begin{align*}
	& Var\left(\tilde{Z}_{2,res}\right) \\
	& \hspace{20pt} = 1 + Var\left(\tilde{U}_{2}\right) + 2 \cdot Cov\left(\tilde{X}_{res}^{*}, \tilde{U}_{2}\right) \\
	& \hspace{20pt} = 1 + \left[1 + Var\left(Z_{2,res}\right) - 2 \cdot \dfrac{Cov\left(Z_{2,res}, Y_{res}\right)}{\tilde{\beta}}\right] + 2 \cdot \left[ \dfrac{Cov\left(Z_{2,res}, Y_{res}\right)}{\tilde{\beta}} - 1 \right] \\
	& \hspace{40pt} \text{by Equations \eqref{EqCandU2Var} and \eqref{EqCandXU2Cov}} \\
	& \hspace{20pt} = Var\left(Z_{2,res}\right),
\end{align*}
implying that Equation \eqref{EqCandVarZ2} is valid.

Based on Equation \eqref{EqCandidateCovZ1Y}, we have that
\begin{align*}
	& Cov\left(\tilde{Z}_{1,res}, \tilde{Y}_{res}\right) \\
	& \hspace{20pt} = \tilde{\beta} + \tilde{\beta} \cdot Cov\left(\tilde{X}_{res}^{*}, \tilde{U}_{1}\right) \\
	& \hspace{20pt} = \tilde{\beta} + \tilde{\beta} \cdot \left[ \dfrac{Cov\left(Z_{1,res}, Y_{res}\right)}{\tilde{\beta}} - 1 \right] \\
	& \hspace{40pt} \text{by Equation \eqref{EqCandXU1Cov}} \\
	& \hspace{20pt} = Cov\left(Z_{1,res}, Y_{res}\right),
\end{align*}
implying that Equation \eqref{EqCandCovZ1Y} is valid.

Based on Equation \eqref{EqCandidateCovZ2Y}, we have that
\begin{align*}
	& Cov\left(\tilde{Z}_{2,res}, \tilde{Y}_{res}\right) \\
	& \hspace{20pt} = \tilde{\beta} + \tilde{\beta} \cdot Cov\left(\tilde{X}_{res}^{*}, \tilde{U}_{2}\right) \\
	& \hspace{20pt} = \tilde{\beta} + \tilde{\beta} \cdot \left[ \dfrac{Cov\left(Z_{2,res}, Y_{res}\right)}{\tilde{\beta}} - 1 \right] \\
	& \hspace{40pt} \text{by Equation \eqref{EqCandXU2Cov}} \\
	& \hspace{20pt} = Cov\left(Z_{2,res}, Y_{res}\right),
\end{align*}
implying that Equation \eqref{EqCandCovZ2Y} is valid.

Based on Equation \eqref{EqCandidateCovZ1Z2}, we have that
\begin{align*}
	& Cov\left(\tilde{Z}_{1,res}, \tilde{Z}_{2,res}\right) \\
	& \hspace{20pt} = 1 + Cov\left(\tilde{X}_{res}^{*}, \tilde{U}_{1}\right) + Cov\left(\tilde{X}_{res}^{*}, \tilde{U}_{2}\right) + Cov\left(\tilde{U}_{1}, \tilde{U}_{2}\right) \\
	& \hspace{20pt} = 1 + \left[ \dfrac{Cov\left(Z_{1,res}, Y_{res}\right)}{\tilde{\beta}} - 1 \right]  + \left[ \dfrac{Cov\left(Z_{2,res}, Y_{res}\right)}{\tilde{\beta}} - 1 \right] \\
	& \hspace{40pt} + \left[ 1 + Cov\left(Z_{1,res}, Z_{2,res}\right) - \left[\dfrac{Cov\left(Z_{1,res}, Y_{res}\right) + Cov\left(Z_{2,res}, Y_{res}\right)}{\tilde{\beta}} \right] \right] \\
	& \hspace{60pt} \text{by Equations \eqref{EqCandXU1Cov}-\eqref{EqCandU1U2Cov}} \\
	& = Cov\left(Z_{1,res}, Z_{2,res}\right),
\end{align*}
implying that Equation \eqref{EqCandCovZ1Z2} is valid.

This concludes the proof for the case $\tilde{\beta} > 0$ because we have constructed a valid distribution for the candidate latent variables that satisfies the model restrictions and the data restrictions.

\subsubsection{Case 2: $\tilde{\beta} = 0$}\label{AppCase2}

First, we note that $\tilde{\beta} = 0$ implies that $LB = 0$ because $LB \geq 0$ by Assumption \ref{AsDataRestrictions}. Consequently, we have that
\begin{equation}\label{EqCovZ1Yzero}
	Cov\left(Z_{1,res}, Y_{res}\right) = 0
\end{equation}
and
\begin{equation}\label{EqCovZ2Yzero}
	Cov\left(Z_{2,res}, Y_{res}\right) = 0.
\end{equation}

Now, we need to create a distribution for the candidate latent variables $\left(\tilde{X}_{res}^{*}, \tilde{\epsilon}_{res}, \tilde{U}_{1}, \tilde{U}_{2}\right)$. The candidate observable variables $\left(\tilde{Y}_{res}, \tilde{Z}_{1,res}, \tilde{Z}_{1,res}\right)$ are created based on Equations \eqref{EqCandY}-\eqref{EqCandZ2}.

We impose that $$\left(\tilde{X}_{res}^{*}, \tilde{\epsilon}_{res}, \tilde{U}_{1}, \tilde{U}_{2}\right) \text{ are jointly distributed with mean } \boldsymbol{\mu} \text{ and variance } \boldsymbol{\Omega}.$$

The vector $\boldsymbol{\mu}$ is given by
\begin{align}
	\label{EqCandXexpZero} \mathbb{E}\left[\tilde{X}_{res}^{*}\right] & \coloneqq 0 \\
	\label{EqCandEpsexpZero} \mathbb{E}\left[\tilde{\epsilon}_{res} \right] & \coloneqq 0 \\
	\label{EqCandU1expZero} \mathbb{E}\left[\tilde{U}_{1} \right] & \coloneqq \mathbb{E}\left[Z_{1,res} \right] \\
	\label{EqCandU2expZero} \mathbb{E}\left[\tilde{U}_{2} \right] & \coloneqq \mathbb{E}\left[Z_{2,res} \right]
\end{align}
and the covariance matrix $\boldsymbol{\Omega}$ is given by
\begin{align}
	\label{EqCandXVarZero}Var\left(\tilde{X}_{res}^{*} \right) & \coloneqq 1 \\
	\label{EqCandEpsVarZero} Var\left(\tilde{\epsilon}_{res} \right) & \coloneqq Var\left(Y_{res}\right) \\
	\label{EqCandU1VarZero} Var\left(\tilde{U}_{1} \right) & \coloneqq 1 + Var\left(Z_{1,res}\right) \\
	\label{EqCandU2VarZero} Var\left(\tilde{U}_{2} \right) & \coloneqq 1 + Var\left(Z_{2,res}\right) \\
	Cov\left(\tilde{X}_{res}^{*},\tilde{\epsilon}_{res}\right) & \coloneqq 0 \\
	Cov\left(\tilde{U}_{1},\tilde{\epsilon}_{res}\right) & \coloneqq 0 \\
	\label{EqCandEpsU2CovZero} Cov\left(\tilde{U}_{2},\tilde{\epsilon}_{res}\right) & \coloneqq 0 \\
	\label{EqCandXU1CovZero} Cov\left(\tilde{X}_{res}^{*},\tilde{U}_{1}\right) & \coloneqq -1 \\
	\label{EqCandXU2CovZero} Cov\left(\tilde{X}_{res}^{*},\tilde{U}_{2}\right) & \coloneqq -1 \\
	\label{EqCandU1U2CovZero} Cov\left(\tilde{U}_{1},\tilde{U}_{2}\right) & \coloneqq Cov\left(Z_{1,res}, Z_{2,res}\right) + 1.
\end{align}

First, we must show that the above distribution is a valid distribution, i.e., we must show that
\begin{align}
	\label{EqRes11} Var\left(\tilde{\epsilon}_{res} \right) & \geq 0 \\
	\label{EqRes12} Var\left(\tilde{U}_{1} \right) & \geq 0 \\
	\label{EqRes13} Var\left(\tilde{U}_{2} \right) & \geq 0 \\
	\label{EqRes14} Var\left(\tilde{\epsilon}_{res} \right) & \geq \left[ Cov\left(\tilde{X}_{res}^{*},\tilde{\epsilon}_{res}\right) \right]^{2} \\
	\label{EqRes15}  Var\left(\tilde{U}_{1} \right) & \geq \left[ Cov\left(\tilde{X}_{res}^{*},\tilde{U}_{1}\right) \right]^{2} \\
	\label{EqRes16} Var\left(\tilde{U}_{2} \right) & \geq \left[ Cov\left(\tilde{X}_{res}^{*},\tilde{U}_{2}\right) \right]^{2} \\
	\label{EqRes17} Var\left(\tilde{U}_{1} \right) \cdot Var\left(\tilde{\epsilon}_{res} \right) & \geq \left[ Cov\left(\tilde{U}_{1},\tilde{\epsilon}_{res}\right) \right]^{2} \\
	\label{EqRes18} Var\left(\tilde{U}_{2} \right) \cdot Var\left(\tilde{\epsilon}_{res} \right) & \geq \left[ Cov\left(\tilde{U}_{2},\tilde{\epsilon}_{res}\right) \right]^{2} \\
	\label{EqRes19} Var\left(\tilde{U}_{1} \right) \cdot Var\left(\tilde{U}_{2} \right) & \geq \left[ Cov\left(\tilde{U}_{1},\tilde{U}_{2}\right) \right]^{2},
\end{align}
and
\begin{equation}\label{EqDet11tilde}
	det(\tilde{A}) = \tilde{a} \cdot (\tilde{d} \cdot \tilde{f} - \tilde{e}^2) + \tilde{b} \cdot (\tilde{c} \cdot \tilde{e} - \tilde{b} \cdot \tilde{f}) + \tilde{c} \cdot (\tilde{b} \cdot \tilde{e} - \tilde{d} \cdot \tilde{c}) \geq 0,
\end{equation}
where
\begin{equation}\label{EqMatrixA}
	\tilde{A} \coloneqq \left[\begin{matrix}
		\tilde{a} & \tilde{b} & \tilde{c} \\
		\tilde{b} & \tilde{d} & \tilde{e} \\
		\tilde{c} & \tilde{e} & \tilde{f}
	\end{matrix}\right]
\end{equation}
and
\begin{align}
	\tilde{a} & \coloneqq Var\left(\tilde{X}_{res}^{*}\right) \\
	\tilde{b} & \coloneqq Cov\left(\tilde{X}_{res}^{*}, \tilde{U}_{1}\right)\\
	\tilde{c} & \coloneqq Cov\left(\tilde{X}_{res}^{*}, \tilde{U}_{2}\right)\\
	\tilde{d} & \coloneqq Var\left(\tilde{U}_{1}\right) \\
	\tilde{e} & \coloneqq Cov\left(\tilde{U}_{1}, \tilde{U}_{2}\right)\\
	\tilde{f} & \coloneqq Var\left(\tilde{U}_{2}\right). \label{EqMatrixf}
\end{align}

Observe that Inequalities \eqref{EqRes11}-\eqref{EqRes14}, \eqref{EqRes17} and \eqref{EqRes18} are trivially satisfied due to Equations \eqref{EqCandXVarZero}-\eqref{EqCandEpsU2CovZero}.

Note also that Inequalities \eqref{EqRes15} and \eqref{EqRes16} are satisfied because their right-hand sides are equal to 1 while their left-hand side must be greater than or equal to 1 according to Equations \eqref{EqCandU1VarZero}, \eqref{EqCandU2VarZero}, \eqref{EqCandXU1CovZero} and \eqref{EqCandXU2CovZero}.

Additionally, observe that Inequality \eqref{EqRes19} is satisfied according to the argument in Footnote \ref{FootCS}.

Lastly, Inequality \eqref{EqDet11tilde} is satisfied strictly because $Var\left(Z_{1,res}\right) \cdot Var\left(Z_{2,res}\right) > \left[Cov\left(Z_{1,res}, Z_{2,res}\right)\right]^{2}$ according to the Cauchy–Schwarz inequality and Assumption \ref{AsDataRestrictions}. To prove this claim, use Equation \eqref{EqDet2} and plug the values Equations \eqref{EqCandXVarZero}-\eqref{EqCandU1U2CovZero} and \eqref{EqMatrixA}-\eqref{EqMatrixf}.

Now, we must show that the candidate latent variables satisfy Equations \eqref{EqCandCovXeps}-\eqref{EqCandCovU1U2}. Notice that Restrictions \eqref{EqCandCovXeps}-\eqref{EqCandCovU2eps} are satisfied by construction. Observe also that Restriction \eqref{EqCandCovU1U2} is satisfied due to Assumption \ref{AsDataRestrictions}.

Lastly, we must show that the data restrictions (Equations \eqref{EqCandExpY}-\eqref{EqCandCovZ1Z2}) are valid. Since we have shown that candidate latent variables satisfy Equations \eqref{EqCandCovXeps}-\eqref{EqCandCovU1U2}, we can follow the same steps used to prove Lemma \ref{LemmaDataRestrictions} to show that
\begin{align}
	\label{EqCandidateExpYZero} \mathbb{E}\left[\tilde{Y}_{res}\right] & = 0 \\
	\label{EqCandidateExpZ1Zero} \mathbb{E}\left[\tilde{Z}_{1,res}\right] & = \mathbb{E}\left[\tilde{U}_{1}\right] \\
	\label{EqCandidateExpZ2Zero} \mathbb{E}\left[\tilde{Z}_{2,res}\right] & = \mathbb{E}\left[\tilde{U}_{2}\right] \\
	\label{EqCandidateVarYZero} Var\left(\tilde{Y}_{res}\right) & = \tilde{\beta}^{2} + Var\left(\tilde{\epsilon}_{res}\right) \\
	\label{EqCandidateVarZ1Zero} Var\left(\tilde{Z}_{1,res}\right) & = 1 + Var\left(\tilde{U}_{1}\right) + 2 \cdot Cov\left(\tilde{X}_{res}^{*}, \tilde{U}_{1}\right) \\
	\label{EqCandidateVarZ2Zero} Var\left(\tilde{Z}_{2,res}\right) & = 1 + Var\left(\tilde{U}_{2}\right) + 2 \cdot Cov\left(\tilde{X}_{res}^{*}, \tilde{U}_{2}\right) \\
	\label{EqCandidateCovZ1YZero} Cov\left(\tilde{Z}_{1,res}, \tilde{Y}_{res}\right) & = \tilde{\beta} + \tilde{\beta} \cdot Cov\left(\tilde{X}_{res}^{*}, \tilde{U}_{1}\right) \\
	\label{EqCandidateCovZ2YZero} Cov\left(\tilde{Z}_{2,res}, \tilde{Y}_{res}\right) & = \tilde{\beta} + \tilde{\beta} \cdot Cov\left(\tilde{X}_{res}^{*}, \tilde{U}_{2}\right) \\
	\label{EqCandidateCovZ1Z2Zero} Cov\left(\tilde{Z}_{1,res}, \tilde{Z}_{2,res}\right) & = 1 + Cov\left(\tilde{X}_{res}^{*}, \tilde{U}_{1}\right) + Cov\left(\tilde{X}_{res}^{*}, \tilde{U}_{2}\right) + Cov\left(\tilde{U}_{1}, \tilde{U}_{2}\right).
\end{align}

Based on Equation \eqref{EqCandidateExpYZero}, we have that
\begin{align*}
	\mathbb{E}\left[\tilde{Y}_{res}\right] & = 0 = \mathbb{E}\left[Y_{res}\right]
\end{align*}
according to Equation \eqref{EqExpY}, implying that Equation \eqref{EqCandExpY} is valid.

Based on Equation \eqref{EqCandidateExpZ1Zero}, we have that
\begin{align*}
	\mathbb{E}\left[\tilde{Z}_{1,res}\right] & = \mathbb{E}\left[\tilde{U}_{1}\right] = \mathbb{E}\left[Z_{1,res}\right]
\end{align*}
according to Equation \eqref{EqCandU1expZero}, implying that Equation \eqref{EqCandExpZ1} is valid.

Based on Equation \eqref{EqCandidateExpZ2Zero}, we have that
\begin{align*}
	\mathbb{E}\left[\tilde{Z}_{2,res}\right] & = \mathbb{E}\left[\tilde{U}_{2}\right] = \mathbb{E}\left[Z_{2,res}\right]
\end{align*}
according to Equation \eqref{EqCandU2expZero}, implying that Equation \eqref{EqCandExpZ2} is valid.

Based on Equation \eqref{EqCandidateVarYZero}, we have that
\begin{align*}
	Var\left(\tilde{Y}_{res}\right) & = Var\left(\tilde{\epsilon}_{res}\right) = Var\left(Y_{res}\right)
\end{align*}
by Equation \eqref{EqCandEpsVarZero}, implying that Equation \eqref{EqCandVarY} is valid.

Based on Equation \eqref{EqCandidateVarZ1Zero}, we have that
\begin{align*}
	& Var\left(\tilde{Z}_{1,res}\right) \\
	& \hspace{20pt} = 1 + Var\left(\tilde{U}_{1}\right) + 2 \cdot Cov\left(\tilde{X}_{res}^{*}, \tilde{U}_{1}\right) \\
	& \hspace{20pt} = 1 + \left[1 + Var\left(Z_{1,res}\right) \right] + 2 \cdot \left[ - 1 \right] \\
	& \hspace{40pt} \text{by Equations \eqref{EqCandU1VarZero} and \eqref{EqCandXU1CovZero}} \\
	& \hspace{20pt} = Var\left(Z_{1,res}\right),
\end{align*}
implying that Equation \eqref{EqCandVarZ1} is valid.

Based on Equation \eqref{EqCandidateVarZ2Zero}, we have that
\begin{align*}
	& Var\left(\tilde{Z}_{2,res}\right) \\
	& \hspace{20pt} = 1 + Var\left(\tilde{U}_{2}\right) + 2 \cdot Cov\left(\tilde{X}_{res}^{*}, \tilde{U}_{2}\right) \\
	& \hspace{20pt} = 1 + \left[1 + Var\left(Z_{2,res}\right) \right] + 2 \cdot \left[ - 1 \right] \\
	& \hspace{40pt} \text{by Equations \eqref{EqCandU2VarZero} and \eqref{EqCandXU2CovZero}} \\
	& \hspace{20pt} = Var\left(Z_{2,res}\right),
\end{align*}
implying that Equation \eqref{EqCandVarZ2} is valid.

Based on Equation \eqref{EqCandidateCovZ1YZero}, we have that
\begin{align*}
	Cov\left(\tilde{Z}_{1,res}, \tilde{Y}_{res}\right) & = 0 = Cov\left(Z_{1,res}, Y_{res}\right)
\end{align*}
by Equation \eqref{EqCovZ1Yzero}, implying that Equation \eqref{EqCandCovZ1Y} is valid.

Based on Equation \eqref{EqCandidateCovZ2YZero}, we have that
\begin{align*}
	Cov\left(\tilde{Z}_{2,res}, \tilde{Y}_{res}\right) & = 0 = Cov\left(Z_{2,res}, Y_{res}\right)
\end{align*}
by Equation \eqref{EqCovZ2Yzero}, implying that Equation \eqref{EqCandCovZ2Y} is valid.

Based on Equation \eqref{EqCandidateCovZ1Z2}, we have that
\begin{align*}
	& Cov\left(\tilde{Z}_{1,res}, \tilde{Z}_{2,res}\right) \\
	& \hspace{20pt} = 1 + Cov\left(\tilde{X}_{res}^{*}, \tilde{U}_{1}\right) + Cov\left(\tilde{X}_{res}^{*}, \tilde{U}_{2}\right) + Cov\left(\tilde{U}_{1}, \tilde{U}_{2}\right) \\
	& \hspace{20pt} = 1 + \left[ - 1 \right]  + \left[ - 1 \right] + \left[ 1 + Cov\left(Z_{1,res}, Z_{2,res}\right) \right] \\
	& \hspace{60pt} \text{by Equations \eqref{EqCandXU1CovZero}-\eqref{EqCandU1U2CovZero}} \\
	& = Cov\left(Z_{1,res}, Z_{2,res}\right),
\end{align*}
implying that Equation \eqref{EqCandCovZ1Z2} is valid.

This concludes the proof for the case $\tilde{\beta} = 0$ because we have constructed a distribution for the candidate latent variables that satisfies the model restrictions and the data restrictions.

\phantom{a}

Combining Parts 1 and 2 of Appendix \ref{AppProofAllInfo}, we can conclude that, if $UB \geq LB$, then the interval $\left[LB,UB \right]$ exhausts all the information contained in the first two moments of the data distribution.

\pagebreak

\section{No Prior Knowledge of the Target Parameter's Sign} \label{AppNoSign}

\setcounter{table}{0}
\renewcommand\thetable{B.\arabic{table}}

\setcounter{figure}{0}
\renewcommand\thefigure{B.\arabic{figure}}

\setcounter{equation}{0}
\renewcommand\theequation{B.\arabic{equation}}

\setcounter{theorem}{0}
\renewcommand\thetheorem{B.\arabic{theorem}}

\setcounter{proposition}{0}
\renewcommand\theproposition{B.\arabic{proposition}}

\setcounter{corollary}{0}
\renewcommand\thecorollary{B.\arabic{corollary}}

\setcounter{assumption}{0}
\renewcommand\theassumption{B.\arabic{assumption}}

\setcounter{definition}{0}
\renewcommand\thedefinition{B.\arabic{definition}}

\setcounter{Lemma}{0}
\renewcommand\theLemma{B.\arabic{Lemma}}

In this appendix, we propose a partial identification strategy that requires no prior knowledge about the sign of the true target parameter $\beta$. To do so, we keep Assumptions \ref{AsExogeneous}, \ref{AsScale}, \ref{AsMEexogenous} and \ref{AsNonClassical}, and modify Assumptions \ref{AsPosEff} and \ref{AsDataRestrictions}. For convenience, we list all the necessary assumptions here.

\begin{assumption}[Exogeneity]\label{AsAppExogeneous}
	After conditioning on the covariate vector $W$, the true unobserved measure $X^{*}$ is exogenous, i.e., $Cov\left(X_{res}^{*}, \epsilon_{res}\right) = 0$.
\end{assumption}

\begin{assumption}[Scale Normalization]\label{AsAppScale}
	The scale of the true unobserved measure is normalized to have unit variance, i.e., $Var\left(X_{res}^{*}\right) = 1$.
\end{assumption}

\begin{assumption}[Reduced-Form Coefficients]\label{AsAppPosEff}
	The reduced-form coefficients identify the sign of the true target parameter, i.e., $sign\left\lbrace Cov\left(Z_{1}, Y_{res}\right) \right\rbrace = sign\left\lbrace Cov\left(Z_{2}, Y_{res}\right) \right\rbrace = sign\left\lbrace \beta \right\rbrace$.\footnote{The $sign$ function equals one if its argument is a strictly positive number, zero if its argument is zero, and minus one if its argument is a strictly negative number.}
\end{assumption}

\begin{assumption}[Exogeneity of the Measurement Errors]\label{AsAppMEexogenous}
	The measurement error variables are exogenous, i.e., $Cov\left(U_{1}, \epsilon_{res}\right) = 0$ and $Cov\left(U_{2}, \epsilon_{res}\right) = 0$.
\end{assumption}

\begin{assumption}[Nonclassical Measurement Error]\label{AsAppNonClassical}
	The measurement error terms are positively correlated with each other, i.e., $Cov\left(U_{1}, U_{2}\right) \geq 0$.
\end{assumption}

\begin{assumption}[Well-behaved Data Distributions]\label{AsAppDataRestrictions}
	The variance of the outcome and proxy variables are strictly positive, the covariance of the two proxy variables is strictly greater than the negative of the variable of interest's variance, and the proxies are not linearly dependent, i.e., $Var\left(Y_{res}\right) > 0$, $Var\left(Z_{1,res}\right) > 0$, $Var\left(Z_{2,res}\right) > 0$, $Cov\left(Z_{1,res}, Z_{2,res}\right) > -1$, and $Var\left(Z_{1,res}\right) \cdot Var\left(Z_{2,res}\right) \neq \left[Cov\left(Z_{1,res}, Z_{2,res}\right)\right]^{2}$.
\end{assumption}

Before stating our identification result, define
\begin{equation}\label{EqLBb}
	LB_{B} \coloneqq \max \left\lbrace \begin{matrix}
		\dfrac{Cov\left(Z_{1,res}, Y_{res}\right) + Cov\left(Z_{2,res}, Y_{res}\right)}{1 + Cov\left(Z_{1,res}, Z_{2,res}\right)}, \hspace{10pt} \dfrac{- B + \sqrt{B^{2} - 4 \cdot A \cdot C}}{- 2 \cdot C}, \\
		\\
		- \sqrt{\dfrac{\left\lbrace \begin{matrix}
					Cov(Z_{1,res}, - Y_{res})^2 \cdot Var(Z_{2,res}) + Cov(Z_{2,res}, - Y_{res})^2 \cdot Var(Z_{1,res}) \\
					- 2\cdot Cov(Z_{1,res}, - Y_{res}) \cdot Cov(Z_{2,res}, - Y_{res}) \cdot Cov(Z_{1,res},Z_{2,res})
				\end{matrix} \right\rbrace}{Var(Z_{1,res}) \cdot Var(Z_{2,res}) - Cov(Z_{1,res},Z_{2,res})^2}}
	\end{matrix}  \right\rbrace,
\end{equation}
where
\begin{align*}
	A & \coloneqq -\left[Cov\left(Z_{1,res}, - Y_{res}\right) - Cov\left(Z_{2,res}, - Y_{res}\right)\right]^{2}, \\
	B & \coloneqq -2 \cdot \left\lbrace\begin{matrix}
		Cov\left(Z_{1,res}, - Y_{res}\right) \cdot \left[Var\left(Z_{2,res}\right) - Cov\left(Z_{1,res}, Z_{2,res}\right)\right] \\
		\hspace{10pt} + Cov\left(Z_{2,res}, - Y_{res}\right) \cdot \left[Var\left(Z_{1,res}\right) - Cov\left(Z_{1,res}, Z_{2,res}\right)\right]
	\end{matrix}  \right\rbrace, \text{ and} \\
	C & \coloneqq \left[Var\left(Z_{1,res}\right) + 1\right] \cdot \left[Var\left(Z_{2,res}\right) + 1\right] - \left[Cov\left(Z_{1,res}, Z_{2,res}\right) + 1\right]^{2},
\end{align*}

We now state the main result of this appendix: partial identification of $\beta$ (Equation \eqref{EqModel}) using two proxies (Equation \eqref{EqProxiesCovariates}) with nonclassical measurement error and no prior knowledge of target parameter's sign.
\begin{proposition}\label{PropAppMainId}
	Under Assumptions \ref{AsAppExogeneous}-\ref{AsAppDataRestrictions}, we have that:
	\begin{equation}\label{EqAppBoundsPositive}
		\beta \in \left\lbrace \begin{array}{cl}
			\left\lbrace \emptyset \right\rbrace & \text{ if } UB < LB \\
			\left[LB, UB\right] & \text{ if } UB \geq LB
		\end{array} \right.
	\end{equation}
	if \hspace{5pt} $Cov\left(Z_{1}, Y_{res}\right) > 0$, $$\beta = 0 \text{ if } \hspace{5pt} Cov\left(Z_{1}, Y_{res}\right) = 0$$ and
	\begin{equation}\label{EqAppBoundsNegative}
		\beta \in \left\lbrace \begin{array}{cl}
			\left\lbrace \emptyset \right\rbrace & \text{ if } LB_{B} < -UB \\
			\left[-UB, LB_{B}\right] & \text{ if } LB_{B} \geq -UB
		\end{array} \right.
	\end{equation}
	if \hspace{5pt} $Cov\left(Z_{1}, Y_{res}\right) < 0$.

	Moreover, these bounds exhaust all the information contained in the first two moments of the observable variables $\left(Y_{res}, Z_{1,res},Z_{2,res}\right)$.
\end{proposition}
\begin{proof}
	See Appendix \ref{AppProofAppMainId}.
\end{proof}

The main drawback of Proposition \ref{PropAppMainId} in comparison to Proposition \ref{PropMainId} is that this appendix's result requires testing the sign of $Cov\left(Z_{1}, Y_{res}\right)$ and $Cov\left(Z_{2}, Y_{res}\right)$ before estimating the bounds around $\beta$. This pre-testing step causes some challenges in controlling size when conducting inference.

\subsection{Proof of Proposition \ref{PropAppMainId}}\label{AppProofAppMainId}

When $Cov\left(Z_{1}, Y_{res}\right) > 0$, Assumption \ref{AsAppPosEff} implies that $\beta > 0$. We can then follow the steps described in Appendix \ref{AppProofMainId} to derive Equation \eqref{EqAppBoundsPositive} and to show that these bounds exhaust all the information contained in the first two moments of the observable variables $\left(Y_{res}, Z_{1,res},Z_{2,res}\right)$.

When $Cov\left(Z_{1}, Y_{res}\right) = 0$, Assumption \ref{AsAppPosEff} implies that $\beta = 0$ and we have point-identification.

When $Cov\left(Z_{1}, Y_{res}\right) < 0$, Assumption \ref{AsAppPosEff} implies that $\beta < 0$. To derive Equation \eqref{EqAppBoundsNegative}, we start by rewriting Equation \eqref{EqModel} as $$\check{Y}_{res} = \check{\beta} \cdot X_{res}^{*} + \check{\epsilon}_{res},$$ where $\check{Y}_{res} \coloneqq - Y_{res}$, $\check{\beta} \coloneqq - \beta$ and $\check{\epsilon}_{res} \coloneqq - \epsilon_{res}$. Since $\check{\beta} > 0$, we can follow the steps described in Appendix \ref{AppProofMainId} to derive Equation \eqref{EqAppBoundsNegative} and to show that these bounds exhaust all the information contained in the first two moments of the observable variables $\left(Y_{res}, Z_{1,res},Z_{2,res}\right)$.

\pagebreak

\section{Endogenous Measurement Error Terms} \label{AppEndoME}

\setcounter{table}{0}
\renewcommand\thetable{C.\arabic{table}}

\setcounter{figure}{0}
\renewcommand\thefigure{C.\arabic{figure}}

\setcounter{equation}{0}
\renewcommand\theequation{C.\arabic{equation}}

\setcounter{theorem}{0}
\renewcommand\thetheorem{C.\arabic{theorem}}

\setcounter{proposition}{0}
\renewcommand\theproposition{C.\arabic{proposition}}

\setcounter{corollary}{0}
\renewcommand\thecorollary{C.\arabic{corollary}}

\setcounter{assumption}{0}
\renewcommand\theassumption{C.\arabic{assumption}}

\setcounter{definition}{0}
\renewcommand\thedefinition{C.\arabic{definition}}

\setcounter{Lemma}{0}
\renewcommand\theLemma{C.\arabic{Lemma}}

In this appendix, we propose a partial identification strategy that allows the proxies' measurement error terms $\left(U_{1}, U_{2} \right)$ to be endogenous. To do so, we keep Assumptions \ref{AsExogeneous}, \ref{AsScale}, \ref{AsPosEff}, and \ref{AsNonClassical}, weaken Assumption \ref{AsMEexogenous}, and strengthen Assumption \ref{AsDataRestrictions}.\footnote{It is possible to prove Proposition \ref{PropAppEndoMEMainId} using Assumption \ref{AsDataRestrictions} instead of Assumption \ref{AsAppEndoMEDataRestrictions}. However, its proof would be much longer because it would require to show that the matrix $\Omega$  in Case 2 of Appendix \ref{AppProofEndoMEAllInfo} is positive semi-definite instead of positive definite.} For convenience, we list all the necessary assumptions here.

\begin{assumption}[Exogeneity]\label{AsAppEndoMEExogeneous}
	After conditioning on the covariate vector $W$, the true unobserved measure $X^{*}$ is exogenous, i.e., $Cov\left(X_{res}^{*}, \epsilon_{res}\right) = 0$.
\end{assumption}

\begin{assumption}[Scale Normalization]\label{AsAppEndoMEScale}
	The scale of the true unobserved measure is normalized to have unit variance, i.e., $Var\left(X_{res}^{*}\right) = 1$.
\end{assumption}

\begin{assumption}[Known Direction of the True Effect]\label{AsAppEndoMEPosEff}
	The direction of the effect of the true unobservable variable of interest on the outcome variable is known, i.e., $\beta \geq 0$.
\end{assumption}

\begin{assumption}[Exogeneity of the Measurement Errors]\label{AsAppEndoMEMEexogenous}
	The measurement error variables are possibly endogenous, i.e., $Cov\left(U_{1}, \epsilon_{res}\right) \in \mathbb{R}$ and $Cov\left(U_{2}, \epsilon_{res}\right) \in \mathbb{R}$.\footnote{Since Assumption \ref{AsAppEndoMEMEexogenous} does not impose any restrictions on $Cov\left(U_{1}, \epsilon_{res}\right)$ and $Cov\left(U_{1}, \epsilon_{res}\right)$, it is not technically an assumption. Despite this technicality, we state it as an assumption to clearly distinguish this appendix's results from those in the main text.}\textsuperscript{,}\footnote{The results in this appendix hold even if we impose $Cov\left(U_{1}, \epsilon_{res}\right) \geq 0$ and $Cov\left(U_{2}, \epsilon_{res}\right) \geq 0$. This alternative assumption still allows for the measurement error terms to be endogenous. In this appendix, we opt for imposing no restrictions on $Cov\left(U_{1}, \epsilon_{res}\right)$ and $Cov\left(U_{1}, \epsilon_{res}\right)$ to improve interpretability.}
\end{assumption}

\begin{assumption}[Nonclassical Measurement Error]\label{AsAppEndoMENonClassical}
	The measurement error terms are positively correlated with each other, i.e., $Cov\left(U_{1}, U_{2}\right) \geq 0$.
\end{assumption}

\begin{assumption}[Well-behaved Data Distributions]\label{AsAppEndoMEDataRestrictions}
	The variance of the outcome and proxy variables are strictly positive, the covariances between each proxy and the outcome variable are positive, the covariance of the two proxy variables is strictly greater than the negative of the variable of interest's variance, and there is no perfectly linear relationship among the proxies and the outcome variable, i.e., $Var\left(Y_{res}\right) > 0$, $Var\left(Z_{1,res}\right) > 0$, $Var\left(Z_{2,res}\right) > 0$, $Cov\left(Z_{1,res}, Y_{res}\right) \geq 0$, $Cov\left(Z_{2,res}, Y_{res}\right) \geq 0$, $Cov\left(Z_{1,res}, Z_{2,res}\right) > -1$, and $Var\left(Z_{1,res}\right) \cdot Var\left(Z_{2,res}\right) \neq \left[Cov\left(Z_{1,res}, Z_{2,res}\right)\right]^{2}$, $Var\left(Z_{1,res}\right) \cdot Var\left(Y\right) \neq \left[Cov\left(Z_{1,res}, Y\right)\right]^{2}$, $Var\left(Z_{2,res}\right) \cdot Var\left(Y\right) \neq \left[Cov\left(Z_{2,res}, Y\right)\right]^{2}$, and $$det\left(\left[\begin{matrix}
	Var(Y_{res}) & Cov\left(Z_{1,res}, Y_{res}\right) & Cov\left(Z_{2,res}, Y_{res}\right) \\
	Cov\left(Z_{1,res}, Y_{res}\right) & Var\left(Z_{1,res}\right) & Cov\left(Z_{1,res}, Z_{2,res}\right) \\
	Cov\left(Z_{2,res}, Y_{res}\right) & Cov\left(Z_{1,res}, Z_{2,res}\right) & Var\left(Z_{2,res}\right)
	\end{matrix}\right]\right) > 0.$$
\end{assumption}

We now state our main result: partial identification of $\beta$ (Equation \eqref{EqModel}) using two proxies (Equation \eqref{EqProxiesCovariates}) with nonclassical and possibly endogenous measurement error.

\begin{proposition}\label{PropAppEndoMEMainId}
	Under Assumptions \ref{AsAppEndoMEExogeneous}-\ref{AsAppEndoMEDataRestrictions}, we have that:
	\begin{equation}\label{EqAppEndoMEBounds}
		\beta \in \left[LB_{C}, UB\right],
	\end{equation}
	where
	\begin{equation}\label{EqAppEndoMELB}
		LB_{C} \coloneqq 0
	\end{equation}
	and
	\begin{equation}\label{EqAppEndoMEUB}
		UB \coloneqq \sqrt{Var\left(Y_{res}\right)}.
	\end{equation}

	Moreover, the bounds in Equation \eqref{EqAppEndoMEBounds} exhaust all the information contained in the first two moments of the observable variables $\left(Y_{res}, Z_{1,res},Z_{2,res}\right)$.
\end{proposition}

\begin{proof}
	See Appendix \ref{AppProofAppEndoMEMainId}.
\end{proof}

The only difference between the bounds in Proposition \ref{PropAppEndoMEMainId} and the bounds in Proposition \ref{PropMainId} is the lower bound. When the measurement error terms are possibly endogenous, the lower bound becomes uninformative and entirely defined by prior knowledge of the direction of the true effect (Assumption \ref{AsAppEndoMEPosEff}). Consequently, we have that $LB_{C} \leq LB$, i.e., the bounds in Proposition \ref{PropAppEndoMEMainId} are weakly wider than the bounds in Proposition \ref{PropMainId}.

\subsection{Proof of Proposition \ref{PropAppEndoMEMainId}}\label{AppProofAppEndoMEMainId}

We break the proof of Proposition \ref{PropAppEndoMEMainId} in three parts:
\begin{enumerate}
	\item Section \ref{AppProofEndoMEDataRestriction} derives the restrictions imposed by the first two moments of the observable data on the underlying data-generating process.

	\item Section \ref{AppProofEndoMELbBetaUB} shows that $LB_{C} \leq \beta \leq UB$.

	\item Section \ref{AppProofEndoMEAllInfo} defines the meaning of exhausting all the information contained in the first two moments of the data distribution and shows that the interval $\left[LB_{C},UB \right]$ exhausts all the information contained in the first two moments of the data distribution.
\end{enumerate}

\subsubsection{Data Restrictions}\label{AppProofEndoMEDataRestriction}

In this section, we derive the restrictions imposed by the first two moments of the observable data distribution $\left(Y_{res}, Z_{1,res}, Z_{2,res}\right)$ on the distribution of the latent variables $\left(X_{res}^{*}, \epsilon_{res},U_{1},U_{2}\right)$. First, we state all nine data restrictions in a lemma. Afterwards, we prove them.

\begin{Lemma}[Data Restrictions]\label{AppEndoMELemmaDataRestrictions}
	Under Assumptions \ref{AsAppEndoMEExogeneous}, \ref{AsAppEndoMEScale} and \ref{AsAppEndoMEMEexogenous}, we have that
	\begin{align}
		\label{AppEndoMEEqExpY} \mathbb{E}\left[Y_{res}\right] & = 0 \\
		\label{AppEndoMEEqExpZ1} \mathbb{E}\left[Z_{1,res}\right] & = \mathbb{E}\left[U_{1}\right] \\
		\label{AppEndoMEEqExpZ2} \mathbb{E}\left[Z_{2,res}\right] & = \mathbb{E}\left[U_{2}\right] \\
		\label{AppEndoMEEqVarY} Var\left(Y_{res}\right) & = \beta^{2} + Var\left(\epsilon_{res}\right) \\
		\label{AppEndoMEEqVarZ1} Var\left(Z_{1,res}\right) & = 1 + Var\left(U_{1}\right) + 2 \cdot Cov\left(X_{res}^{*}, U_{1}\right) \\
		\label{AppEndoMEEqVarZ2} Var\left(Z_{2,res}\right) & = 1 + Var\left(U_{2}\right) + 2 \cdot Cov\left(X_{res}^{*}, U_{2}\right) \\
		\label{AppEndoMEEqCovZ1Y} Cov\left(Z_{1,res}, Y_{res}\right) & = \beta + \beta \cdot Cov\left(X_{res}^{*}, U_{1}\right) + Cov\left(U_{1}, \epsilon_{res}\right) \\
		\label{AppEndoMEEqCovZ2Y} Cov\left(Z_{2,res}, Y_{res}\right) & = \beta + \beta \cdot Cov\left(X_{res}^{*}, U_{2}\right) + Cov\left(U_{2}, \epsilon_{res}\right) \\
		\label{AppEndoMEEqCovZ1Z2} Cov\left(Z_{1,res}, Z_{2,res}\right) & = 1 + Cov\left(X_{res}^{*}, U_{1}\right) + Cov\left(X_{res}^{*}, U_{2}\right) + Cov\left(U_{1}, U_{2}\right).
	\end{align}
\end{Lemma}

\begin{proof}
	The proofs of Equations \eqref{AppEndoMEEqExpY}, \eqref{AppEndoMEEqExpZ1}, \eqref{AppEndoMEEqExpZ2}, \eqref{AppEndoMEEqVarY}, \eqref{AppEndoMEEqVarZ1}, \eqref{AppEndoMEEqVarZ2} and \eqref{AppEndoMEEqCovZ1Z2} are identical to proofs detailed in Appendix \ref{AppProofDataRestriction}.

	To prove Equations \eqref{AppEndoMEEqCovZ1Y} and \eqref{AppEndoMEEqCovZ2Y}, fix $k \in \left\lbrace 1, 2 \right\rbrace$ arbitrarily and note that
	\begin{align*}
		& Cov\left(Z_{k,res},Y_{res}\right) \\
		& \hspace{20pt} = Cov\left(X_{res}^{*} + U_{k}, \beta \cdot X_{res}^{*} + \epsilon_{res}\right) \\
		& \hspace{40pt} \text{by Equations \eqref{EqModel} and \eqref{EqProxiesCovariates}} \\
		& \hspace{20pt} = \beta \cdot Var\left(X_{res}^{*}\right) + Cov\left(X_{res}^{*}, \epsilon_{res}\right) + \beta \cdot Cov\left(X_{res}^{*}, U_{k}\right) + Cov\left(U_{k}, \epsilon_{res}\right) \\
		& \hspace{40pt} \text{by the properties of the covariance} \\
		& \hspace{20pt} = \beta + \beta \cdot Cov\left(X_{res}^{*}, U_{k}\right) + Cov\left(U_{k}, \epsilon_{res}\right) \\
		& \hspace{40pt} \text{by Assumptions \ref{AsAppEndoMEExogeneous}, \ref{AsAppEndoMEScale} and \ref{AsAppEndoMEMEexogenous}.}
	\end{align*}

\end{proof}

\subsubsection{$LB_{C} \leq \beta \leq UB$}\label{AppProofEndoMELbBetaUB}

In this section, we prove that (i) $\beta \leq UB$ and (ii) $\beta \geq LB_{C}$.

\textbf{Part 1: $\beta \leq UB$}

This proof is identical to the proof in Part 1 of Appendix \ref{AppProofLbBetaUB}.

Rearranging Equation \eqref{AppEndoMEEqVarY}, we have that
\begin{align*}
	\beta^{2} & = Var\left(Y_{res}\right) - Var\left(\epsilon_{res}\right) \\
	& \leq Var\left(Y_{res}\right)
\end{align*}
because $Var\left(\epsilon_{res}\right) \geq 0$, implying that $$\beta \leq \sqrt{Var\left(Y_{res}\right)} = UB$$ by Assumption \ref{AsAppEndoMEPosEff}.

~

\textbf{Part 2: $\beta \geq LB_{C}$}

We have that $\beta \geq LB_{C}$ because $LB_{C} = 0$ according to Equation \eqref{EqAppEndoMELB} and $\beta \geq 0$ according to Assumption \ref{AsAppEndoMEPosEff}.

\subsubsection{$\left[LB_{C},UB \right]$ exhausts all the information contained in the first two moments of the data distribution}\label{AppProofEndoMEAllInfo}

First, we define the meaning of exhausting all the information contained in the first two moments of the data distribution. Second, we show that the interval $\left[LB_{C},UB \right]$ exhausts all the information contained in the first two moments of the data distribution.

\begin{definition}
	A set $\mathcal{B} \subseteq \mathbb{R}$ exhausts all the information contained in the first two moments of the data distribution if, for any candidate target parameter $\tilde{\beta} \in \mathcal{B}$, there exist candidate latent variables $\left(\tilde{X}_{res}^{*}, \tilde{\epsilon}_{res}, \tilde{U}_{1}, \tilde{U}_{2}\right)$ and candidate observable variables $\left(\tilde{Y}_{res}, \tilde{Z}_{1,res}, \tilde{Z}_{1,res}\right)$ such that
	\begin{enumerate}
		\item the candidate observable variables are induced by the candidate latent variables according to the model restrictions imposed by Equations \eqref{EqModel} and \eqref{EqProxiesCovariates}, i.e.,
		\begin{align}
			\label{AppEndoMEEqCandY} \tilde{Y}_{res} & \coloneqq \tilde{\beta} \cdot \tilde{X}_{res}^{*} + \tilde{\epsilon}_{res} \\
			\label{AppEndoMEEqCandZ1} \tilde{Z}_{1,res} & \coloneqq \tilde{X}_{res}^{*} + \tilde{U}_{1} \\
			\label{AppEndoMEEqCandZ2} \tilde{Z}_{2,res} & \coloneqq \tilde{X}_{res}^{*} + \tilde{U}_{2};
		\end{align}

		\item the candidate latent variables and the candidate target parameter satisfy the model restrictions imposed by Assumptions \ref{AsAppEndoMEExogeneous}-\ref{AsAppEndoMENonClassical}, i.e.,
		\begin{align}
			\label{AppEndoMEEqCandCovXeps} Cov\left(\tilde{X}_{res}^{*}, \tilde{\epsilon}_{res}\right) & = 0 \\
			\label{AppEndoMEEqCandVarX} Var\left(\tilde{X}_{res}^{*}\right) & = 1 \\
			\label{AppEndoMEEqCandBeta} \tilde{\beta} & \geq 0 \\
			\label{AppEndoMEEqCandCovU1U2} Cov\left(\tilde{U}_{1}, \tilde{U}_{2}\right) & \geq 0
		\end{align}

		\item the first two moments of the candidate observable variables are equal to the first two moments of the true observable variables, i.e.,
		\begin{align}
			\label{AppEndoMEEqCandExpY} \mathbb{E}\left[\tilde{Y}_{res}\right] & = \mathbb{E}\left[Y_{res}\right] \\
			\label{AppEndoMEEqCandExpZ1} \mathbb{E}\left[\tilde{Z}_{1,res}\right] & = \mathbb{E}\left[Z_{1,res}\right] \\
			\label{AppEndoMEEqCandExpZ2} \mathbb{E}\left[\tilde{Z}_{2,res}\right] & = \mathbb{E}\left[Z_{2,res}\right] \\
			\label{AppEndoMEEqCandVarY} Var\left(\tilde{Y}_{res}\right) & = Var\left(Y_{res}\right) \\
			\label{AppEndoMEEqCandVarZ1} Var\left(\tilde{Z}_{1,res}\right) & = Var\left(Z_{1,res}\right) \\
			\label{AppEndoMEEqCandVarZ2} Var\left(\tilde{Z}_{2,res}\right) & = Var\left(Z_{2,res}\right) \\
			\label{AppEndoMEEqCandCovZ1Y} Cov\left(\tilde{Z}_{1,res}, \tilde{Y}_{res}\right) & = Cov\left(Z_{1,res}, Y_{res}\right) \\
			\label{AppEndoMEEqCandCovZ2Y} Cov\left(\tilde{Z}_{2,res}, \tilde{Y}_{res}\right) & = Cov\left(Z_{2,res}, Y_{res}\right) \\
			\label{AppEndoMEEqCandCovZ1Z2} Cov\left(\tilde{Z}_{1,res}, \tilde{Z}_{2,res}\right) & = Cov\left(Z_{1,res}, Z_{2,res}\right).
		\end{align}
	\end{enumerate}
\end{definition}

Now, we show that the interval $\left[LB_{C},UB \right]$ exhausts all the information contained in the first two moments of the data distribution. This proof has three parts. First, we show that there exists a distribution for the candidate latent variables $\left(\tilde{X}_{res}^{*}, \tilde{\epsilon}_{res}, \tilde{U}_{1}, \tilde{U}_{2}\right)$ that is consistent with $\beta = UB$. Then, we show that there exists a distribution for the candidate latent variables $\left(\tilde{X}_{res}^{*}, \tilde{\epsilon}_{res}, \tilde{U}_{1}, \tilde{U}_{2}\right)$ that is consistent with $\beta = LB_{C}$. Lastly, we use mixtures of these two distributions to show that there exists a distribution for the candidate latent variables $\left(\tilde{X}_{res}^{*}, \tilde{\epsilon}_{res}, \tilde{U}_{1}, \tilde{U}_{2}\right)$ that is consistent with $\beta \in \left(LB_{C}, UB \right)$.

\phantom{a}

\textbf{Case 1: $\tilde{\beta} = UB$}

Fix $\tilde{\beta} = UB$. The proof is by construction.

We need to create a distribution for the candidate latent variables $\left(\tilde{X}_{res}^{*}, \tilde{\epsilon}_{res}, \tilde{U}_{1}, \tilde{U}_{2}\right)$. The candidate observable variables $\left(\tilde{Y}_{res}, \tilde{Z}_{1,res}, \tilde{Z}_{1,res}\right)$ are created based on Equations \eqref{AppEndoMEEqCandY}-\eqref{AppEndoMEEqCandZ2}.

We impose that $\left(\tilde{X}_{res}^{*}, \tilde{\epsilon}_{res}, \tilde{U}_{1}, \tilde{U}_{2}\right)$ follows the same distribution proposed in Case 1 in Appendix \ref{AppProofAllInfo}. According to the same steps described in Case 1 in Appendix \ref{AppProofAllInfo}, this distribution is a valid distribution,  satisfies Assumptions \ref{AsAppEndoMEExogeneous}-\ref{AsAppEndoMENonClassical} and satisfies the data restrictions. Additionally, this distribution is consistent with $\tilde{\beta} = UB$.

\phantom{a}

\textbf{Case 2: $\beta = LB_{C}$}

Fix $\tilde{\beta} = LB_{C} = 0$. The proof is by construction.

We need to create a distribution for the candidate latent variables $\left(\tilde{X}_{res}^{*}, \tilde{\epsilon}_{res}, \tilde{U}_{1}, \tilde{U}_{2}\right)$. The candidate observable variables $\left(\tilde{Y}_{res}, \tilde{Z}_{1,res}, \tilde{Z}_{1,res}\right)$ are created based on Equations \eqref{AppEndoMEEqCandY}-\eqref{AppEndoMEEqCandZ2}.

We impose that $$\left(\tilde{X}_{res}^{*}, \tilde{\epsilon}_{res}, \tilde{U}_{1}, \tilde{U}_{2}\right) \text{ are jointly distributed with mean } \boldsymbol{\mu} \text{ and variance } \boldsymbol{\Omega}.$$

The vector $\boldsymbol{\mu}$ is given by
\begin{align}
	\label{AppEndoMEEqCandXexp} \mathbb{E}\left[\tilde{X}_{res}^{*}\right] & \coloneqq 0 \\
	\label{AppEndoMEEqCandEpsexp} \mathbb{E}\left[\tilde{\epsilon}_{res} \right] & \coloneqq 0 \\
	\label{AppEndoMEEqCandU1exp} \mathbb{E}\left[\tilde{U}_{1} \right] & \coloneqq \mathbb{E}\left[Z_{1,res} \right] \\
	\label{AppEndoMEEqCandU2exp} \mathbb{E}\left[\tilde{U}_{2} \right] & \coloneqq \mathbb{E}\left[Z_{2,res} \right]
\end{align}
and the covariance matrix $\boldsymbol{\Omega}$ is given by
\begin{align}
	\label{AppEndoMEEqCandXVar} Var\left(\tilde{X}_{res}^{*} \right) & \coloneqq 1 \\
	\label{AppEndoMEEqCandEpsVar} Var\left(\tilde{\epsilon}_{res} \right) & \coloneqq Var\left(Y_{res}\right) \\
	\label{AppEndoMEEqCandU1Var} Var\left(\tilde{U}_{1} \right) & \coloneqq 1 + Var\left(Z_{1,res}\right) \\
	\label{AppEndoMEEqCandU2Var} Var\left(\tilde{U}_{2} \right) & \coloneqq 1 + Var\left(Z_{2,res}\right) \\
	Cov\left(\tilde{X}_{res}^{*},\tilde{\epsilon}_{res}\right) & \coloneqq 0 \\
	\label{AppEndoMEEqCandU1EpsilonCov} Cov\left(\tilde{U}_{1},\tilde{\epsilon}_{res}\right) & \coloneqq Cov\left(Z_{1,res}, Y_{res}\right) \\
	\label{AppEndoMEEqCandU2EpsilonCov} Cov\left(\tilde{U}_{2},\tilde{\epsilon}_{res}\right) & \coloneqq Cov\left(Z_{2,res}, Y_{res}\right) \\
	\label{AppEndoMEEqCandXU1Cov} Cov\left(\tilde{X}_{res}^{*},\tilde{U}_{1}\right) & \coloneqq - 1 \\
	\label{AppEndoMEEqCandXU2Cov} Cov\left(\tilde{X}_{res}^{*},\tilde{U}_{2}\right) & \coloneqq  - 1 \\
	\label{AppEndoMEEqCandU1U2Cov} Cov\left(\tilde{U}_{1},\tilde{U}_{2}\right) & \coloneqq 1 + Cov\left(Z_{1,res}, Z_{2,res}\right)
\end{align}

First, we must show that the above distribution is a valid distribution. To do so, we will show that
\begin{align}
	\Omega & = \left[\begin{matrix}
		Var\left(\tilde{X}_{res}^{*} \right) & Cov\left(\tilde{X}_{res}^{*},\tilde{U}_{1}\right) & Cov\left(\tilde{X}_{res}^{*},\tilde{U}_{2}\right) & Cov\left(\tilde{X}_{res}^{*},\tilde{\epsilon}_{res}\right) \\
		Cov\left(\tilde{X}_{res}^{*},\tilde{U}_{1}\right) & Var\left(\tilde{U}_{1} \right) & Cov\left(\tilde{U}_{1},\tilde{U}_{2}\right) & Cov\left(\tilde{U}_{1},\tilde{\epsilon}_{res}\right) \\
		Cov\left(\tilde{X}_{res}^{*},\tilde{U}_{2}\right) & Cov\left(\tilde{U}_{1},\tilde{U}_{2}\right) & Var\left(\tilde{U}_{2} \right) & Cov\left(\tilde{U}_{2},\tilde{\epsilon}_{res}\right) \\
		Cov\left(\tilde{X}_{res}^{*},\tilde{\epsilon}_{res}\right) & Cov\left(\tilde{U}_{1},\tilde{\epsilon}_{res}\right) & Cov\left(\tilde{U}_{2},\tilde{\epsilon}_{res}\right) & Var\left(\tilde{\epsilon}_{res}\right)
	\end{matrix}\right] \nonumber \\
	\label{EqOmega} & = \left[\begin{matrix}
		1 & -1 & -1 & 0 \\
		-1 & 1 + Var\left(Z_{1,res}\right) & 1 + Cov\left(Z_{1,res}, Z_{2,res}\right) & Cov\left(Z_{1,res}, Y_{res}\right) \\
		-1 & 1 + Cov\left(Z_{1,res}, Z_{2,res}\right) & 1 + Var\left(Z_{2,res}\right) & Cov\left(Z_{2,res}, Y_{res}\right) \\
		0 & Cov\left(Z_{1,res}, Y_{res}\right) & Cov\left(Z_{2,res}, Y_{res}\right) & Var\left(Y_{res}\right)
	\end{matrix}\right]
\end{align}
is positive definite, i.e., we need to check that the four leading principal minors of matrix $\Omega$ have strictly positive determinants.

\textbf{First Leading Principal Minor has strictly positive determinant:}

The first leading principal minor is $Var\left(\tilde{X}_{res}^{*} \right) = 1$ according to Equation \eqref{AppEndoMEEqCandXVar}.

\textbf{Second Leading Principal Minor has strictly positive determinant:}

The second leading principal minor is given by $$\Omega_{2} \coloneqq \left[\begin{matrix}
1 & -1 \\
-1 & 1 + Var\left(Z_{1,res}\right)
\end{matrix}\right],$$
whose determinant satisfies
\begin{align*}
	det\left(\Omega_{2}\right) = 1 + Var\left(Z_{1,res}\right) - 1 = Var\left(Z_{1,res}\right),
\end{align*}
which is strictly positive according to Assumption \ref{AsAppEndoMEDataRestrictions}.

\textbf{Third Leading Principal Minor has strictly positive determinant:}

The third leading principal minor is given by $$\Omega_{3} \coloneqq \left[\begin{matrix}
1 & -1 & -1 \\
-1 & 1 + Var\left(Z_{1,res}\right) & 1 + Cov\left(Z_{1,res}, Z_{2,res}\right) \\
-1 & 1 + Cov\left(Z_{1,res}, Z_{2,res}\right) & 1 + Var\left(Z_{2,res}\right)
\end{matrix}\right],$$
whose determinant is identical to the determinant in Equation \eqref{EqDet11tilde}. As explained in Appendix \ref{AppCase2}, this determinant is strictly positive.

\textbf{Fourth Leading Principal Minor has strictly positive determinant:}

The fourth leading principal minor is the full matrix $\Omega$. After some algebra, we can show that
\begin{align*}
	det\left(\Omega\right) & = Var(Y_{res}) \cdot [Var(Z_{1,res}) \cdot Var(Z_{2,res}) - Cov(Z_{1,res}, Z_{2,res})^2] \\
	& \hspace{20pt} - Cov(Z_{2,res}, Y_{res})^2 \cdot Var(Z_{1,res}) - Cov(Z_{1,res}, Y_{res})^2 \cdot Var(Z_{2,res}) \\
	& \hspace{20pt} + 2 \cdot Cov(Z_{1,res}, Y_{res}) \cdot Cov(Z_{2,res}, Y_{res}) \cdot Cov(Z_{1,res}, Z_2).
\end{align*}
Additionally, we have that $$det\left(\Omega\right) = det\left(\left[\begin{matrix}
Var(Y_{res}) & Cov\left(Z_{1,res}, Y_{res}\right) & Cov\left(Z_{2,res}, Y_{res}\right) \\
Cov\left(Z_{1,res}, Y_{res}\right) & Var\left(Z_{1,res}\right) & Cov\left(Z_{1,res}, Z_{2,res}\right) \\
Cov\left(Z_{2,res}, Y_{res}\right) & Cov\left(Z_{1,res}, Z_{2,res}\right) & Var\left(Z_{2,res}\right)
\end{matrix}\right]\right),$$
where the right-hand side is strictly positive according to Assumption \ref{AsAppEndoMEDataRestrictions}.

\phantom{a}

Hence, we can conclude that the proposed distribution for $\left(\tilde{X}_{res}^{*}, \tilde{\epsilon}_{res}, \tilde{U}_{1}, \tilde{U}_{2}\right)$ is a valid distribution.

Second, we must show that the candidate latent variables satisfy Equations \eqref{AppEndoMEEqCandCovXeps}-\eqref{AppEndoMEEqCandCovU1U2}. Since Restrictions \eqref{AppEndoMEEqCandCovXeps}-\eqref{AppEndoMEEqCandBeta} are satisfied by construction, we only have to show that Restriction \eqref{AppEndoMEEqCandCovU1U2} is satisfied. To do so, note that $Cov\left(\tilde{U}_{1},\tilde{U}_{2}\right) = 1 + Cov\left(Z_{1,res}, Z_{2,res}\right) > 0$ according to Equation \eqref{AppEndoMEEqCandU1U2Cov} and Assumption \ref{AsAppEndoMEDataRestrictions}. Consequently, all the model restrictions are satisfied.

Lastly, we must show that the data restrictions (Equations \eqref{AppEndoMEEqCandExpY}-\eqref{AppEndoMEEqCandCovZ1Z2}) are valid. Since we have shown that candidate latent variables satisfy Equations \eqref{AppEndoMEEqCandCovXeps}-\eqref{AppEndoMEEqCandCovU1U2}, we can follow the same steps used to prove Lemma \ref{AppEndoMELemmaDataRestrictions} to show that
\begin{align}
	\label{AppEndoMEEqCandidateExpY} \mathbb{E}\left[\tilde{Y}_{res}\right] & = 0 \\
	\label{AppEndoMEEqCandidateExpZ1} \mathbb{E}\left[\tilde{Z}_{1,res}\right] & = \mathbb{E}\left[\tilde{U}_{1}\right] \\
	\label{AppEndoMEEqCandidateExpZ2} \mathbb{E}\left[\tilde{Z}_{2,res}\right] & = \mathbb{E}\left[\tilde{U}_{2}\right] \\
	\label{AppEndoMEEqCandidateVarY} Var\left(\tilde{Y}_{res}\right) & = Var\left(\tilde{\epsilon}_{res}\right) \\
	\label{AppEndoMEEqCandidateVarZ1} Var\left(\tilde{Z}_{1,res}\right) & = 1 + Var\left(\tilde{U}_{1}\right) + 2 \cdot Cov\left(\tilde{X}_{res}^{*}, \tilde{U}_{1}\right) \\
	\label{AppEndoMEEqCandidateVarZ2} Var\left(\tilde{Z}_{2,res}\right) & = 1 + Var\left(\tilde{U}_{2}\right) + 2 \cdot Cov\left(\tilde{X}_{res}^{*}, \tilde{U}_{2}\right) \\
	\label{AppEndoMEEqCandidateCovZ1Y} Cov\left(\tilde{Z}_{1,res}, \tilde{Y}_{res}\right) & = Cov\left(\tilde{U}_{1}, \tilde{\epsilon}_{res}\right) \\
	\label{AppEndoMEEqCandidateCovZ2Y} Cov\left(\tilde{Z}_{2,res}, \tilde{Y}_{res}\right) & = Cov\left(\tilde{U}_{2}, \tilde{\epsilon}_{res}\right) \\
	\label{AppEndoMEEqCandidateCovZ1Z2} Cov\left(\tilde{Z}_{1,res}, \tilde{Z}_{2,res}\right) & = 1 + Cov\left(\tilde{X}_{res}^{*}, \tilde{U}_{1}\right) + Cov\left(\tilde{X}_{res}^{*}, \tilde{U}_{2}\right) + Cov\left(\tilde{U}_{1}, \tilde{U}_{2}\right).
\end{align}

Based on Equation \eqref{AppEndoMEEqCandidateExpY}, we have that
\begin{align*}
	\mathbb{E}\left[\tilde{Y}_{res}\right] & = 0 = \mathbb{E}\left[Y_{res}\right]
\end{align*}
according to Equation \eqref{AppEndoMEEqExpY}, implying that Equation \eqref{AppEndoMEEqCandExpY} is valid.

Based on Equation \eqref{AppEndoMEEqCandidateExpZ1}, we have that
\begin{align*}
	\mathbb{E}\left[\tilde{Z}_{1,res}\right] & = \mathbb{E}\left[\tilde{U}_{1}\right] = \mathbb{E}\left[Z_{1,res}\right]
\end{align*}
according to Equation \eqref{AppEndoMEEqCandU1exp}, implying that Equation \eqref{AppEndoMEEqCandExpZ1} is valid.

Based on Equation \eqref{AppEndoMEEqCandidateExpZ2}, we have that
\begin{align*}
	\mathbb{E}\left[\tilde{Z}_{2,res}\right] & = \mathbb{E}\left[\tilde{U}_{2}\right] = \mathbb{E}\left[Z_{2,res}\right]
\end{align*}
according to Equation \eqref{AppEndoMEEqCandU2exp}, implying that Equation \eqref{AppEndoMEEqCandExpZ2} is valid.

Based on Equation \eqref{AppEndoMEEqCandidateVarY}, we have that
\begin{align*}
	Var\left(\tilde{Y}_{res}\right) & = Var\left(\tilde{\epsilon}_{res}\right) \\
	& = Var\left(Y_{res}\right) \\
	& \hspace{20pt} \text{by Equation \eqref{AppEndoMEEqCandEpsVar}} \\
	& = Var\left(Y_{res}\right),
\end{align*}
implying that Equation \eqref{AppEndoMEEqCandVarY} is valid.

Based on Equation \eqref{AppEndoMEEqCandidateVarZ1}, we have that
\begin{align*}
	& Var\left(\tilde{Z}_{1,res}\right) \\
	& \hspace{20pt} = 1 + Var\left(\tilde{U}_{1}\right) + 2 \cdot Cov\left(\tilde{X}_{res}^{*}, \tilde{U}_{1}\right) \\
	& \hspace{20pt} = 1 + \left[1 + Var\left(Z_{1,res}\right)\right] + 2 \cdot \left[ - 1 \right] \\
	& \hspace{40pt} \text{by Equations \eqref{AppEndoMEEqCandU1Var} and \eqref{AppEndoMEEqCandXU1Cov}} \\
	& = Var\left(Z_{1,res}\right),
\end{align*}
implying that Equation \eqref{AppEndoMEEqCandVarZ1} is valid.

Based on Equation \eqref{AppEndoMEEqCandidateVarZ2}, we have that
\begin{align*}
	& Var\left(\tilde{Z}_{2,res}\right) \\
	& \hspace{20pt} = 1 + Var\left(\tilde{U}_{2}\right) + 2 \cdot Cov\left(\tilde{X}_{res}^{*}, \tilde{U}_{2}\right) \\
	& \hspace{20pt} = 1 + \left[1 + Var\left(Z_{2,res}\right)\right] + 2 \cdot \left[  - 1 \right] \\
	& \hspace{40pt} \text{by Equations \eqref{AppEndoMEEqCandU2Var} and \eqref{AppEndoMEEqCandXU2Cov}} \\
	& \hspace{20pt} = Var\left(Z_{2,res}\right),
\end{align*}
implying that Equation \eqref{AppEndoMEEqCandVarZ2} is valid.

Based on Equation \eqref{AppEndoMEEqCandidateCovZ1Y}, we have that
\begin{align*}
	& Cov\left(\tilde{Z}_{1,res}, \tilde{Y}_{res}\right) \\
	& \hspace{20pt} = Cov\left(\tilde{U}_{1}, \tilde{\epsilon}_{res}\right) \\
	& \hspace{20pt} = Cov\left(Z_{1,res}, Y_{res}\right)\\
	& \hspace{40pt} \text{by Equation \eqref{AppEndoMEEqCandU1EpsilonCov}},
\end{align*}
implying that Equation \eqref{AppEndoMEEqCandCovZ1Y} is valid.

Based on Equation \eqref{AppEndoMEEqCandidateCovZ2Y}, we have that
\begin{align*}
	& Cov\left(\tilde{Z}_{2,res}, \tilde{Y}_{res}\right) \\
	& \hspace{20pt} = Cov\left(\tilde{U}_{2}, \tilde{\epsilon}_{res}\right) \\
	& \hspace{20pt} = Cov\left(Z_{2,res}, Y_{res}\right)\\
	& \hspace{40pt} \text{by Equation \eqref{AppEndoMEEqCandU2EpsilonCov}},
\end{align*}
implying that Equation \eqref{AppEndoMEEqCandCovZ2Y} is valid.

Based on Equation \eqref{AppEndoMEEqCandidateCovZ1Z2}, we have that
\begin{align*}
	& Cov\left(\tilde{Z}_{1,res}, \tilde{Z}_{2,res}\right) \\
	& \hspace{20pt} = 1 + Cov\left(\tilde{X}_{res}^{*}, \tilde{U}_{1}\right) + Cov\left(\tilde{X}_{res}^{*}, \tilde{U}_{2}\right) + Cov\left(\tilde{U}_{1}, \tilde{U}_{2}\right) \\
	& \hspace{20pt} = 1 + \left[ - 1 \right]  + \left[ - 1 \right] + \left[ 1 + Cov\left(Z_{1,res}, Z_{2,res}\right) \right] \\
	& \hspace{60pt} \text{by Equations \eqref{AppEndoMEEqCandXU1Cov}-\eqref{AppEndoMEEqCandU1U2Cov}} \\
	& = Cov\left(Z_{1,res}, Z_{2,res}\right),
\end{align*}
implying that Equation \eqref{AppEndoMEEqCandCovZ1Z2} is valid.

This concludes the proof of Case 2 because we have constructed a valid distribution of the candidate random variables that satisfies Assumptions \ref{AsAppEndoMEExogeneous}-\ref{AsAppEndoMENonClassical}, satisfies the data restrictions, and is consistent with $\tilde{\beta} = LB_{C}$.

\phantom{a}

\textbf{Case 3: $\tilde{\beta} \in \left(LB_{C}, UB\right)$}

Fix $\tilde{\beta} \in \left(LB_{C}, UB\right)$. The proof is by construction.

We need to create a distribution for the candidate latent variables $\left(\tilde{X}_{res}^{*}, \tilde{\epsilon}_{res}, \tilde{U}_{1}, \tilde{U}_{2}\right)$. The candidate observable variables $\left(\tilde{Y}_{res}, \tilde{Z}_{1,res}, \tilde{Z}_{1,res}\right)$ are created based on Equations \eqref{AppEndoMEEqCandY}-\eqref{AppEndoMEEqCandZ2}.

To create this distribution, \begin{enumerate}
	\item define the vector of random variables $\tilde{L}_{U}$ according to the distribution of $\left(\tilde{X}_{res}^{*}, \tilde{\epsilon}_{res}, \tilde{U}_{1}, \tilde{U}_{2}\right)$ in Case 1 in this appendix.

	\item define the vector of random variables $\tilde{L}_{L}$ according to the distribution of $\left(\tilde{X}_{res}^{*}, \tilde{\epsilon}_{res}, \tilde{U}_{1}, \tilde{U}_{2}\right)$ in Case 2 in this appendix.

	\item define $\tilde{\alpha} \in \left(0,1\right)$ such that $\tilde{\beta} = \tilde{\alpha} \cdot LB_{C} + \left(1 - \tilde{\alpha}\right) \cdot UB$.

	\item define $\tilde{L}$ as a Bernoulli random variable with $\mathbb{P}\left[\tilde{L} = 1\right] = \tilde{\alpha}$ that is independent of $\tilde{L}_{L}$ and $\tilde{L}_{U}$.
\end{enumerate}

We impose that $$\left(\tilde{X}_{res}^{*}, \tilde{\epsilon}_{res}, \tilde{U}_{1}, \tilde{U}_{2}\right) = \tilde{L} \cdot \tilde{L}_{L} + \left(1 - \tilde{L}\right) \cdot \tilde{L}_{U}.$$

This concludes the proof of Case 3 because we have constructed a valid distribution of the candidate random variables that satisfies Assumptions \ref{AsAppEndoMEExogeneous}-\ref{AsAppEndoMENonClassical}, satisfies the data restrictions, and is consistent with $\tilde{\beta} \in \left(LB_{C}, UB\right)$.

\pagebreak

\section{Unrestricted Covariance between the Measurement Error Terms} \label{AppFreeCov}

\setcounter{table}{0}
\renewcommand\thetable{D.\arabic{table}}

\setcounter{figure}{0}
\renewcommand\thefigure{D.\arabic{figure}}

\setcounter{equation}{0}
\renewcommand\theequation{D.\arabic{equation}}

\setcounter{theorem}{0}
\renewcommand\thetheorem{D.\arabic{theorem}}

\setcounter{proposition}{0}
\renewcommand\theproposition{D.\arabic{proposition}}

\setcounter{corollary}{0}
\renewcommand\thecorollary{D.\arabic{corollary}}

\setcounter{assumption}{0}
\renewcommand\theassumption{D.\arabic{assumption}}

\setcounter{definition}{0}
\renewcommand\thedefinition{D.\arabic{definition}}

\setcounter{Lemma}{0}
\renewcommand\theLemma{D.\arabic{Lemma}}

In this appendix, we propose a partial identification strategy that does not restrict the covariance between the proxies's measurement error terms $\left(U_{1}, U_{2} \right)$. To do so, we keep Assumptions \ref{AsExogeneous}, \ref{AsScale}, \ref{AsPosEff} and \ref{AsMEexogenous}, completely remove Assumption \ref{AsNonClassical}, and modify Assumption \ref{AsDataRestrictions}. For convenience, we list all the necessary assumptions here.

\begin{assumption}[Exogeneity]\label{AsAppFreeCovExogeneous}
	After conditioning on the covariate vector $W$, the true unobserved measure $X^{*}$ is exogenous, i.e., $Cov\left(X_{res}^{*}, \epsilon_{res}\right) = 0$.
\end{assumption}

\begin{assumption}[Scale Normalization]\label{AsAppFreeCovScale}
	The scale of the true unobserved measure is normalized to have unit variance, i.e., $Var\left(X_{res}^{*}\right) = 1$.
\end{assumption}

\begin{assumption}[Known Direction of the True Effect]\label{AsAppFreeCovPosEff}
	The direction of the effect of the true unobservable variable of interest on the outcome variable is known, i.e., $\beta \geq 0$.
\end{assumption}

\begin{assumption}[Exogeneity of the Measurement Errors]\label{AsAppFreeCovMEexogenous}
	The measurement error variables are exogenous, i.e., $Cov\left(U_{1}, \epsilon_{res}\right) = 0$ and $Cov\left(U_{2}, \epsilon_{res}\right) = 0$
\end{assumption}

\begin{assumption}[Well-behaved Data Distributions]\label{AsAppFreeCovDataRestrictions}
	The variance of the outcome variable is strictly positive and the covariances between each proxy and the outcome variable are positive, i.e., $Var\left(Y_{res}\right) > 0$, $Cov\left(Z_{1,res}, Y_{res}\right) \geq 0$ and $Cov\left(Z_{2,res}, Y_{res}\right) \geq 0.$
\end{assumption}

We now state the main result of this appendix: partial identification of $\beta$ (Equation \eqref{EqModel}) using two proxies (Equation \eqref{EqProxiesCovariates}) with nonclassical measurement error and unrestricted covariance between the measurement error terms.

\begin{proposition}\label{PropAppFreeCovMainId}
	Under Assumptions \ref{AsAppFreeCovExogeneous}-\ref{AsAppFreeCovDataRestrictions}, we have that:
	\begin{equation}\label{EqAppFreeCovBounds}
		\beta \in \left\lbrace \begin{array}{cl}
			\left\lbrace \emptyset \right\rbrace & \text{ if } UB < LB_{D} \\
			\left[LB_{D}, UB\right] & \text{ if } UB \geq LB_{D}
		\end{array} \right.,
	\end{equation}
	where
	\begin{equation}\label{EqAppFreeCovLB}
		LB_{D} \coloneqq \max \left\lbrace \begin{matrix}
			\dfrac{- B + \sqrt{B^{2} - 4 \cdot A \cdot C}}{2 \cdot C}, \\
			\\
			\sqrt{\dfrac{\left\lbrace \begin{matrix}
						Cov(Z_{1,res},Y_{res})^2 \cdot Var(Z_{2,res}) + Cov(Z_{2,res},Y_{res})^2 \cdot Var(Z_{1,res}) \\
						- 2\cdot Cov(Z_{1,res},Y_{res}) \cdot Cov(Z_{2,res},Y_{res}) \cdot Cov(Z_{1,res},Z_{2,res})
					\end{matrix} \right\rbrace}{Var(Z_{1,res}) \cdot Var(Z_{2,res}) - Cov(Z_{1,res},Z_{2,res})^2}}
		\end{matrix}  \right\rbrace
	\end{equation}
	and $UB$, $A$, $B$ and $C$ are defined in Section \ref{SidResults}. Note that $LB_{D} \geq 0$ according to Assumption \ref{AsAppFreeCovDataRestrictions}.

	Moreover, the bounds in Equation \eqref{EqAppFreeCovBounds} exhaust all the information contained in the first two moments of the observable variables $\left(Y_{res}, Z_{1,res},Z_{2,res}\right)$.
\end{proposition}

\begin{proof}
	See Appendix \ref{AppProofAppFreeCovMainId}.
\end{proof}

The only difference between the bounds in Proposition \ref{PropAppFreeCovMainId} and the bounds in Proposition \ref{PropMainId} is the lower bound. Since the term $\dfrac{Cov\left(Z_{1,res}, Y_{res}\right) + Cov\left(Z_{2,res}, Y_{res}\right)}{1 + Cov\left(Z_{1,res}, Z_{2,res}\right)}$ captures the identifying power of Assumption \ref{AsNonClassical}, it is not present in $LB_{D}$. Consequently, we have that that $LB_{D} \leq LB$, i.e., the bounds in Proposition \ref{PropAppFreeCovMainId} are weakly wider than the bounds in Proposition \ref{PropMainId}.

\subsection{Proof of Proposition \ref{PropAppFreeCovMainId}}\label{AppProofAppFreeCovMainId}

The proof Proposition \ref{PropAppFreeCovMainId} is very similar to the proof of Proposition \ref{PropMainId} in Appendix \ref{AppProofMainId}. The only difference is that we must skip the steps related to Assumption \ref{AsNonClassical}. Specifically, we must skip
\begin{enumerate}
	\item Step ``Third Term in the Lower Bound'' in Appendix in Part 3 in Appendix \ref{AppProofLbBetaUB},
	\item Equation \eqref{EqCandCovU1U2},  and
	\item Equation \eqref{EqCovU1U2}.
\end{enumerate}

\pagebreak

\section{Estimation and Inference: Detailed Explanation}\label{AppEstimation}

\setcounter{table}{0}
\renewcommand\thetable{E.\arabic{table}}

\setcounter{figure}{0}
\renewcommand\thefigure{E.\arabic{figure}}

\setcounter{equation}{0}
\renewcommand\theequation{E.\arabic{equation}}

\setcounter{theorem}{0}
\renewcommand\thetheorem{E.\arabic{theorem}}

\setcounter{proposition}{0}
\renewcommand\theproposition{E.\arabic{proposition}}

\setcounter{corollary}{0}
\renewcommand\thecorollary{E.\arabic{corollary}}

\setcounter{assumption}{0}
\renewcommand\theassumption{E.\arabic{assumption}}

\setcounter{definition}{0}
\renewcommand\thedefinition{E.\arabic{definition}}

\setcounter{Lemma}{0}
\renewcommand\theLemma{E.\arabic{Lemma}}

In this appendix, we propose a simple parametric estimator for the bounds described in Proposition \ref{PropMainId} (Appendix \ref{AppEstimator}) and a method to compute $\alpha$-confidence sets that contain the identified set with a probability at least as large as the chosen confidence level $\alpha \in \left(0, 1\right)$ (Appendix \ref{AppInference}).

Throughout this appendix, we assume that we observe an independently distributed sample $\left\lbrace Y_{i}, W_{1,i}, \ldots, W_{J,i}, Z_{1,i}, Z_{2,i} \right\rbrace_{i = 1}^{N}$, where $N \in \mathbb{N}$ is the sample size.

\subsection{Estimation}\label{AppEstimator}

In this section, we propose a simple parametric estimator for the bounds described in Proposition \ref{PropMainId}. Algorithm \ref{AlgEstimator} describes how to implement this estimator in nine steps.

\begin{algorithm}[Parametric Estimator for the Identified Set]\label{AlgEstimator} \phantom{a}

	\begin{enumerate}
		\item Run the OLS regression $$Y_{i} = a_{0}^{Y} + \sum_{j=1}^{J} a_{j}^{Y} \cdot W_{j,i} + \epsilon_{i}^{Y},$$ where $\widehat{a}^{Y} \coloneqq \left( \widehat{a}_{0}^{Y},  \widehat{a}_{1}^{Y}, \cdots, \widehat{a}_{J}^{Y}\right)$ are the OLS estimators, and save the residuals $$\widehat{Y}_{res,i} \coloneqq Y_{i} - \widehat{a}_{0}^{Y} - \sum_{j=1}^{J} \widehat{a}_{j}^{Y} \cdot W_{j,i}.$$

		\item Run the OLS regression $$Z_{1,i} = a_{0}^{1} + \sum_{j=1}^{J} a_{j}^{1} \cdot W_{j,i} + \epsilon_{i}^{1},$$ where $\widehat{a}^{1} \coloneqq \left( \widehat{a}_{0}^{1},  \widehat{a}_{1}^{1}, \cdots, \widehat{a}_{J}^{1}\right)$ are the OLS estimators, and save the residuals $$\widehat{Z}_{1,res,i} \coloneqq Z_{1,i} - \widehat{a}_{0}^{1} - \sum_{j=1}^{J} \widehat{a}_{j}^{1} \cdot W_{j,i}.$$

		\item Run the OLS regression $$Z_{2,i} = a_{0}^{2} + \sum_{j=1}^{J} a_{j}^{2} \cdot W_{j,i} + \epsilon_{i}^{2},$$ where $\widehat{a}^{2} \coloneqq \left( \widehat{a}_{0}^{2},  \widehat{a}_{2}^{2}, \cdots, \widehat{a}_{J}^{2}\right)$ are the OLS estimators, and save the residuals $$\widehat{Z}_{2,res,i} \coloneqq Z_{2,i} - \widehat{a}_{0}^{2} - \sum_{j=1}^{J} \widehat{a}_{j}^{2} \cdot W_{j,i}.$$

		\item Estimate
		\begin{equation}\label{EqLBfirstterm}
			\widehat{l}_{1} \coloneqq \dfrac{Cov\left(\widehat{Z}_{1,res,i}, \widehat{Y}_{res,i}\right) + Cov\left(\widehat{Z}_{2,res,i}, \widehat{Y}_{res,i}\right)}{1 + Cov\left(\widehat{Z}_{1,res,i}, \widehat{Z}_{2,res,i}\right)}.
		\end{equation}

		\item Estimate
		\begin{equation}\label{EqLBfourthterm}
			\widehat{l}_{2} \coloneqq \dfrac{- \widehat{B} + \sqrt{\widehat{B}^{2} - 4 \cdot \widehat{A} \cdot \widehat{C}}}{2 \cdot \widehat{C}},
		\end{equation}
		where
		\begin{align*}
			\widehat{A} & \coloneqq -\left[Cov\left(\widehat{Z}_{1,res}, \widehat{Y}_{res}\right) - Cov\left(\widehat{Z}_{2,res}, \widehat{Y}_{res}\right)\right]^{2}, \\
			\widehat{B} & \coloneqq -2 \cdot \left\lbrace\begin{matrix}
				Cov\left(\widehat{Z}_{1,res}, \widehat{Y}_{res}\right) \cdot \left[Var\left(\widehat{Z}_{2,res}\right) - Cov\left(\widehat{Z}_{1,res}, \widehat{Z}_{2,res}\right)\right] \\
				\hspace{10pt} + Cov\left(\widehat{Z}_{2,res}, \widehat{Y}_{res}\right) \cdot \left[Var\left(\widehat{Z}_{1,res}\right) - Cov\left(\widehat{Z}_{1,res}, \widehat{Z}_{2,res}\right)\right]
			\end{matrix}  \right\rbrace, \text{ and} \\
			\widehat{C} & \coloneqq \left[Var\left(\widehat{Z}_{1,res}\right) + 1\right] \cdot \left[Var\left(\widehat{Z}_{2,res}\right) + 1\right] - \left[Cov\left(\widehat{Z}_{1,res}, \widehat{Z}_{2,res}\right) + 1\right]^{2}.
		\end{align*}

		\item Estimate
		\begin{equation}\label{EqLBfifthterm}
			\widehat{l}_{3} \coloneqq \sqrt{\dfrac{\left\lbrace \begin{matrix}
						Cov(\widehat{Z}_{1,res},\widehat{Y}_{res})^2 \cdot Var(\widehat{Z}_{2,res}) + Cov(\widehat{Z}_{2,res},\widehat{Y}_{res})^2 \cdot Var(\widehat{Z}_{1,res}) \\
						- 2\cdot Cov(\widehat{Z}_{1,res},\widehat{Y}_{res}) \cdot Cov(\widehat{Z}_{2,res},\widehat{Y}_{res}) \cdot Cov(\widehat{Z}_{1,res},\widehat{Z}_{2,res})
					\end{matrix} \right\rbrace}{Var(\widehat{Z}_{1,res}) \cdot Var(\widehat{Z}_{2,res}) - Cov(\widehat{Z}_{1,res},\widehat{Z}_{2,res})^2}}.
		\end{equation}

		\item Estimate
		\begin{equation}\label{EqLBestimator}
			\widehat{LB} \coloneqq \max\left\lbrace \widehat{l}_{1}, \widehat{l}_{2}, \widehat{l}_{3} \right\rbrace.
		\end{equation}

		\item Estimate
		\begin{equation}\label{EqUBestimator}
			\widehat{UB} \coloneqq \sqrt{Var\left(\widehat{Y}_{res}\right)}.
		\end{equation}

		\item If $\widehat{UB} < \widehat{LB}$, then the estimator for the identified set in Proposition \ref{PropMainId} is given by $\emptyset$. If $\widehat{UB} \geq \widehat{LB}$, then the estimator for the identified set in Proposition \ref{PropMainId} is given by $\left[\widehat{LB}, \widehat{UB}\right]$.
	\end{enumerate}
\end{algorithm}

\subsection{Inference}\label{AppInference}

In this section,  we address the sample uncertainty behind the estimation of our bounds. In summary, we propose one method to construct an $\alpha$-confidence interval around the identified set, $\left[LB, UB\right]$.  Specifically, this interval contains the true identified set with known probability, where the probability is computed based on the thought exercise of an infinitely large number of samples.

To construct our $\alpha$-confidence interval around the identified set, we combine the confidence bounds proposed by \cite{Chernozhukov2013} with a Bonferroni-style correction. Since the upper bound does not rely on intersection bounds, we use a simple bootstrap procedure to compute a $\frac{\left(1 + \alpha\right)}{2}$-confidence bound above it. Since the lower bound relies on intersection bounds, we use the method proposed by \cite{Chernozhukov2013} to compute a $\frac{\left(1 + \alpha\right)}{2}$-confidence bound below it. By connecting the lower confidence bound below our estimated lower bound to the upper confidence bound above our estimated upper bound, we construct an $\alpha$-confidence interval around the identified set. Similar procedures were used by \cite{Flores2013} and \cite{Possebom2024}.

Algorithm \ref{AlgInference} describes how to construct our confidence intervals in detail.

\begin{algorithm}[$\alpha$-Confidence Interval  around the Identified Set]\label{AlgInference} \phantom{a}

	\begin{enumerate}
		\item Upper $\frac{\left(1 + \alpha\right)}{2}$-Confidence Bound
		\begin{enumerate}
			\item Using the empirical bootstrap, collect $B \in \mathbb{N}$ bootstrapped estimates of the upper bound, $\left\lbrace \widehat{UB}_{b}\right\rbrace_{b = 1}^{B}$.

			\item Obtain the $\frac{\left(1 + \alpha\right)}{2}$-quantile of $\left\lbrace \left\vert \widehat{UB}_{b} - \widehat{UB} \right\vert \right\rbrace_{b = 1}^{B}$ and denote it by $c_{UB}\left(\frac{1 + \alpha}{2}\right)$.

			\item Construct the Upper $\frac{\left(1 + \alpha\right)}{2}$-Confidence Bound as $\widehat{C}_{UB}\left(\frac{1 + \alpha}{2}\right) \coloneqq \widehat{UB} + c_{UB}\left(\frac{1 + \alpha}{2}\right)$.
		\end{enumerate}

		\item Lower $\frac{\left(1 + \alpha\right)}{2}$-Confidence Bound
		\begin{enumerate}
			\item Define the vector of estimated bounding functions,  $\widehat{l} \coloneqq \left(\widehat{l}_{1}, \widehat{l}_{2}, \widehat{l}_{3}\right)^{\prime}$.

			\item Using the empirical bootstrap,  collect $B \in \mathbb{N}$ bootstrapped estimates of the bounding functions, $\left\lbrace \widehat{l}_{b}^{*} \right\rbrace_{b = 1}^{B}$.

			\item Estimate the variance-covariance matrix of $\sqrt{N} \cdot \left(\widehat{l} - l \right)$ using the variance-covariance matrix of $\left\lbrace \sqrt{N} \cdot\widehat{l}_{b}^{*} \right\rbrace_{b = 1}^{B}$. Denote the estimated variance-covariance matrix by $\widehat{\Omega}$.

			\item Let $\widehat{g}_{k}^{\prime}$ denote the $k$-th row of $\widehat{\Omega}^{\sfrac{1}{2}}$ for $k \in \left\lbrace 1, 2, 3 \right\rbrace$.

			\item Estimate $\widehat{s}_{k} \coloneqq \frac{\left\vert \left\vert \widehat{g}_{k}  \right\vert \right\vert}{\sqrt{N}}$ where $\left\vert \left\vert \cdot \right\vert \right\vert$ is the Euclidean norm.

			\item Simulate $R \in \mathbb{N}$ draws from the multivariate standard normal of dimension 3 and denote them by $Z_{1}, \ldots, Z_{R}$ as row vectors.

			\item Calculate $Z_{r}^{*}\left(k\right) \coloneqq \frac{\widehat{g}_{k}^{\prime} \cdot Z_{r}}{\left\vert \left\vert \widehat{g}_{k}  \right\vert \right\vert}$ for $r \in \left\lbrace 1, \ldots, R \right\rbrace$.

			\item Let $Q_{p}\left(A\right)$ denote the $p$-th quantile of a random variable A and, following \cite{Chernozhukov2013}, define $c_{N} \coloneqq 1 - \frac{0.1}{\ln{N}}$.

			\item Compute $\kappa \coloneqq Q_{c_{N}}\left(\max\left\lbrace Z_{r}^{*}\left(1\right), Z_{r}^{*}\left(2\right), Z_{r}^{*}\left(3\right) \right\rbrace; r = 1, \ldots, R\right)$. In plain English, we calculate the maximum of $Z_{r}^{*}\left(1\right), Z_{r}^{*}\left(2\right)$, and $Z_{r}^{*}\left(3\right)$ for each replication $r$, and take the $c_{N}$-th quantile of those $R$ values.

			\item Compute $\widehat{K} \coloneqq \left\lbrace k \in \left\lbrace 1, 2, 3 \right\rbrace \colon \widehat{l}_{k} \geq \left(\max\limits_{\tilde{k} \in \left\lbrace 1, 2, 3 \right\rbrace} \left\lbrace \widehat{l}_{\tilde{k}} - \kappa \cdot \widehat{s}_{\tilde{k}} \right\rbrace - 2 \cdot \kappa \cdot \widehat{s}_{k} \right)\right\rbrace$.

			\item Compute $\widehat{\kappa} \coloneqq Q_{\frac{\left(1 + \alpha\right)}{2}}\left(\max\limits_{k \in \widehat{K}}\left\lbrace Z_{r}^{*}\left(k\right) \right\rbrace; r = 1, \ldots, R\right)$.

			\item Construct the Lower $\frac{\left(1 + \alpha\right)}{2}$-Confidence Bound as $$\widehat{C}_{LB}\left(\frac{1 + \alpha}{2}\right) \coloneqq \max\limits_{k \in \left\lbrace 1, 2, 3 \right\rbrace} \left\lbrace \widehat{l}_{k} - \widehat{\kappa} \cdot \widehat{s}_{k} \right\rbrace.$$
		\end{enumerate}

		\item $\alpha$-Confidence Interval
		\begin{enumerate}
			\item Construct a $\alpha$-Confidence Interval using $\left[\widehat{C}_{LB}, \widehat{C}_{UB}\right]$. This interval contains the true identified set $\left[LB, UB\right]$ with probability $\alpha \in \left(0.5, 1\right)$.
		\end{enumerate}
	\end{enumerate}
\end{algorithm}

\pagebreak

\section{Sources of Measurement Error in FAO-GAEZ data}\label{AppSourceError}

\setcounter{table}{0}
\renewcommand\thetable{F.\arabic{table}}

\setcounter{figure}{0}
\renewcommand\thefigure{F.\arabic{figure}}

\setcounter{equation}{0}
\renewcommand\theequation{F.\arabic{equation}}

\setcounter{theorem}{0}
\renewcommand\thetheorem{F.\arabic{theorem}}

\setcounter{proposition}{0}
\renewcommand\theproposition{F.\arabic{proposition}}

\setcounter{corollary}{0}
\renewcommand\thecorollary{F.\arabic{corollary}}

\setcounter{assumption}{0}
\renewcommand\theassumption{F.\arabic{assumption}}

\setcounter{definition}{0}
\renewcommand\thedefinition{F.\arabic{definition}}

\setcounter{Lemma}{0}
\renewcommand\theLemma{F.\arabic{Lemma}}

In this appendix, we discuss potential sources of measurement error in the FAO-GAEZ data. Our main goal is arguing the plausibility of Assumption \ref{AsNonClassical}.

Consider the following model describing how FAO-GAEZ version $v$ proxies for agricultural productivity of crop $c$
\begin{equation}
	Z_{v} = f_v(M+\eta_v,\theta+\zeta_{v})
\end{equation}
where $f_{v}$ is a known function, $M$ denotes a vector of predictors (e.g., true precipitation and true temperature), and $\theta$ is a vector of parameters. Importantly, $M$ and $\theta$ are permanent objects in the sense that they do not change across versions. Since these permanent objects are not perfectly observed, we include two error terms that vary across versions: $\zeta_v$ captures calibration error and $\eta_v$ captures measurement error in the predictors.

Thus, the measurement error in version $v$ can be decomposed as:
\begin{equation}
	\begin{split}
		U_v &= f_v(M+\eta_v,\theta+\zeta_v) -X^* \\
		&= \underbrace{f_v(M+\eta_v,\theta+\zeta_v) - f_v(M,\theta)}_{\text{imperfect data term}} + \underbrace{f_v(M,\theta) - X^*}_{\text{functional form misspecification}}
	\end{split}
\end{equation}
Intuitively, the first term $\left(f_v(M+\eta_v,\theta+\zeta_v) - f_v(M,\theta)\right)$ captures measurement error arising from imprecise input data. This term will be positively correlated across versions if the input data presents worse quality in developing regions over time. For example, developing regions may have a smaller number of weather stations and may rely more heavily on interpolation methods to estimate the predictors $M$, implying that $\eta_{v}$ is correlated across versions of the FAO-GAEZ project.

The second term $\left(f_v(M,\theta) - X^*\right)$ captures measurement error arising from functional form misspecification. This type of error occurs because researchers do not know the true model $f^{*}$ connecting predictors and parameters to crop productivity $X^{*} = f^{*}\left(M,\theta\right)$. This term will be positively correlated across versions if the proxy model $f_{v}$ is not updated heavily across versions of the FAO-GAEZ project. In particular, this term will be perfectly correlated across versions if $f_{v}$ is the same function for different values of $v$.

Lastly, this decomposition illustrates why Assumption \ref{AsNonClassical} is plausible. A sufficient condition for $Cov(U_1,U_2) \geq 0$ is (i) persistence on both the imperfect data term and the misspecification term across versions and (ii) a non-negative covariance between these two terms across versions. According to the last two paragraphs, condition (i) is plausible in our setting. Furthermore, condition (ii) may arise if predictor measurement and functional-form assumptions are less accurate in developing regions. The latter is plausible in our context because scientific investment in developing regions is smaller.

\pagebreak

\section{Applications Using Other Datasets}\label{AppOther}

\setcounter{table}{0}
\renewcommand\thetable{G.\arabic{table}}

\setcounter{figure}{0}
\renewcommand\thefigure{G.\arabic{figure}}

\setcounter{equation}{0}
\renewcommand\theequation{G.\arabic{equation}}

\setcounter{theorem}{0}
\renewcommand\thetheorem{G.\arabic{theorem}}

\setcounter{proposition}{0}
\renewcommand\theproposition{G.\arabic{proposition}}

\setcounter{corollary}{0}
\renewcommand\thecorollary{G.\arabic{corollary}}

\setcounter{assumption}{0}
\renewcommand\theassumption{G.\arabic{assumption}}

\setcounter{definition}{0}
\renewcommand\thedefinition{G.\arabic{definition}}

\setcounter{Lemma}{0}
\renewcommand\theLemma{G.\arabic{Lemma}}

In this appendix, we discuss other empirical contexts where our partial identification strategy may be useful. To illustrate the broad applicability of our methodology, we will briefly analyze seven empirical examples where the variable of interest is measured with error, and there might exist two possible proxies.

First, many recent studies \citep{Olken2009,Adena2015} focus on the impact of media access on economic behavior. To do so, they frequently measure media access by combining an irregular terrain model and the location of TV or radio transmitters. Since the most commonly used irregular terrain model is in its third version \citep{Hufford1995}, this setting is very similar to the FAO-GAEZ productivity measures discussed in Sections \ref{Scontext} and \ref{Sresults}. For example, we can use our proposed tool to analyze a reduced-form version of Panel B in Table 11 by \cite{Olken2009} and reanalyze Equation (1) by \cite{Adena2015}.

Second, multiple articles \citep{Montalvo2005,Desmet2017,Desmet2025} analyze the impact of social heterogeneity on socioeconomic outcomes, including the likelihood of civil conflict and economic growth. Importantly, measures of social heterogeneity may vary on many dimensions: (i) fractionalization or polarization, (ii) levels of aggregation, and (iii) the weight assigned to the distance between different groups. These differences matter because different measures of ethnolinguistic fractionalization suggest different effects of social heterogeneity on economic growth \citep[Table 6]{Desmet2025}. In this setting, we can use our proposed tool by treating each measure of ethnolinguistic fractionalization as a proxy for the true latent ethnolinguistic fractionalization.

Third, many articles \citep{Post2013,Gan2015,Bissonnette2017} discuss the impact of subjective beliefs about life expectancy on economic behavior (e.g., savings, investment, bequests). To measure this type of belief, many surveys ask individuals about their subjective probability of attaining various ages. However, individual responses frequently present measurement issues, such as rounding and focal answers. To address these problems, many authors adopt different imputation methods that approximate the true subjective probability of attaining a certain age. By treating the results of each imputation model as a different proxy, our proposed method may be used in this empirical setting. For example, we can reanalyze Equation (4-3) by \cite{Post2013}.

Fourth, multiple studies \citep{Hsieh2009,Hsieh2010} are interested in the impact of total factor productivity on economic outcomes, such as plant exit and income levels. To measure total factor productivity, different authors adopt different models \citep{Olley1996,Hsieh2009,Ackerberg2015}. By treating the measures produced by each model as a different proxy for the true latent total factor productivity, our partial identification strategy may be used in this empirical context. For example, we can reanalyze Table VIII by \cite{Hsieh2009}.

Fifth, economic historians \citep{Sala-i-Martin1996,Hartley2021} frequently focus on understanding GDP growth and fluctuations in the long run. To do so, they need measures of income levels over long periods of time, including moments when national accounts were not collected in a systematic way. To access this type of variable, most authors use the Maddison Project Data \citep{Bolt2025}, which is in its fifth version. Each version varies either because new countries were included in the sample or, more importantly to us, new information was collected about existing countries and their historical income data were updated. Since different versions of these variables may have different measurement error terms when compared to the true income level of each country, this setting is very similar to the FAO-GAEZ productivity measures discussed in Sections \ref{Scontext} and \ref{Sresults}. For example, we can use our proposed tool to reanalyze Equations (1), (3) and (5) by \cite{Sala-i-Martin1996} and Equations (1) and (2) by \cite{Hartley2021}.

Sixth, \citet[Tables 1 and 2]{Rodrik2005} analyze the impact of democratic transitions on economic outcomes. To measure their treatment variable, they use the Polity IV index. This expert-coded measure of democracy levels is in its fourth version and is based on experts' interpretation of political regimes in each country and each year. Since different experts might interpret the same political regime differently, there is measurement error in each version of this index. Although this setting is similar to the FAO-GAEZ productivity measures (Sections \ref{Scontext} and \ref{Sresults}), we believe that Assumption \ref{AsNonClassical} may not be appropriate in this context. Since political experts might overreact to reanalyses of historical phenomena, it may be the case that the measurement error terms in different versions of the Polity Index are negatively correlated. In this case, the bounds proposed in Appendix \ref{AppFreeCov} represent a more appropriate approach than the bounds in the main text.

Seventh, multiple articles \citep{Alfaro2008,Rajkumar2008,Aksoy2025} are interested in the impact of a country's governance quality on economic outcomes, such as health quality, education quality, political trust and international capital flows. To measure governance quality, \cite{Alfaro2008} and \cite{Rajkumar2008} use a variety of measures from the Political Risk Services Group, while \cite{Aksoy2025} uses the V-DEM index and one measure from the Worldwide Governance Indicators. By treating these different measures as different proxies for the true unobservable governance quality, our partial identification strategy may be applied in this empirical context. For example, we can reanalyze Tables 5, 9 and 10 (Panel C) by \cite{Alfaro2008}, Equations (1) and (2) by \cite{Rajkumar2008}, and Equation (2) by \cite{Aksoy2025}.

\newpage

\section{Simple Examples} \label{AppExample}

\setcounter{table}{0}
\renewcommand\thetable{H.\arabic{table}}

\setcounter{figure}{0}
\renewcommand\thefigure{H.\arabic{figure}}

\setcounter{equation}{0}
\renewcommand\theequation{H.\arabic{equation}}

\setcounter{theorem}{0}
\renewcommand\thetheorem{H.\arabic{theorem}}

\setcounter{proposition}{0}
\renewcommand\theproposition{H.\arabic{proposition}}

\setcounter{corollary}{0}
\renewcommand\thecorollary{H.\arabic{corollary}}

\setcounter{assumption}{0}
\renewcommand\theassumption{H.\arabic{assumption}}

\setcounter{definition}{0}
\renewcommand\thedefinition{H.\arabic{definition}}

\setcounter{Lemma}{0}
\renewcommand\theLemma{H.\arabic{Lemma}}

In this appendix, we propose three simple data-generating processes where alternative methodologies perform poorly when compared to our partial identification strategy. Our first example (Appendix \ref{AppExampleLess}) shows that the reduced-form coefficient (Equation \eqref{EqReducedForm}) may be strictly less than our lower bound, while our second example (Appendix \ref{AppExampleGreater}) shows that the reduced-form coefficient may be strictly greater than our upper bound.\footnote{These two examples illustrate that using the reduced-form coefficients as either a lower bound or an upper bound of the true coefficient is not appropriate.} Lastly, our third example (Appendix \ref{AppExampleLubotsky}) shows that the estimand proposed by \cite{Lubotsky2006} may present a larger bias than our lower bound.

\subsection{Example 1: $b_{1} < LB$} \label{AppExampleLess}

In this section, we create a class of data-generating processes such that $$b_{1} < LB,$$
where $b_{1}$ is defined in Equation \eqref{EqReducedForm}.

To do so, assume that $\beta > 0$, $c_{1} > 0$ and
\begin{equation}\label{EqMesErrExample}
	U_{1} = a_{1} + c_{1} \cdot X_{res}^{*},
\end{equation}
implying that
\begin{equation}\label{EqMesErrExampleZ}
	Z_{1,res} = a_{1} + \left( 1 + c_{1} \right) \cdot X_{res}^{*}
\end{equation}
according to Equation \eqref{EqProxiesCovariates}. In this case, we have that
\begin{align}
	b_{1} & = \dfrac{Cov\left(Z_{1,res},Y_{res}\right)}{Var\left(Z_{1,res}\right)} \nonumber \\
	& \hspace{20pt} \text{by definition} \nonumber \\
	& = \dfrac{Cov\left(a_{1} + \left( 1 + c_{1} \right) \cdot X_{res}^{*}, \hspace{10pt} \beta \cdot X_{res}^{*} + \epsilon_{res} \right)}{Var\left(a_{1} + \left( 1 + c_{1} \right) \cdot X_{res}^{*}\right)} \nonumber \\
	& \hspace{20pt} \text{according to Equations \eqref{EqModel} and \eqref{EqMesErrExampleZ}} \nonumber \\
	& = \dfrac{\beta \cdot \left( 1 + c_{1} \right)}{\left( 1 + c_{1} \right)^{2}} \label{EqRedFormCoeff} \\
	& \hspace{20pt} \text{according to Assumptions \ref{AsExogeneous} and \ref{AsScale}}. \nonumber
\end{align}

Moreover, our lower bound satisfies:
\begin{align}
	LB & \geq \dfrac{Cov\left(Z_{1,res}, Y_{res}\right)}{\sqrt{Var\left(Z_{1,res}\right)}} \nonumber \\
	& \hspace{20pt} \text{according to Equations \eqref{EqLB} and \eqref{EqBetaZ1}} \nonumber \\
	& \geq 2 \cdot \dfrac{Cov\left(Z_{1,res}, Y_{res}\right)}{1 + Var\left(Z_{1,res}\right)} \nonumber \\
	& \hspace{20pt} \text{according to Equation \eqref{EqNB2}} \nonumber \\
	& = 2 \cdot \dfrac{\beta \cdot \left( 1 + c_{1} \right)}{1 + \left( 1 + c_{1} \right)^{2}} \label{EqExampleLB} \\
	& \hspace{20pt} \text{according to the same steps used to derived Equation \eqref{EqRedFormCoeff}.} \nonumber
\end{align}

Now, note that
\begin{equation*}
	\begin{array}{rrcl}
		& c_{1} & > & 0 \hspace{10pt} \text{by assumption} \\
		\Leftrightarrow & c_{1} \cdot \left(c_{1} + 2 \right) & > & 0  \hspace{10pt} \text{because } c_{1} > 0 \\
		\Leftrightarrow & c_{1}^{2} + 2 \cdot c_{1} & > & 0 \\
		\Leftrightarrow & c_{1}^{2} + 2 \cdot c_{1} + c_{1}^{2} + 2 \cdot c_{1} + 2 & > & c_{1}^{2} + 2 \cdot c_{1} + 2 \hspace{10pt} \text{because } c_{1} > 0 \\
		\Leftrightarrow & 2 \cdot \left(1 + 2 \cdot c_{1} + c_{1}^{2}\right) & > & 1 + \left(1 + 2 \cdot c_{1} + c_{1}^{2}\right) \\
		\Leftrightarrow & 2 \cdot \left(1 + c_{1}\right)^{2} & > & 1 + \left(1 + c_{1}\right)^{2} \\
		\Leftrightarrow & 2 \cdot \dfrac{\beta \cdot \left(1 + c_{1}\right)}{1 + \left(1 + c_{1}\right)^{2}} & > & \dfrac{\beta \cdot \left(1 + c_{1}\right)}{\left(1 + c_{1}\right)^{2}} \hspace{10pt} \text{because } \beta > 0 \text{ and } c_{1} > 0 \\
		\Rightarrow & \beta \geq LB & > & b_{1} \hspace{10pt} \text{ by Equations \eqref{EqRedFormCoeff} and \eqref{EqExampleLB}}.
	\end{array}
\end{equation*}

Intuitively, the last result implies that, if the covariance between the true unobservable variable $X_{res}^{*}$ and its proxy $Z_{1,res}$ is sufficiently large, then the reduced-form coefficient will be more biased than our lower bound.

Therefore, we can conclude that there exist simple data-generating processes where our partial identification strategy's performance compares positively against the reduce-form approach.

\subsection{Example 2: $b_{1} > UB$} \label{AppExampleGreater}

In this section, we create a class of data-generating processes such that $$b_{1} > UB,$$ where $b_{1}$ is defined in Equation \eqref{EqReducedForm}.

To do so, assume that $\beta = UB \coloneqq \sqrt{Var\left(Y\right)}$, $\epsilon_{res} = 0$ and $$\left[\begin{matrix}
X_{res}^{*} \\
U_{1}
\end{matrix}\right] \sim N
\left(\left[\begin{matrix}
0 \\
0
\end{matrix}\right], \left[\begin{matrix}
1 & Cov\left(X_{res}^{*}, U_{1}\right) \\
Cov\left(X_{res}^{*}, U_{1}\right) & Var\left(U_{1}\right)
\end{matrix}\right] \right),$$
where $Var\left(U_{1}\right) \in \left(0, 1 \right)$ and $Cov\left(X_{res}^{*}, U_{1}\right) \in \left(-\dfrac{1+Var\left(U_{1}\right)}{2}, - Var\left(U_{1}\right)\right)$.

In this case, we have that
\begin{align*}
	b_{1} & = \dfrac{Cov\left(Z_{1,res},Y_{res}\right)}{Var\left(Z_{1,res}\right)} \\
	& \hspace{20pt} \text{by definition} \\
	& = \dfrac{Cov\left(X_{res}^{*} + U_{1},\beta \cdot X_{res}^{*} + \epsilon_{res}\right)}{Var\left(X_{res}^{*} + U_{1}\right)} \\
	& \hspace{20pt} \text{by Equations \eqref{EqModel} and \eqref{EqProxiesCovariates}} \\
	& = \dfrac{Cov\left(X_{res}^{*} + U_{1},UB \cdot X_{res}^{*}\right)}{Var\left(X_{res}^{*} + U_{1}\right)} \\
	& \hspace{20pt} \text{by assumption} \\
	& = \dfrac{UB + UB \cdot Cov\left(X_{res}^{*}, U_{1}\right)}{1 + Var\left(U_{1}\right) + 2 \cdot Cov\left(X_{res}^{*}, U_{1}\right)} \\
	& \hspace{20pt} \text{by Assumption \ref{AsScale}} \\
	& = UB \cdot \left[\dfrac{1 + Cov\left(X_{res}^{*}, U_{1}\right)}{1 + Var\left(U_{1}\right) + 2 \cdot Cov\left(X_{res}^{*}, U_{1}\right)}\right] \\
	& \hspace{20pt} \text{where } \dfrac{1 + Cov\left(X_{res}^{*}, U_{1}\right)}{1 + Var\left(U_{1}\right) + 2 \cdot Cov\left(X_{res}^{*}, U_{1}\right)} > 0 \\
	& \hspace{20pt} \text{because }  Cov\left(X_{res}^{*}, U_{1}\right) > -\dfrac{1+Var\left(U_{1}\right)}{2} > -1 \\
	& > UB \cdot \left[\dfrac{1 + Cov\left(X_{res}^{*}, U_{1}\right)}{1 - Cov\left(X_{res}^{*}, U_{1}\right) + 2 \cdot Cov\left(X_{res}^{*}, U_{1}\right)}\right] \\
	& \hspace{20pt} \text{because }  Cov\left(X_{res}^{*}, U_{1}\right) < -Var\left(U_{1}\right)  \\
	& = UB = \beta.
\end{align*}

Intuitively, the last result implies that, if the covariance between the true unobservable variable $X_{res}^{*}$ and the measurement error $U_{1}$ is neither too large nor too small in comparison to the variance of the measurement error, then the reduce-form coefficient will be more biased than our upper bound.

Therefore, we can conclude that there exist simple data-generating processes where our partial identification strategy's performance compares positively against the reduce-form approach.

\subsection{Example 3: Comparing our method against the one proposed by \cite{Lubotsky2006}} \label{AppExampleLubotsky}

In this section, we compare our method to the closest alternative tool in the literature: the bias minimizing linear regression estimand proposed by \cite{Lubotsky2006}. We start by explaining the method developed by these authors and, then, describe a data-generating process where our lower bound is closer to the target parameter than the estimand proposed by \cite{Lubotsky2006}.

Similarly to us, \cite{Lubotsky2006} observe two proxies for an unobservable variable of interest, $X^{*}$, and are interested in the coefficient of the linear regression model described by Equation \eqref{EqModel}. Differently from us, they do not restrict the covariance between the error terms, i.e., they do not impose Assumption \ref{AsNonClassical}. For this reason, we compare their procedure against the method described in Appendix \ref{AppFreeCov}, which does not impose this assumption either. In the current appendix, we impose Assumptions \ref{AsExogeneous}, \ref{AsScale}, \ref{AsPosEff}, \ref{AsMEexogenous} and a weaker version of Assumption \ref{AsDataRestrictions}. Using these assumptions, Proposition \ref{PropAppFreeCovMainId} and Equations \eqref{EqNB2} and \eqref{EqNB1} imply that
\begin{equation}\label{EqLBlubotsky}
	\beta \geq LB_{D} \geq \dfrac{Cov\left(Z_{1,res}, Y_{res}\right)}{\sqrt{Var\left(Z_{1,res}\right)}} \geq 2 \cdot \dfrac{Cov\left(Z_{1,res}, Y_{res}\right)}{1 + Var\left(Z_{1,res}\right)}.
\end{equation}

\cite{Lubotsky2006} impose Assumptions \ref{AsExogeneous}, \ref{AsMEexogenous} and an additional restriction on the covariance between the error terms and the true unobservable variable of interest. Specifically, they impose the following condition.

\begin{assumption}[Individually Classical Measurement Errors]\label{AsAppLubotsky}
	Each measurement error term is individually classic, i.e., $Cov\left(X^{*}_{res},U_{k}\right) = 0$ for any $k \in \left\lbrace 1, 2 \right\rbrace$.
\end{assumption}

Note that this assumption does not impose that the measurement error terms are jointly classical since $Cov\left(U_{1}, U_{2}\right)$ might be different from zero. As a consequence, using $Z_{2,res}$ as an instrument for $Z_{1,res}$ does not identify the target parameter.

Moreover, observe that \cite{Lubotsky2006} do not restrict neither the variance of the true unobservable variable nor the sign of the true coefficient. However, to simplify the exposition, we will impose Assumptions \ref{AsScale} and \ref{AsPosEff} in this appendix.

To overcome the identification challenges posed by measurement error, \cite{Lubotsky2006} propose to use convex combinations of the two proxies in a linear regression model. Specifically, they define $$Z\left(\delta\right) \coloneqq \delta \cdot Z_{1,res} + \left(1 - \delta\right) \cdot Z_{2,res},$$ where $\delta \in \left[0,1\right]$, and analyze $$Y_{res} = b\left(\delta\right) \cdot X^{*}_{res} + V_{\delta},$$ where $b\left(\delta\right) \coloneqq \frac{Cov\left(Z\left(\delta\right), Y_{res}\right)}{Var\left(Z\left(\delta\right)\right)}$ and $V_{\delta}$ is the residual. Due to individually classic measurement error terms and Assumption \ref{AsPosEff}, we have that $$\beta \geq b\left(\delta\right)$$ for any $\delta \in \left[0,1\right]$.

To minimize bias, \cite{Lubotsky2006} propose to focus on the convex combination of proxy variables that generates the OLS coefficient that is as close as possible to the target parameter. In other words, they want to identify $$b^{*} \coloneqq \max\limits_{\delta\in \left[0,1\right]} b\left(\delta\right).$$ Interestingly, they show that one can identify $b^{*}$ according to $$b^{*} = b_{1} + b_{2},$$ where $b_{1}$ and $b_{2}$ are defined as the OLS coefficients of the following linear regression model: $$Y_{res} = b_{1} \cdot Z_{1,res} + b_{2} \cdot Z_{2,res} + V.$$

Under Assumptions  \ref{AsExogeneous}, \ref{AsScale}, \ref{AsMEexogenous} and \ref{AsAppLubotsky}, \citet[Equation (4)]{Lubotsky2006} shows that
\begin{equation}\label{EqBstar}
	b^{*} = \beta \cdot \left[\frac{Var\left(U_{1}\right) + Var\left(U_{2}\right) - 2 \cdot Cov\left(U_{1},U_{2}\right)}{Var\left(U_{1}\right) + Var\left(U_{2}\right) - 3 \cdot Cov\left(U_{1},U_{2}\right) + Var\left(U_{1}\right) \cdot Var\left(U_{2}\right)}\right]
\end{equation}
Moreover, under Assumptions  \ref{AsExogeneous}, \ref{AsScale}, \ref{AsPosEff}, \ref{AsMEexogenous}, \ref{AsDataRestrictions} and \ref{AsAppLubotsky}, our lower bound satisfies
\begin{align}
	LB_{D} & \geq 2 \cdot \dfrac{Cov\left(Z_{1,res}, Y_{res}\right)}{1 + Var\left(Z_{1,res}\right)} \nonumber \\
	& \hspace{20pt} \text{according to Equation \eqref{EqLBlubotsky}} \nonumber \\
	& \label{EqLBcClassic} = \beta \cdot \left[\frac{2}{2 + Var\left(U_{1}\right)}\right]
\end{align}
according to Lemma \ref{LemmaDataRestrictions}.

Now, we show that there exist data-generating processes that satisfy Assumptions  \ref{AsExogeneous}, \ref{AsScale}, \ref{AsPosEff}, \ref{AsMEexogenous}, \ref{AsDataRestrictions} and \ref{AsAppLubotsky} such that $\beta \geq LB_{D} > b^{*}$.

Assume that $Var\left(U_{1}\right) = 1$, $Var\left(U_{2}\right) > 1$ and $Var\left(U_{2}\right) \geq \frac{3 \cdot Cov\left(U_{1}, U_{2}\right) - 1}{2}$. Notice that Equation \eqref{EqLBcClassic} implies that
\begin{equation}\label{EqLBcClassic2}
	LB_{D} \geq \frac{2}{3} \cdot \beta
\end{equation}
and Equation \eqref{EqBstar} implies that
\begin{equation}\label{EqBstar2}
	b^{*} = \beta \cdot \left[\frac{1 + Var\left(U_{2}\right) - 2 \cdot Cov\left(U_{1},U_{2}\right)}{1 + 2\cdot Var\left(U_{2}\right) - 3 \cdot Cov\left(U_{1},U_{2}\right)}\right].
\end{equation}
Observe also that
\begin{equation*}
	\begin{array}{rrcl}
		& Var\left(U_{2}\right) & > & 1 \\
		\Leftrightarrow & 2 + 4 \cdot Var\left(U_{2}\right) - 6 \cdot Cov\left(U_{1},U_{2}\right) & > & 3 + 3 \cdot Var\left(U_{2}\right) - 6 \cdot Cov\left(U_{1},U_{2}\right) \\
		\Leftrightarrow & \frac{2}{3} \cdot \left[ 1 + 2 \cdot Var\left(U_{2}\right) - 3 \cdot Cov\left(U_{1},U_{2}\right) \right] & > & 1 + 1 \cdot Var\left(U_{2}\right) - 2 \cdot Cov\left(U_{1},U_{2}\right) \\
		\Leftrightarrow & \frac{2}{3} & > & \frac{1 + 1 \cdot Var\left(U_{2}\right) - 2 \cdot Cov\left(U_{1},U_{2}\right)}{1 + 2 \cdot Var\left(U_{2}\right) - 3 \cdot Cov\left(U_{1},U_{2}\right)}.
	\end{array}
\end{equation*}
because $Var\left(U_{2}\right) \geq \frac{3 \cdot Cov\left(U_{1}, U_{2}\right) - 1}{2}$.

Combining the last equation with Equations \eqref{EqLBcClassic2}-\eqref{EqBstar2} and Assumption \ref{AsPosEff}, we find that $\beta \geq LB_{D} > b^{*}$. In plain English, in this data-generating process, our lower bound presents a smaller bias than the estimand proposed by \cite{Lubotsky2006}.

Intuitively, we achieve a smaller bias than $b^{*}$ because our lower bound does not restrict itself to the class of OLS estimands based on convex combinations of the proxy variables.

Therefore, we can conclude that there exist simple data-generating processes that satisfy Assumptions  \ref{AsExogeneous}, \ref{AsScale}, \ref{AsPosEff}, \ref{AsMEexogenous}, \ref{AsDataRestrictions} and \ref{AsAppLubotsky} and create a scenario where our partial identification strategy's performance compares positively against the procedure recommended by \cite{Lubotsky2006}.

\pagebreak

\section{A Discussion on Partial Identification}\label{AppPartialIdentification}

\setcounter{table}{0}
\renewcommand\thetable{I.\arabic{table}}

\setcounter{figure}{0}
\renewcommand\thefigure{I.\arabic{figure}}

\setcounter{equation}{0}
\renewcommand\theequation{I.\arabic{equation}}

\setcounter{theorem}{0}
\renewcommand\thetheorem{I.\arabic{theorem}}

\setcounter{proposition}{0}
\renewcommand\theproposition{I.\arabic{proposition}}

\setcounter{corollary}{0}
\renewcommand\thecorollary{I.\arabic{corollary}}

\setcounter{assumption}{0}
\renewcommand\theassumption{I.\arabic{assumption}}

\setcounter{definition}{0}
\renewcommand\thedefinition{I.\arabic{definition}}

\setcounter{Lemma}{0}
\renewcommand\theLemma{I.\arabic{Lemma}}

In this appendix, we intuitively explain the concept of partial identification and why it is a useful tool for applied researchers. For deeper reviews about partial identification, the interested reader may check the work by \cite{Tamer2010}, \cite{Manski2011}, \cite{Canay2017}, and \cite{Lewbel2018}. To understand the importance of partial identification, we first generically define a data-generating process, a target parameter and point identification. Then, we intuitively define model uncertainty and partial identification, explaining how the latter addresses the former. Lastly, we discuss sample uncertainty and confidence intervals in a context of partially identified target parameters.

A data-generating process is a statistical model. It defines the distribution of latent variables and the functions that connect latent variables to observable variables. These functions may be parametric or non-parametric. For example, in our context, the data-generating process includes the coefficients of a linear regression model, the true unobservable productivity, the economic shock in the outcome equation, and the measurement error terms in the two versions of the FAO-GAEZ proxies.

Importantly, the data-generating process includes the model assumptions through restrictions in the distribution of the latent variables. In our context, it imposes that the economic shock is uncorrelated with the true unobservable productivity and with the measurement error terms. It also imposes that the two measurement error terms are positively correlated with each other.

A target parameter is a function of the data-generating process. In our context, it is the coefficient of a linear regression and captures the effect of increasing agricultural productivity by one standard deviation.

A point-identification result shows that a target parameter is equal to a function of the distribution of observable random variables.\footnote{To fix ideas, imagine a randomized control trial with perfect compliance. The data-generating process defines the potential outcomes, the treatment assignment variable and the usual definition of the observed outcome. The target parameter is the expected value of the difference between the treated and untreated potential outcomes. In particular, this target parameter is identified by the difference between the expected observed outcome of treated individuals and the expected observed outcome of control individuals.}
Importantly, all objects here are defined at the population level. In our context, we could have derived a point-identification result by imposing that there is no measurement error. In this case, we would observe the true productivity and could point-identify its effect by regressing the outcome variable on our measure of agricultural productivity.

The point-identification approach usually requires strong restrictions on the distribution of the latent variables \citep{Manski2011}. However, researchers may be unsure about the true model of the data-generating process. Two equally reasonable researchers may disagree about the plausibility of the assumptions included in the data-generating process. They may even believe in different sets of assumptions \citep{Tamer2010}.

Partial identification tools address this type of model uncertainty. A scalar target parameter is partially identified if there is an interval that contains the target parameter and whose boundary points are functions of the distribution of observable random variables. A researcher may adopt a weaker set of assumptions, which is not sufficient for point-identification but is sufficient for partial identification. The length of this identified interval is an intuitive measure of model uncertainty. If there is less model uncertainty, the interval is shorter and may even collapse to a point.

Moreover, a different researcher may adopt an even weaker set of assumptions, which still partially identifies the target parameter. However, due to imposing even weaker assumptions, the identified set is wider. This extra length is intuitively connected with the additional model uncertainty, which is represented by weaker assumptions.

Our approach allows for both types of researchers. The first researcher (Section \ref{Sidentification}) is not willing to assume that measurement error does not exist but is willing to impose restrictions on the sign of the true effect, on the correlation between the measurement error and the outcome shock, and on the sign of the correlation between the measurement error terms. The second researcher (Appendices \ref{AppNoSign}-\ref{AppFreeCov}) is not willing to make one of these assumptions and finds a wider identified set.

Lastly, bounds in partial identification results are defined at the population level. Consequently, they do not take into account sample uncertainty, i.e., they do not consider that a finite sample is a random draw from a superpopulation. Similarly to the point-identification approach, the partial identification approach takes sample uncertainty into account when estimating the bounds of the identified set. It does so by estimating confidence intervals.

There are two types of confidence intervals in the partial identification approach. The first type of confidence interval contains the true target parameter with a known probability, where the probability is computed based on the thought experiment of an infinitely large number of samples. The second type of confidence interval contains the entire identified set with a known probability. The trade-off between these types of confidence intervals is that the second confidence interval is usually wider but easier to estimate. In our context, Section \ref{Sestimation} explains how to construct the second type of confidence interval.

\newpage

\section{Assumptions on Model Primitives}\label{AppPrimitive}

\setcounter{table}{0}
\renewcommand\thetable{J.\arabic{table}}

\setcounter{figure}{0}
\renewcommand\thefigure{J.\arabic{figure}}

\setcounter{equation}{0}
\renewcommand\theequation{J.\arabic{equation}}

\setcounter{theorem}{0}
\renewcommand\thetheorem{J.\arabic{theorem}}

\setcounter{proposition}{0}
\renewcommand\theproposition{J.\arabic{proposition}}

\setcounter{corollary}{0}
\renewcommand\thecorollary{J.\arabic{corollary}}

\setcounter{assumption}{0}
\renewcommand\theassumption{J.\arabic{assumption}}

\setcounter{definition}{0}
\renewcommand\thedefinition{J.\arabic{definition}}

\setcounter{Lemma}{0}
\renewcommand\theLemma{J.\arabic{Lemma}}

In this appendix, we describe five assumptions on the primitives of our model that imply Assumptions \ref{AsExogeneous}-\ref{AsNonClassical} in Section \ref{Sassumptions}. The primitives of our model are the random variables $Y$, $W_{1}$, \ldots, $W_{J}$, $X^{*}$, $\epsilon$, $Z_{1}$ and $Z_{2}$, and the following outcome equation:
\begin{equation}\label{EqAppPrimitiveModelCovariates}
	Y = \alpha_{0} + \sum_{j = 1}^{J} \alpha_{j} \cdot W_{j} + \beta \cdot X^{*} + \epsilon,
\end{equation}

Using the random variables in our model primitives, we define five additional random variables:
\begin{equation}\label{EqAppPrimitiveXres}
	X_{res}^{*} \coloneqq X^{*} - W \cdot \left(\mathbb{E}\left[W^{T} \cdot W\right]\right)^{-1} \cdot \mathbb{E}\left[W^{T} \cdot X^{*}\right],
\end{equation}
\begin{equation} \label{EqAppPrimitiveZres}
	Z_{k,res} \coloneqq Z_{k} - W \cdot \left(\mathbb{E}\left[W^{T} \cdot W\right]\right)^{-1} \cdot \mathbb{E}\left[W^{T} \cdot Z_{k}\right]
\end{equation}
and
\begin{equation} \label{EqAppPrimitiveU}
	U_{k} \coloneqq Z_{k,res} - X_{res}^{*}
\end{equation}
for any $k \in \left\lbrace 1, 2 \right\rbrace$, where $W = \left[1, W_{1}, \ldots, W_{J} \right]$ is the row vector of covariates including the constant term.

Our first assumption on the primitives of our model imposes that the economic shock is exogenous.
\begin{assumption}\label{AsAppPrimitiveExogeneous}
	$Cov\left(A,\epsilon\right) = 0$ for any $A \in \left\lbrace W_{1},\ldots,W_{J},X^{*} \right\rbrace$.
\end{assumption}
This assumption implies that
\begin{align*}
	\epsilon_{res} & \coloneqq \epsilon - W \cdot \left(\mathbb{E}\left[W^{T} \cdot W\right]\right)^{-1} \cdot \mathbb{E}\left[W^{T} \cdot \epsilon\right] \\
	& = \epsilon
\end{align*}
and that
\begin{align*}
	Cov\left(X_{res}^{*}, \epsilon_{res}\right) & = Cov\left(X^{*} - W \cdot \left(\mathbb{E}\left[W^{T} \cdot W\right]\right)^{-1} \cdot \mathbb{E}\left[W^{T} \cdot X^{*}\right], \epsilon\right) \\
	& = Cov\left(X^{*}, \epsilon\right) - Cov\left(W , \epsilon\right) \cdot \left(\mathbb{E}\left[W^{T} \cdot W\right]\right)^{-1} \cdot \mathbb{E}\left[W^{T} \cdot X^{*}\right] \\
	& = 0.
\end{align*}
Hence, Assumption \ref{AsAppPrimitiveExogeneous} implies Assumption \ref{AsExogeneous}.

Our second assumption on the primitives of our model is Assumption \ref{AsScale} directly because we see it as modifying the target parameter, as explained in the main text.

Our third assumption on the primitives of our model is Assumption \ref{AsPosEff} directly.

Our fourth assumption on the primitives of our model imposes that the proxy variables are exogenous.
\begin{assumption}\label{AsAppPrimitiveMEexogenous}
	$Cov\left(Z_{k}, \epsilon\right) = 0$ for any $k \in \left\lbrace 1, 2 \right\rbrace$.
\end{assumption}
Fix $k \in \left\lbrace 1, 2 \right\rbrace$ arbitrarily and note that
\begin{align*}
	Cov\left(U_{k}, \epsilon_{res}\right) & = Cov\left(U_{k}, \epsilon\right) \\
	& = Cov\left(Z_{k,res} - X_{res}^{*}, \epsilon\right) \\
	& = Cov\left(Z_{k} - W \cdot \delta_{k} - X^{*} + W \cdot \delta_{X}, \epsilon\right) \\
	& \hspace{20pt} \text{where } \delta_{k} \coloneqq \left(\mathbb{E}\left[W^{T} \cdot W\right]\right)^{-1} \cdot \mathbb{E}\left[W^{T} \cdot Z_{k}\right] \\
	& \hspace{20pt} \text{and } \delta_{X}\coloneqq \left(\mathbb{E}\left[W^{T} \cdot W\right]\right)^{-1} \cdot \mathbb{E}\left[W^{T} \cdot X^{*}\right] \\
	& = Cov\left(Z_{k} - X^{*} - W \cdot \left(\delta_{k} - \delta_{X}\right), \epsilon\right) \\
	& = Cov\left(Z_{k}, \epsilon\right) - Cov\left(X^{*}, \epsilon\right) - Cov\left(W, \epsilon\right) \cdot \left(\delta_{k} - \delta_{X}\right) \\
	= 0.
\end{align*}
Hence, Assumptions \ref{AsAppPrimitiveExogeneous} and \ref{AsAppPrimitiveMEexogenous} implies Assumption \ref{AsMEexogenous}.

Lastly, since $U_{1}$ and $U_{2}$ are defined based on model primitives, our fifth assumption is Assumption \ref{AsNonClassical} directly.

\end{document}